\documentclass[]{pasj01}

\usepackage{pdflscape}
\usepackage{booktabs}
\usepackage{graphicx}
\usepackage{cellspace}
\usepackage{url} 
\usepackage{tablefootnote}
\usepackage{xtab}
\usepackage{makecell}
\usepackage{comment}
\usepackage{longtable}

\Received{2023 November 22}
\Accepted{2024 March 3}
\SetRunningHead{Astronomical Society of Japan}{Usage of \texttt{pasj00.cls}}

\definecolor{MyDarkBlue}{rgb}{0,0.08,0.5}
\definecolor{MyDarkRed}{rgb}{0.7,0.02,0.02}
\definecolor{MyDarkmagenta}{rgb}{0.0,0.7,0.0}

\newcommand{\argmax}{\mathop{\rm argmax}\limits}

\begin{document}

\title{ALMA 2D Super-resolution Imaging of Taurus-Auriga Protoplanetary Disks: Probing Statistical Properties of Disk Substructures}
\altaffiltext{1}{Institute of Astronomy and Astrophysics, Academia Sinica, 11F of AS/NTU
Astronomy-Mathematics Building, No.1, Sec. 4, Roosevelt Rd, Taipei 10617, Taiwan}
\altaffiltext{2}{Department of Astronomy, Graduate School of Science, The University of Tokyo, 7-3-1 Hongo, Bunkyo-ku, Tokyo 113-0033, Japan}
\altaffiltext{3}{National Astronomical Observatory of Japan, 2-21-1 Osawa, Mitaka, Tokyo 181-8588, Japan}
\altaffiltext{4}{Division of Liberal Arts, Kogakuin University, 1-24-2 Nishi-Shinjuku, Shinjuku-ku, Tokyo 163-8677, Japan}
\altaffiltext{5}{Leiden Observatory, Leiden University, P.O. Box 9513, NL-2300 RA Leiden, The Netherlands}
\altaffiltext{6}{Department of Earth and Planetary Sciences, Tokyo Institute of Technology, 2-12-1 Oh-okayama, Meguro-ku, Tokyo 152-8551, Japan}
\altaffiltext{7}{Division of Systems and Information Engineering, Ashikaga University, 268-1 Omae-cho, Ashikagashi, Tochigi 326-8558, Japan}
\altaffiltext{8}{Department of Astronomical Science, School of Physical Sciences, Graduate University for Advanced Studies, 2-21-1 Osawa, Mitaka, Tokyo 181-8588, Japan}
\altaffiltext{9}{The Institute of Statistical Mathematics, 10-3 Midori-cho, Tachikawa, Tokyo 190-8562, Japan}
\altaffiltext{10}{Department of Statistical Science, School of Multidisciplinary Sciences, Graduate University for Advanced Studies, 10-3 Midori-cho, Tachikawa, Tokyo 190-8562, Japan}
\altaffiltext{11}{Astrobiology Center, National Institutes of Natural Sciences, 2-21-1 Osawa, Mitaka, Tokyo 181-8588, Japan}
\author{Masayuki \textsc{Yamaguchi}\altaffilmark{1,2,3,}$^{*}$,
        Takayuki  \textsc{Muto} \altaffilmark{4,5,6},
        Takashi  \textsc{Tsukagoshi}\altaffilmark{7},
        Hideko \textsc{Nomura} \altaffilmark{3,8},
        Naomi \textsc{Hirano} \altaffilmark{1},
        Takeshi \textsc{Nakazato} \altaffilmark{3},
        Shiro \textsc{Ikeda} \altaffilmark{9,10},
        Motohide \textsc{Tamura} \altaffilmark{2,11,3} and
        Ryohei \textsc{Kawabe}\altaffilmark{3}
        } 
\email{myamaguchi@asiaa.sinica.edu.tw}

\maketitle

\begin{abstract}
In the past decade, ALMA observations of protoplanetary disks revealed various substructures including gaps and rings. Their origin may be probed through statistical studies on the physical properties of the substructures. We present the analyses of archival ALMA Band 6 continuum data of 43 disks (39 Class II and 4 Herbig Ae) in the Taurus-Auriga region. We employ a novel 2D super-resolution imaging technique based on sparse modeling to obtain images with high fidelity and spatial resolution. As a result, we have obtained images with spatial resolutions comparable to a few au ($0''.02 - 0''.1$), which is two to three times better than conventional CLEAN methods. All dust disks are spatially resolved, with the radii ranging from 8 to 238 au with a median radius of 45 au. Half of the disks harbor clear gap structures, whose radial locations show a bimodal distribution with peaks at $\lesssim20$ au and $\gtrsim30$ au. We also see structures indicating weak gaps at all the radii in the disk. We find that the widths of these gaps increase with their depths, which is consistent with the model of planet-disk interactions. The inferred planet mass-orbital radius distribution indicates that the planet distribution is analogous to our Solar System. However, planets with Neptune mass or lower may exist in all the radii.

\end{abstract}


\section{Introduction}\label{sec:intro}

Planets are believed to be formed in protoplanetary disks around young stars (e.g., \cite{Hayashi1985, Shu1987}). Recent Atacama Large Millimeter/Submillimeter Array (ALMA) observations have revealed that many disks around Class II stars have small-scale structures such as gaps, rings, and asymmetries, through both survey programs and individual target studies \citep{Andrews2018a, Tsukagoshi2019, Hashimoto2021dmtau, Cieza2021, Orihara2023}.

Various mechanisms have been proposed as the origins of the disk substructures: gravitational disk instabilities \citep{Youdin2011, Takahashi2014}, magneto-hydrodynamics \citep{Flock2015}, planet-disk interaction \citep{Takeuchi1996, Zhu2012}, and chemical processes \citep{Zhang2015, Okuzumi2016}. The recent discovery of protoplanets in the central cavity of PDS~70 \citep{Keppler2018, Benisty2021} and AB Aur \citep{Currie2022} might indicate some cavity structures may be caused by embedded planet(s). However, we do not yet have firm evidence of what the major process is to produce substructures in protoplanetary disks.

Higher spatial resolution is crucial in capturing the substructures in disks. If the substructures are of dynamical origin, disk scale height ($\sim$ sound crossing time over one Keplerian orbit) becomes a key length scale. It is typically $10\%$ of the radial distance from the central star (see \cite{Andrews2020}, and references therein). At the distance of 140 pc (e.g., Taurus-Auriga region; \cite{Galli2018}), the observations with $\sim 0''.1$ or better resolution are required, and indeed, the observations with this level of spatial resolution have found substructures in Taurus-Auriga Region \citep{ALMAPartnership2015, Tang2017, Clarke2018, Long2018, Long2019, Ubeira2019, Facchini2020, Huang2020, Ueda2022}. Many of them are annular gaps residing at $10-100$ au from the central star. Yet, no clear statistical trends of gap properties have been reported so the origin of the substructure is controversial. The samples may still be biased towards mm-bright disks that can be easily resolved with a high signal-to-noise ratio. Furthermore, an observational framework of disk structure analyses is still being developed (e.g., \cite{Huang2018, Marel2019, Bae2022}). In this regard, increasing the sample of disks resolved with spatial resolution of $0''.1$ or better can provide us with valuable insights into the origin of disk substructures.

The increase in spatial resolution does not necessarily require new observations. Recently, we have shown that imaging using sparse modeling (SpM) can yield high-fidelity images with spatial resolution improved by a factor of $\sim 2-3$, using the data of protoplanetary disks taken by ALMA \citep{Yamaguchi2020, Yamaguchi2021}. In the analyses of the disk around T Tau, we achieved the spatial resolution of  $0''.04$, which is a factor of three higher than that of the reconstructed with conventional CLEAN, $0''.12$. Therefore, it is natural to extend the analyses to data of other disks with the nominal CLEAN beam size larger than $0''.1-0''.2$. In this paper, we apply SpM to the archival ALMA Band 6 continuum data of 43 young stellar objects (39 Class II objects and four Herbig stars) in the Taurus-Auriga region (age of $\sim 1-3$ Myr; \cite{Luhman2023}, a distance of $\sim 140$ pc; \cite{Galli2018}). With the SpM technique, we obtain higher spatial resolution images ($0''.02-0''.1$) improved by a factor of $2-3$. We compare the images obtained by SpM and CLEAN and then search for substructures in high-resolution images obtained with SpM. We also show statistical analyses of quantities that characterize the substructures.

This paper is constructed as follows. We describe sample selection in Section \ref{sec:source_selection}.  Section \ref{sec:datadeduction_imaging} outlines the data reduction processes and imaging methods, including both CLEAN and SpM.  We address the quality of SpM images in Section \ref{sec:images} and discuss the total flux values of ALMA observations in Section \ref{sec:totalflux}. We define the categorization of disk substructures in Section \ref{sec:categorize_ra} and then discuss the statistical properties of gaps in Sections \ref{sec:radialdistribution_gap} and \ref{sec:corr_disksub}. Section \ref{sec:discussion} is devoted to a discussion of the origin of structures and Section \ref{sec:conclusion} is for conclusions.

\section{Source Selection} \label{sec:source_selection}

\begin{figure*}[t]
\includegraphics[width=0.98\textwidth]{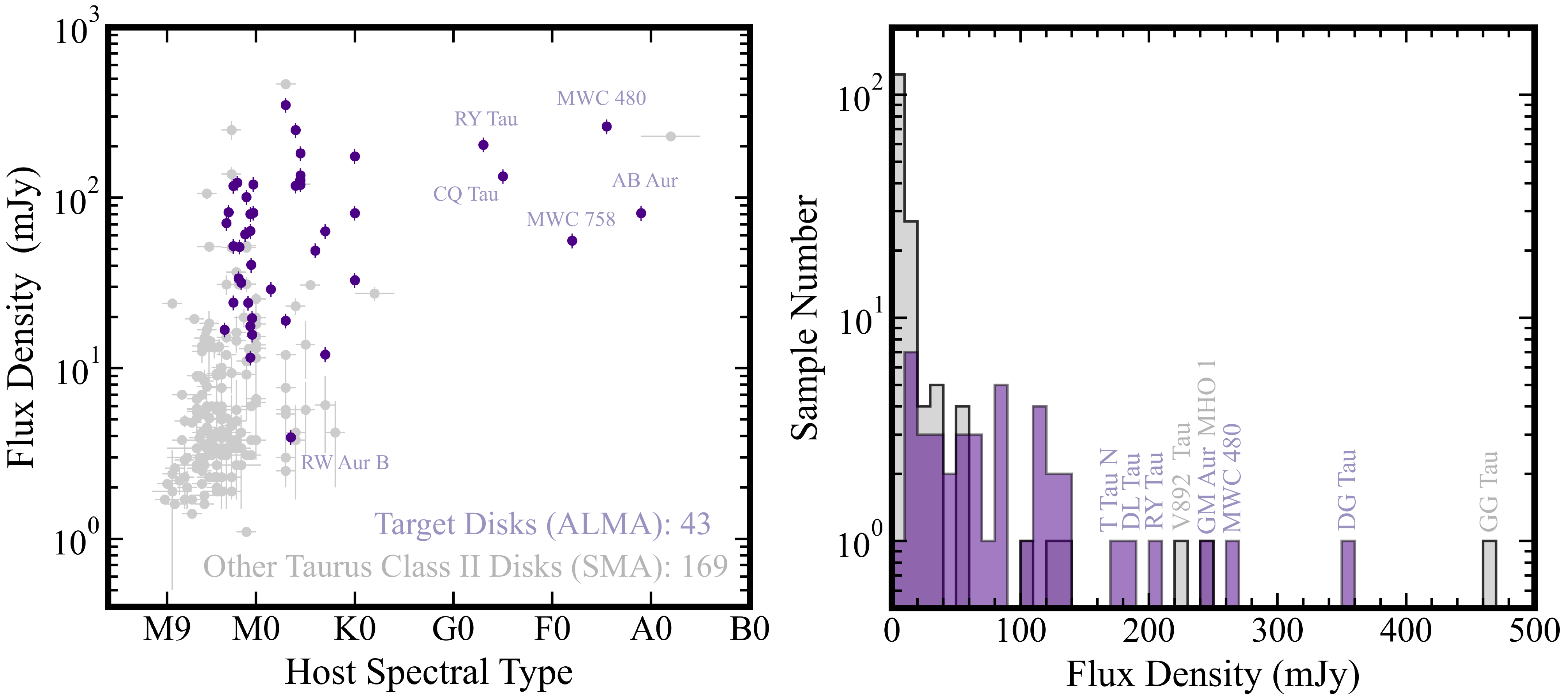}
\caption{Left: Flux density at 1.1-1.3~mm for Class II Taurus-Auriga disks versus the spectral types of their central stars. Right: histogram of the flux density for targets in the left panel. The targets of this study are shown in purple and other Taurus-Auriga disks are in gray. The flux density of our targets is measured from SpM images while that of other targets are from the SMA observations \citep{Andrews2013} are also given.}
\label{fig:disk_spt_vs_flux}
\end{figure*}

We first describe the source selection in this study. We select targets from those in the Taurus-Auriga region that are identified as Class II YSOs or Herbig Ae stars \citep{Rebull2010, Luhman2010, Vioque2018} with existing ALMA Band 6 data at spatial resolutions higher than $0''.2$. We further impose the following criteria:

\begin{enumerate}

\item The spectral type of M3 to A1 \citep{Mora2001, Herczeg2014}: Stars with $\rm M9-M4$ are excluded because the majority of them are binary disks having significantly faint flux densities (a few mJy) at 1.3 mm \citep{Andrews2013} and have not been observed with ALMA.

\item Single stars or the primary stars of binary/multiple systems with separation larger than $0''.4$ ($\sim$50~au): With the latter condition, the tidal interaction effect from the companion (e.g., \cite{Ichikawa2021}) may not be very significant on the size and the structures of the disk.

\item The ALMA data should have Signal-to-Noise Ratios (SNRs) better than 30 in the CLEAN image domain: This is because the data lower than this SNR made it difficult to reconstruct a high-fidelity image \footnote[1]{``Fidelity'' is a measure of how faithfully the brightness distribution on a reconstructed image matches an expected astronomical object. Based on previous ALMA high-sensitivity observations of disk structures, we regard the high-fidelity image of the disks to be that having smooth axisymmetric (e.g., circular rings and gaps) or non-axisymmetric (e.g., spirals or local blobs larger than the spatial resolution) structures. Images having random structures such as many small dot-like structures are considered as low fidelity.} with the SpM imaging in our analysis.

\end{enumerate}

In addition, we have added two circumbinary disk systems around spectroscopic binaries (UZ Tau E and DQ Tau) and one companion star of a binary system (RW Aur B). The former is included because the binary separation is very close (orbital period less than ten days; \cite{Jensen2007, Czekala2016}) so the binarity should not affect the outer disk structures significantly. The latter is included because there is a substantial separation from the primary star with a distance of $\sim 1''.5$ ($\sim 245$ au) and we find the presence of substructures within the disk.

Consequently, we have in total 43 YSOs that comprise 29 single stars, 12 stars within binary or multiple systems with large ($> 50$ au) separation, and two spectroscopic binaries (see Table \ref{tab:stellar_properties} for a complete list). This corresponds to $20\%$ in Taurus-Auriga surveyed by the Submillimeter Array (SMA) \citep{Andrews2013}. The total flux density at Band 6 (1.1-1.3 mm) spans the wide range of $10-300$ mJy. Figure \ref{fig:disk_spt_vs_flux} compares the total sub-mm flux density of our sample and the entire objects in \cite{Andrews2013}. In terms of stellar mass, our sample comprises 38 low-mass stars ($M_{*} \simeq 0.3-1.5~M_{\odot}$, M-K type stars), and five intermediate-mass stars ($M_{*} \simeq 1.5-2.5~M_{\odot}$, F-A types), of which four are Herbig Ae stars \citep{Vioque2018}.

\begin{table*}
\tbl{Host Stellar and Disk Properties.}{
\begin{tabular}{l@{\hspace{0.5cm}}c@{\hspace{0.5cm}}l@{\hspace{0.5cm}}c@{\hspace{0.5cm}}l@{\hspace{0.5cm}}l@{\hspace{0.5cm}}l@{\hspace{0.5cm}}l@{\hspace{0.5cm}}l@{\hspace{0.1cm}}c@{\hspace{0.1cm}}c}
\hline
Name &Dist & Stellar & Spectral  & $M_{*}$ & $L_{*}$ & $r_{\rm disk}$ & Disk Inc & Disk P.A. & Method & Ref \\
 &  (pc) &  Property & Type &  ($M_{\odot}$) & ($L_{\odot}$) & (au) & (deg) & (deg) & & \\
 (1) & (2) & (3) & (4) & (5) & (6) & (7) & (8) & (9) & (10) & (11)\\
\hline
AB Aur & 162.9 & S & A1.0 & $2.15^{+0.36}_{-0.21}$ & 40.7 & $238\pm5$ & $\sim 23$ & $\sim 36$ & R & a, b, a, a \\
GM Aur & 159.6 & S & K6.0 & $1.14^{+0.02}_{-0.02}$ & 0.6 & $212\pm3$ & $53.7\pm0.1$ & $57.8\pm0.1$ & E & c, b, d, b \\
CI Tau & 158.7 & S & K5.5 & $0.90^{+0.02}_{-0.02}$ & 0.8 & $181\pm3$ & $49.2\pm0.4$ & $13.1\pm0.5$ & E & c, b, d, b \\
DL Tau & 159.3 & S & K5.5 & $1.04^{+0.02}_{-0.02}$ & 0.6 & $161\pm10$ & $44.6\pm0.4$ & $51.7\pm1.0$ & E & c, b, d, b \\
DM Tau & 145.1 & S & M3.0 & $0.55^{+0.02}_{-0.02}$ & 0.1 & $154\pm3$ & $35.7\pm0.1$ & $-21.7\pm0.2$ & E & c, b, d, b \\
LkCa 15 & 158.9 & S & K5.5 & $1.14^{+0.03}_{-0.03}$ & 0.8 & $151\pm3$ & $50.6\pm0.1$ & $61.8\pm0.3$ & E & c, b, d, b \\
AA Tau & 137.2 & S & M0.6 & $0.84^{+0.04}_{-0.04}$ & 0.4 & $139\pm3$ & $59.8\pm0.1$ & $94.5\pm0.4$ & E & c, b, d, b \\
GO Tau & 144.6 & S & M2.3 & $0.45^{+0.01}_{-0.01}$ & 0.2 & $137\pm9$ & $52.7\pm0.7$ & $24.1\pm1.2$ & E & c, b, d, b \\
Haro 6-37 C & 195.7 &T& M1.0 & $0.45^{+0.12}_{-0.08}$ & 0.6 & $133\pm14$ & $21.5\pm0.1$ & $18.6\pm0.4$ & E & c, e, e, e \\
MWC 480 & 161.8 & S & A4.5 & $2.16^{+0.22}_{-0.22}$ & 18.6 & $110\pm6$ & $36.9\pm0.1$ & $-31.5\pm0.3$ & E & c, b, f, g \\
MWC 758 & 160.2 & S & A8.0 & $1.56^{+0.11}_{-0.08}$ & 11.0 & $103\pm5$ & $\sim 21$ & $\sim 62$ & R & a, h, a, a \\
IQ Tau & 131.3 & S & M1.1 & $0.42^{+0.05}_{-0.15}$ & 0.2 & $95\pm8$ & $60.8\pm0.1$ & $42.2\pm0.1$ & G & c, b, f, b \\
CQ Tau & 163.1 & S & F5.0 & $1.51^{+0.07}_{-0.07}$ & 7.4 & $86\pm7$ & $\sim 35$ & $\sim 55$ & R & a, k, d, a \\
UZ Tau E & 131.2 & SB+B & M1.9 & $1.21^{+0.12}_{-0.12}$ & 0.4 & $85\pm5$ & $55.6\pm0.2$ & $90.4\pm0.4$ & E & c, b, f, b \\
DS Tau & 159.1 & S & M0.4 & $1.08^{+0.11}_{-0.11}$ & 0.2 & $68\pm8$ & $63.1\pm0.2$ & $-19.4\pm0.5$ & E & c, b, f, b \\
RY Tau & 128.2 & S & F7.0 & $2.04^{+0.30}_{-0.26}$ & 12.3 & $65\pm3$ & $65.4\pm0.1$ & $23.5\pm0.2$ & E & c, b, g, g \\
CY Tau & 128.9 & S & M2.3 & $0.30^{+0.02}_{-0.02}$ & 0.2 & $64\pm13$ & $27.3\pm0.1$ & $156.9\pm0.1$ & G & c, b, d, b \\
CIDA 9 A & 171.9 & B & M1.8 & $0.78^{+0.08}_{-0.08}$ & 0.2 & $64\pm9$ & $45.1\pm0.3$ & $104.6\pm0.8$ & E & c, b, f, b \\
DN Tau & 128.2 & S & M0.3 & $0.87^{+0.15}_{-0.15}$ & 0.7 & $59\pm8$ & $35.6\pm0.7$ & $86.4\pm2.2$ & E & c, b, d, b \\
DG Tau & 121.2 & S & K7.0 & $\sim 0.7$ & 0.4 & $55\pm1$ & $33.5\pm0.1$ & $135.6\pm0.1$ & G & c, b, i, b \\
DR Tau & 195.7 & S & K6.0 & $1.18^{+0.59}_{-0.44}$ & 0.6 & $51\pm6$ & $8.0\pm0.2$ & $18.2\pm1.8$ & G & c, b, f, b \\
UX Tau A & 139.9 &T& K0.0 & $1.30^{+0.30}_{-0.40}$ & 1.6 & $45\pm7$ & $42.7\pm0.3$ & $-13.8\pm0.9$ & E & c, b, j, b \\
CW Tau & 132.4 & S & K3.0 & $0.64^{+0.01}_{-0.01}$ & 0.5 & $44\pm4$ & $58.3\pm0.1$ & $61.7\pm0.1$ & G & c, b, d, b \\
FT Tau & 127.8 & S & M2.8 & $0.35^{+0.04}_{-0.06}$ & 0.2 & $43\pm5$ & $34.2\pm0.2$ & $124.7\pm0.5$ & E & c, b, f, b \\
V710 Tau A & 142.9 & B & M1.7 & $0.58^{+0.06}_{-0.07}$ & 0.3 & $43\pm6$ & $48.8\pm0.1$ & $84.2\pm0.1$ & G & c, b, f, b \\
V409 Tau & 131.4 & S & M0.6 & $0.50^{+0.13}_{-0.10}$ & 0.7 & $42\pm8$ & $67.8\pm0.1$ & $46.2\pm0.1$ & E & c, b, g, b \\
HO Tau & 161.4 & S & M3.2 & $0.44^{+0.05}_{-0.16}$ & 0.1 & $40\pm8$ & $53.6\pm0.1$ & $117.1\pm0.1$ & G & c, b, f, b \\
DQ Tau & 197.7 & SB & M0.6 & $2.85^{+0.77}_{-0.72}$ & 1.2 & $40\pm6$ & $19.6\pm0.1$ & $3.9\pm0.3$ & G & c, b, f, b \\
BP Tau & 129.1 & S & M0.5 & $1.10^{+0.04}_{-0.04}$ & 0.4 & $37\pm6$ & $38.2\pm0.1$ & $151.8\pm0.1$ & E & c, b, d, b \\
IP Tau & 130.6 & S & M0.6 & $0.80^{+0.23}_{-0.22}$ & 0.3 & $35\pm7$ & $45.0\pm0.2$ & $-7.8\pm0.6$ & E & c, b, f, b \\
DO Tau & 139.4 & S & M0.3 & $0.54^{+0.07}_{-0.07}$ & 0.2 & $35\pm6$ & $27.3\pm0.1$ & $170.0\pm0.1$ & E & c, b, f, b \\
Haro 6-13 & 130.5 & S & K5.5 & $0.93^{+0.14}_{-0.14}$ & 0.8 & $34\pm5$ & $42.3\pm0.1$ & $151.7\pm0.1$ & G & c, b, d, b \\
V836 Tau & 169.6 & S & M0.8 & $0.92^{+0.22}_{-0.20}$ & 0.4 & $31\pm7$ & $43.0\pm0.1$ & $117.4\pm0.1$ & E & c, b, f, b \\
HK Tau A & 133.3 & B & M1.5 & $0.49^{+0.06}_{-0.05}$ & 0.3 & $28\pm5$ & $53.8\pm0.1$ & $175.0\pm0.1$ & G & c, b, f, b \\
HP Tau & 177.1 &T& K4.0 & $1.20^{+1.14}_{-0.18}$ & 1.3 & $23\pm5$ & $20.3\pm0.1$ & $53.6\pm0.2$ & G & c, b, g, b \\
GI Tau & 130.5 & B & M0.4 & $0.52^{+0.15}_{-0.12}$ & 0.5 & $22\pm5$ & $43.8\pm0.1$ & $142.2\pm0.1$ & E & c, b, g, b \\
RW Aur A & 163.5 & $\rm B(?)^{*1}$ & K0.0 & $1.40^{+0.28}_{-0.14}$ & 1.0 & $21\pm2$ & $55.2\pm0.1$ & $38.6\pm0.1$ & G & c, b, f, b \\
T Tau N & 143.7 &T& K0.0 & $2.06^{+0.66}_{-0.43}$ & 6.8 & $20\pm3$ & $25.6\pm1.2$ & $91.0\pm3.5$ & E & c, b, f, b \\
DH Tau A & 135.4 & B & M2.3 & $0.37^{+0.13}_{-0.10}$ & 0.2 & $19\pm5$ & $14.1\pm0.1$ & $22.5\pm0.3$ & G & c, b, g, b \\
HN Tau A & 136.6 & B & K3.0 & $1.53^{+0.15}_{-0.15}$ & 0.2 & $16\pm4$ & $64.5\pm0.1$ & $83.4\pm0.1$ & G & c, b, d, b \\
DK Tau A & 128.5 & B & K8.5 & $0.55^{+0.13}_{-0.13}$ & 0.4 & $15\pm4$ & $17.1\pm0.1$ & $176.2\pm0.4$ & G & c, b, d, b \\
RW Aur B & 163.5 & $\rm B(?)^{*1}$ & K6.5 & $0.86^{+0.11}_{-0.10}$ & 0.6 & $15\pm2$ & $64.6\pm0.1$ & $43.4\pm0.4$ & E & c, b, j, b \\
UY Aur A & 155.6 & $\rm B(?)^{*2}$ & K7.0 & $0.65^{+0.17}_{-0.13}$ & 1.0 & $8\pm5$ & $28.6\pm0.2$ & $127.4\pm0.5$ & G & c, b, g, b \\
\hline
\end{tabular}
}
\begin{tabnote}
\textbf{Note.} Column Description: (1) Name of the host star. (2) Distance adopted from $Gaia$ DR2 parallax \citep{Gaia2016, Gaia2018}. (3) Stellar property; ``S'', ``B'', ``SB'', and ``T'' indicate single, binary, spectroscopic binary, and triple, respectively. (4) Spectral type. (5) Stellar mass. We use dynamical stellar masses. For stars not measured dynamical masses, we use stellar masses derived from an evolutionary track model. (6) Stellar luminosity. (7) Dust disk radius measured with curve growth method (see Appendix \ref{appendix:dust_disk_size}). (8) Disk inclination (Inc). $0^{\circ}$ is face-on, and $90^{\circ}$ is edge-on. The measurements without uncertainties are marked with ``$\sim$''. (9) Position Angle (P.A.) of a disk (east of north). (10) Method used to deproject the disk structure: ``E'' indicates the ellipse-fitting for a ring structure, and  ``G'' indicates the 2D Gaussian fitting for encircling a disk. ``R'' indicates the use of references taken from Keplerian rotation with molecular line observations; AB Aur \citep{Tang2017}, MWC 758 \citep{Isella2010mwc758, Boehler2018}, and CQ Tau \citep{Ubeira2019}. (11) References order: Stellar property, spectral type, stellar mass, and luminosity. (a) \cite{Vioque2018}, (b) \cite{Herczeg2014}, (c) \cite{Kenyon2008}, (d) \cite{Simon2019}, (e) \cite{Akeson2019}, (f) \cite{Braun2021}, (g) \cite{Long2019}, (h) \cite{Vieira2003}, (i) \cite{Podio2013}, (j) \cite{Kraus2009}, (k) \cite{Mora2001}. \footnotemark[$*1$]: the RW Aur system may be a triple star (as RW Aur A may be a spectroscopic binary; \cite{Gahm1999, Petrov2001}). \footnotemark[$*2$]: the UY Aur system may be a triple system (as UY Aur A may be a close binary; \cite{Tang2014}).
\end{tabnote}
\label{tab:stellar_properties}
\end{table*}

\section{Data Reduction and Imaging}\label{sec:datadeduction_imaging}

In this Section, we describe methods of data reduction and imaging. We produce CLEAN and SpM images for all the objects in the same manner as detailed below. The complete list of datasets is presented in Table \ref{tab:summary_obs}.

\setlength\tabcolsep{0.04ex}
\begin{table*}[t]
\caption{Summary of ALMA Observations.}
\renewcommand{\arraystretch}{1.1}
\centering
\begin{tabular}[t]{l@{\hspace{0.00cm}}c@{\hspace{0.00cm}}c@{\hspace{0.00cm}}c@{\hspace{0.0cm}}c@{\hspace{0.0cm}}c}
\hline
Name & Freq & BL &MRS & OST  &Project \\
& (GHz) & (m) & (arcsec) & (min) & IDs \\
(1) & (2) & (3) & (4) & (5) & (6)\\
\hline
AB Aur&226& $15-16196$ &5.1&107.2&2015.1.00889.S \\ &&&&&  2019.1.00579.S \\
GM Aur&261& $15-14851$ &2.1&117.4&2017.1.01151.S \\ &&&&&  2018.1.01230.S \\
CI Tau&230& $15-12145$ &4.4&104.2&2016.1.01164.S \\ &&&&&  2016.1.01370.S \\ &&&&&  2018.1.01631.S \\
DL Tau&226& $17-3638$ &2.4&13.2&2015.1.01207.S \\ &&&&&  2016.1.01164.S \\
DM Tau&224& $15-15238$ &1.9&143.2&2013.1.00498.S \\ &&&&&   2018.1.01755.S \\
LkCa 15&224& $15-13894$ &3.0&88.4&2018.1.01255.S \\ &&&&&  2018.1.00945.S \\
AA Tau&241& $41-16196$ &1.0&203.5&2013.1.01070.S \\ &&&&&  2016.1.01205.S \\ &&&&&  2018.1.01829.S \\
GO Tau&226& $17-3638$ &2.2&13.3&2015.1.01207.S \\ &&&&&  2016.1.01164.S \\
Haro 6-37 C&238& $41-2125$ &2.0&1.8&2013.1.00105.S \\
MWC 480&225& $19-3638$ &2.4&42.6&2016.1.00724.S \\ &&&&&  2016.1.01164.S \\
MWC 758&226& $15-16196$ &2.2&361.7&2017.1.00940.S \\
IQ Tau&225& $21-3638$ &1.9&1.4&2016.1.01164.S \\
CQ Tau&225& $21-8548$ &1.3&73.8&2016.A.00026.S \\ &&&&&  2017.1.01404.S \\
UZ Tau E&225& $21-3697$ &1.6&8.5&2016.1.01164.S \\
DS Tau&225& $21-3697$ &1.7&9.7&2016.1.01164.S \\
RY Tau&225& $21-16196$ &1.3&8.7&2016.1.01164.S \\ &&&&&  2017.1.01460.S \\
\hline
  \end{tabular}
  \hfill
  \begin{tabular}[t]{l@{\hspace{0.00cm}}c@{\hspace{0.00cm}}c@{\hspace{0.00cm}}c@{\hspace{0.0cm}}c@{\hspace{0.0cm}}c}
    \hline
Name & Freq & BL &MRS & OST  &Project \\
& (GHz) & (m) & (arcsec) & (min) & IDs \\
(1) & (2) & (3) & (4) & (5) & (6)\\
\hline
CY Tau&225& $43-2270$ &2.0&7.7&2013.1.00498.S \\
CIDA 9 A&225& $21-3697$ &1.6&8.5&2016.1.01164.S \\
DN Tau&226& $15-3638$ &2.8&12.8&2015.1.01207.S \\ &&&&&  2016.1.01164.S \\
DG Tau&233& $17-16196$ &1.2&329.5&2015.1.01268.S \\ &&&&&  2016.1.00846.S \\
DR Tau&225& $21-3638$ &1.6&7.7&2016.1.01164.S \\
UX Tau A&238& $41-2125$ &2.0&1.8&2013.1.00105.S \\
CW Tau&233& $70-8283$ &1.1&91.3&2019.1.01108.S \\
FT Tau&225& $21-3697$ &1.6&8.5&2016.1.01164.S \\
V710 Tau A&225& $21-3638$ &1.7&8.2&2016.1.01164.S \\
V409 Tau&225& $21-3697$ &1.6&8.5&2016.1.01164.S \\
HO Tau&225& $21-3697$ &1.6&8.4&2016.1.01164.S \\
DQ Tau&225& $21-3638$ &1.6&7.7&2016.1.01164.S \\
BP Tau&225& $21-3697$ &1.6&10.1&2016.1.01164.S \\
IP Tau&225& $21-3697$ &1.6&10.1&2016.1.01164.S \\
DO Tau&225& $21-3638$ &1.8&3.9&2016.1.01164.S \\
Haro 6-13&225& $21-3638$ &1.7&4.0&2016.1.01164.S \\
V836 Tau&225& $21-3697$ &1.6&8.5&2016.1.01164.S \\
HK Tau A&225& $21-3697$ &1.6&8.4&2016.1.01164.S \\
HP Tau&225& $21-3638$ &1.7&4.0&2016.1.01164.S \\
GI Tau&225& $21-3697$ &1.6&8.5&2016.1.01164.S \\
RW Aur&225& $21-16196$ &1.2&136.7&2016.1.01164.S \\ &&&&& 2017.1.01631.S \\
T Tau N&225& $21-3638$ &1.7&8.2&2016.1.01164.S \\
DH Tau A&225& $21-3697$ &1.6&9.9&2016.1.01164.S \\
HN Tau A&225& $21-3638$ &1.7&8.2&2016.1.01164.S \\
DK Tau A&225& $21-3697$ &1.6&8.4&2016.1.01164.S \\
UY Aur A&225& $21-3697$ &1.8&9.6&2016.1.01164.S \\
\hline
  \end{tabular}
\begin{tabnote}
\textbf{Note.} Column Description: (1) Target name. (2) Observing central frequency in GHz. All target sources are taken from Band 6 observations. (3) Range of baseline lengths (BL) in meters from minimum to maximum lengths. (4) Maximum recoverable scale (MRS) in arc-second of data sets. The calculation is based on the definition of the ALMA technical handbook. (5) Total on-source time (OST) in minutes. (6) ALMA project IDs.
\end{tabnote}
\label{tab:summary_obs}
\end{table*}


\subsection{Data Reduction and CLEAN Imaging}\label{sec:clean_imagigng}

We used the Common Astronomy Software Applications package ($\tt CASA$; \cite{CASA2022}) for the calibration. The data were initially calibrated with the $\tt CASA$ pipeline reduction scripts provided by the ALMA Regional Centers. Given that the data were obtained over a period of several years, different versions of $\tt CASA$ were used for these calibrations. We check the maximum recoverable scale (MRS) given by $=0.983~\lambda/D_{5}$ for long baseline data, where $\lambda$ is the observing wavelength and $D_{5}$ is the 5th percentile of $uv$-distance (see ALMA Technical Handbook in \cite{Cortes2023}). If MRS is smaller than the radii of the known gap and ring features of the objects, we combine the data of the compact array configuration.

When combining the data, sets from different observations, the position offset between the phase center and the target source was adjusted as follows. We use the CLEAN images of each dataset and determine the phase center as the local peak of the emission around the center of the image.  For disks with ring-like structures (so that there is no local peak), we determine the phase center as the center of the ellipse structure. After the phase center for each dataset is shifted by the $\tt CASA$ task $\tt fixvis$, the datasets from different observations are assigned to a common phase center using task $\tt fixplanets$. After the coordinates of each dataset were corrected, the visibility of long baseline data was re-scaled using the DSHARP $\tt rescale \_ flux$ function\footnote[2]{$\tt rescale \_ flux$ function is publicly available at \url{https://almascience.eso.org/almadata/lp/DSHARP/scripts/reduction_utils.py}} \citep{Andrews2018a}.

We then applied self-calibration to improve SNR \citep{Richards2022} using a CLEAN model as a reference calibrator in the $\tt CASA$ version 6.1.0 environment. The CLEAN model was generated through the $\tt tclean$ task (hereafter referred to as CLEAN). We consistently employed multi-frequency synthesis ($\tt nterm = 1$; \cite{Rau2011}) and applied Briggs weighting with $\tt robust = 0.5$. Here, we used different CLEAN algorithms based on the disk structure: the Cotton–Schwab algorithm \citep{Schwab1984} for disks lacking substructures, such as gaps or rings, and a multi-scale approach \citep{Cornwell2008} with scale parameters of $[0, 1, 3] \theta_{\rm cl}$ (where $\theta_{\rm cl}$ represents the CLEAN beam size) for disks with substructures.

For the data with an initial SNR higher than 100 on the CLEAN image, we conducted two rounds of phase calibration ($\tt calmode = p$) with different integration times: $\tt solint = inf$ and $\tt solint =$ OST/5, where OST indicates on-source time We then performed one round of amplitude and phase calibration ($\tt calmode = ap$) with an integration time of $\tt solint $=OST/5. For data with initial SNR less than 100, we applied one round of phase calibration with an integration time of $\tt solint = inf$, followed by one round of amplitude and phase self-calibration with an integration time of $\tt solint =$ OST/5. We choose the image with the best SNR among those obtained in the course of self-calibration processes as the final CLEAN image.

After self-calibration, we have seen an improvement of SNR with a factor of $\sim 1.5$ (for data with the maximum baseline of $\sim 10~\rm M\lambda$) to several (for data with the maximum baseline of $\sim 2~\rm M\lambda$). We see that the self-calibration process mitigated patchy artifacts outside the region of source emission. The final CLEAN beam size $\theta_{\rm cl}$, peak intensity, and RMS noise level $\sigma_{\rm cl}$ (collected noise value from the dust emission-free area) are listed in Table \ref{tab:fullsample}.

\begin{table*}
\tbl{Image Properties of CLEAN and SpM}{
\begin{tabular}{l@{\hspace{2mm}}l@{\hspace{2mm}}l@{\hspace{2mm}}lll@{\hspace{2mm}}l@{\hspace{2mm}}lc}
\hline
Name & $\theta_{\rm cl}$  & $\theta_{\rm eff}$ & $\sigma_{\rm cl}$ &  $I_{\rm DT}$ &  Peak $I_{\nu}$  & Peak $I_{\nu}$ & $F_{\nu}$ & $ \log (\Lambda_l, \Lambda_{tsv})$ \\
  & (CLEAN) & (SpM)& (CLEAN)  & (SpM) & (CLEAN) &  (SpM) & (SpM) & (SpM) \\
  & (mas, PA) & (mas, PA) & ($\rm \mu Jy~bm^{-1}$)  &  ($\rm mJy~asec^{-2}$) & ($\rm m Jy~bm^{-1}$)   & ($\rm mJy~asec^{-2}$) & (mJy) &  \\
  &           &           & ($\rm mJy~asec^{-2}$)   &                        & ($\rm mJy~asec^{-2}$)  &                       &       &  \\
 (1) & (2) & (3) & (4) & (5) & (6) & (7) & (8) & (9) \\
 \hline
AB Aur & $49\times 24~(22.3^{\circ}) $ & $60\times 40~(18^{\circ}) $ & 12~(8.81) & 14.44 & 0.99~(727.85) & 314.03 & 81.2$\pm$8.1 & 4, 15 \\
GM Aur & $52\times 30~(3.3^{\circ}) $ & $30\times 20~(3^{\circ}) $ & 15~(8.84) & 21.78 & 1.10~(634.42) & 677.76 & 249.7$\pm$25.0 & 5, 14 \\
CI Tau & $58\times 42~(26.2^{\circ}) $ & $40\times 30~(16^{\circ}) $ & 10~(3.54) & 25.11 & 2.52~(925.85) & 1338.19 & 119.4$\pm$11.9 & 5, 14 \\
DL Tau & $151\times 119~(-1.5^{\circ}) $ & $100\times 80~(-1^{\circ}) $ & 46~(2.25) & 19.71 & 14.3~(704.18) & 1108.53 & 182.3$\pm$18.2 & 4, 12 \\
DM Tau & $34\times 21~(34.4^{\circ}) $ & $30\times 20~(32^{\circ}) $ & 10~(12.76) & 72.19 & 0.34~(433.11) & 486.91 & 104.9$\pm$10.5 & 5, 14 \\
LkCa 15 & $54\times 42~(7.9^{\circ}) $ & $40\times 30~(4^{\circ}) $ & 16~(6.41) & 31.88 & 0.64~(250.83) & 275.40 & 126.7$\pm$12.7 & 5, 14 \\
AA Tau & $36\times 19~(15.5^{\circ}) $ & $30\times 20~(8^{\circ}) $ & 10~(13.0) & 27.98 & 0.84~(1059.78) & 1366.86 & 80.3$\pm$8.0 & 5, 14 \\
GO Tau & $156\times 118~(-12.6^{\circ}) $ & $100\times 80~(-19^{\circ}) $ & 37~(1.78) & 10.19 & 8.21~(393.87) & 755.92 & 52.1$\pm$5.2 & 5, 12 \\
Haro 6-37 C & $184\times 159~(7.8^{\circ}) $ & $110\times 100~(-3^{\circ}) $ & 106~(3.19) & 31.02 & 18.72~(561.90) & 1072.36 & 100.9$\pm$10.1 & 4, 10 \\
MWC 480 & $223\times 155~(19.3^{\circ}) $ & $80\times 60~(26^{\circ}) $ & 85~(2.17) & 55.70 & 43.92~(1118.17) & 2080.21 & 262.0$\pm$26.2 & 5, 11 \\
MWC 758 & $40\times 27~(-4.9^{\circ}) $ & $40\times 30~(5^{\circ}) $ & 8~(6.43) & 9.83 & 0.36~(296.24) & 299.08 & 56.0$\pm$3.2 & 5, 15 \\
IQ Tau & $155\times 104~(-27.3^{\circ}) $ & $110\times 90~(-27^{\circ}) $ & 77~(4.23) & 24.68 & 5.24~(287.96) & 348.60 & 61.0$\pm$6.1 & 4, 12 \\
CQ Tau & $82\times 63~(3.1^{\circ}) $ & $60\times 50~(11^{\circ}) $ & 15~(2.58) & 18.0 & 2.3~(389.41) & 432.15 & 133.5$\pm$13.4 & 4, 14 \\
UZ Tau E & $133\times 123~(3.1^{\circ}) $ & $60\times 60~(-11^{\circ}) $ & 49~(2.64) & 30.60 & 9.63~(520.10) & 786.21 & 122.7$\pm$12.3 & 5, 11 \\
DS Tau & $157\times 105~(-18.0^{\circ}) $ & $100\times 70~(-21^{\circ}) $ & 38~(2.03) & 19.60 & 2.94~(156.84) & 327.87 & 19.7$\pm$2.0 & 5, 12 \\
RY Tau & $51\times 30~(20.5^{\circ}) $ & $30\times 20~(22^{\circ}) $ & 38~(22.16) & 117.26 & 2.75~(1595.05) & 1864.24 & 204.6$\pm$20.5 & 5, 13 \\
CY Tau & $236\times 162~(-0.7^{\circ}) $ & $170\times 130~(-2^{\circ}) $ & 74~(1.7) & 16.06 & 21.32~(492.19) & 749.71 & 117.2$\pm$11.7 & 4, 11 \\
CIDA 9 A & $128\times 98~(4.6^{\circ}) $ & $90\times 80~(-11^{\circ}) $ & 43~(3.0) & 21.70 & 2.59~(182.50) & 259.03 & 33.6$\pm$3.4 & 5, 12 \\
DN Tau & $155\times 126~(2.0^{\circ}) $ & $100\times 80~(-1^{\circ}) $ & 44~(2.0) & 14.33 & 14.61~(662.04) & 1231.69 & 81.6$\pm$8.2 & 5, 12 \\
DG Tau & $36\times 27~(-5.7^{\circ}) $ & $20\times 20~(-2^{\circ}) $ & 12~(11.31) & 124.65 & 6.42~(5877.71) & 8773.88 & 350.5$\pm$35.0 & 6, 13 \\
DR Tau & $131\times 99~(43.7^{\circ}) $ & $50\times 40~(50^{\circ}) $ & 47~(3.20) & 95.59 & 20.03~(1371.26) & 2616.28 & 117.9$\pm$11.8 & 5, 10 \\
UX Tau A & $181\times 158~(3.1^{\circ}) $ & $90\times 70~(-4^{\circ}) $ & 111~(3.43) & 84.69 & 10.93~(337.74) & 743.42 & 81.4$\pm$8.1 & 4, 9 \\
CW Tau & $90\times 60~(20.8^{\circ}) $ & $50\times 40~(10^{\circ}) $ & 16~(2.56) & 34.04 & 6.99~(1136.36) & 1753.04 & 63.7$\pm$6.4 & 5, 13 \\
FT Tau & $149\times 129~(-17.8^{\circ}) $ & $70\times 60~(22^{\circ}) $ & 56~(2.55) & 28.47 & 13.46~(617.47) & 1093.16 & 82.3$\pm$8.2 & 5, 11 \\
V710 Tau A & $163\times 132~(44.5^{\circ}) $ & $80\times 60~(41^{\circ}) $ & 52~(2.15) & 40.46 & 10.26~(423.22) & 723.06 & 51.5$\pm$5.2 & 5, 11 \\
V409 Tau & $136\times 116~(-2.4^{\circ}) $ & $100\times 80~(-1^{\circ}) $ & 44~(2.44) & 24.68 & 4.47~(249.17) & 372.60 & 17.7$\pm$1.8 & 5, 12 \\
HO Tau & $149\times 130~(-56.1^{\circ}) $ & $90\times 80~(-58^{\circ}) $ & 51~(2.33) & 15.08 & 5.37~(244.83) & 388.64 & 16.8$\pm$1.7 & 5, 12 \\
DQ Tau & $131\times 98~(42.0^{\circ}) $ & $50\times 40~(32^{\circ}) $ & 46~(3.18) & 115.62 & 21.68~(1494.48) & 3370.42 & 63.9$\pm$6.4 & 5, 10 \\
BP Tau & $145\times 107~(-22.2^{\circ}) $ & $90\times 70~(-21^{\circ}) $ & 42~(2.38) & 20.01 & 4.85~(275.23) & 307.69 & 40.4$\pm$4.0 & 5, 12 \\
IP Tau & $136\times 104~(-19.5^{\circ}) $ & $80\times 70~(-13^{\circ}) $ & 38~(2.35) & 19.70 & 1.30~(81.26) & 142.51 & 11.5$\pm$1.2 & 5, 12 \\
DO Tau & $142\times 104~(-20.6^{\circ}) $ & $70\times 60~(-16^{\circ}) $ & 50~(3.0) & 44.30 & 22.19~(1332.32) & 1929.26 & 119.9$\pm$12.0 & 5, 11 \\
Haro 6-13 & $139\times 111~(-4.4^{\circ}) $ & $60\times 50~(1^{\circ}) $ & 50~(2.84) & 137.91 & 33.30~(1902.20) & 3970.04 & 135.2$\pm$13.5 & 5, 10 \\
V836 Tau & $172\times 130~(-30.3^{\circ}) $ & $70\times 50~(-17^{\circ}) $ & 55~(2.2) & 34.04 & 10.51~(414.81) & 639.98 & 24.2$\pm$2.4 & 5, 11 \\
HK Tau A & $132\times 115~(-2.2^{\circ}) $ & $70\times 60~(-7^{\circ}) $ & 49~(2.84) & 43.86 & 12.08~(698.09) & 1469.66 & 31.7$\pm$3.2 & 5, 11 \\
HP Tau & $138\times 112~(-4.1^{\circ}) $ & $50\times 50~(-6^{\circ}) $ & 51~(2.89) & 66.77 & 23.26~(1324.81) & 2766.40 & 49.1$\pm$4.9 & 5, 10 \\
GI Tau & $135\times 119~(-4.3^{\circ}) $ & $70\times 50~(-19^{\circ}) $ & 49~(2.67) & 49.39 & 4.85~(266.15) & 406.04 & 15.8$\pm$1.6 & 5, 11 \\
RW Aur A & $40\times 22~(14.6^{\circ}) $ & $20\times 10~(14^{\circ}) $ & 9~(9.24) & 66.79 & 3.55~(3573.07) & 6295.08 & 32.8$\pm$3.3 & 6, 13 \\
T Tau N & $140\times 100~(34.1^{\circ}) $ & $40\times 30~(45^{\circ}) $ & 41~(2.56) & 272.20 & 63.81~(3952.76) & 9145.97 & 174.8$\pm$17.5 & 5, 9 \\
DH Tau A & $146\times 118~(-30.7^{\circ}) $ & $70\times 60~(-27^{\circ}) $ & 53~(2.70) & 40.25 & 9.90~(508.33) & 783.80 & 24.3$\pm$2.4 & 5, 11 \\
HN Tau A & $145\times 101~(38.7^{\circ}) $ & $70\times 40~(38^{\circ}) $ & 39~(2.31) & 144.95 & 6.93~(415.69) & 1083.77 & 12.0$\pm$1.2 & 5, 11 \\
DK Tau A & $137\times 115~(7.0^{\circ}) $ & $50\times 40~(2^{\circ}) $ & 47~(2.63) & 116.03 & 13.37~(747.64) & 1520.14 & 29.1$\pm$2.9 & 5, 10 \\
RW Aur B & $40\times 22~(14.6^{\circ}) $ & $20\times 10~(14^{\circ}) $ & 9~(9.24) & 39.47 & 0.50~(505.23) & 754.80 & 3.9$\pm$0.4 & 6, 13 \\
UY Aur A & $154\times 100~(-16.3^{\circ}) $ & $50\times 30~(-14^{\circ}) $ & 47~(2.72) & 144.59 & 16.18~(932.55) & 5490.32 & 19.0$\pm$1.9 & 5, 9 \\
\hline
\end{tabular}}
\begin{tabnote}
\textbf{Note.} Column description: (1) Name of the host star. The names are ordered by decreasing au-scale size of the disks from top to bottom. (2) CLEAN beam $\theta_{\rm cl}$. $\texttt{Briggs~robust}$ parameters are fixed to be 0.5 for all images. (3) SpM beam $\theta_{\rm eff}$. The value is obtained by the point source injection method (see Section \ref{sec:imaging_spm}). (4) RMS noise $\sigma_{\rm cl}$ of the CLEAN image in the unit of $\rm \mu Jy~beam^{-1}$ and $\rm mJy~arcsec^{-2}$ (denoted in the parentheses). The value is calculated from the emission-free area. (5) Detection threshold $I_{\rm DT}$ of the SpM image. The value is the maximum value of the emission-free area. (6) and (7) Peak intensity of each CLEAN and SpM image, respectively. The unit of the CLEAN value is expressed in both $\rm \mu Jy~beam^{-1}$ and $\rm mJy~arcsec^{-2}$ (denoted in the parentheses). (8) Flux density $F_{\nu}$ of SpM image. The value is obtained by the curve-growth method (see Appendix \ref{appendix:dust_disk_size}). The uncertainty is the $10\%$ absolute calibration error for ALMA observations. (9) The employed two hyper-parameters of SpM image ($\Lambda_l$, $\Lambda_{tsv}$) in logarithmic scale.

\end{tabnote}
\label{tab:fullsample}
\end{table*}

\subsection{Imaging with Sparse Modeling}\label{sec:imaging_spm}

We apply SpM to produce an image from the self-calibrated visibility data. We use $\tt PRIISM$ \footnote[3]{$\tt PRIISM$ (Python Module for Radio Interferometry Imaging with Sparse Modeling) is an imaging tool for ALMA based on the sparse modeling technique, and is publicly available at \url{https://github.com/tnakazato/priism}}, ver. 0.7.2 \citep{Nakazato2020} in conjunction with $\tt CASA$ to perform $\ell _1$+TSV imaging and the cross-validation (CV) scheme as illustrated in \citet{Yamaguchi2021}. The image was reconstructed by minimizing a cost function, which is the weighted chi-square error between the visibility model derived by the model image and observed visibility, accompanied by two regularization terms, namely $\ell_1$-norm and the total squared variation (TSV). Each regularization term is multiplied by hyper-parameters ($\Lambda_{l}$ for $\ell_1$ norm and $\Lambda_{tsv}$ for TSV), which control the relative weighting of the regularization terms to the observations. The SpM algorithm does not involve the beam convolution processes to obtain the final image, unlike the CLEAN algorithm. This potentially leads to images with higher spatial resolution compared to beam-convoluted CLEAN images, i.e., ``super-resolution'' images.

The $\ell _1$ norm (so-called LASSO regression; \cite{Tibshirani1996}) serves as a sparse regularization function in the image domain, regulating the total flux density in the brightness distribution and simultaneously adjusting low-brightness noise intensity in the emission-free region (e.g., \cite{Honma2014}). $\Lambda_{l}$ controls the degree of sparsity of the image. The image with larger $\Lambda_{l}$ is more sparse, that is, lower noise level in the background while the total flux is lower.

The TSV regularization controls the smoothness of the brightness distribution and uses the $\ell _2$ norm (so-called Ridge regression; \cite{Hoerl1970}) in the gradient domain. This regularization suppresses steep brightness variation and thus prefers edge-smoothed distribution \citep{Akiyama2017ApJ, Kuramochi2018}. This regularization and the hyper-parameter $\Lambda_{tsv}$ also play an important role in determining the effective spatial resolution of the reconstructed images. With higher values of $\Lambda_{tsv}$, smoother (smaller gradient), or low spatial resolution, images are reconstructed.

We obtain one model image that minimizes the cost function for a given set of $(\Lambda_{l}, \Lambda_{tsv})$, and we have a set of model images with a range of values of $(\Lambda_{l}, \Lambda_{tsv})$. If the weights of the regularization terms are either excessively strong or weak, the model visibility derived from the reconstructed image deviates from the observed one. To select a pair of $(\Lambda_{l}, \Lambda_{tsv})$ for the optimal image, we employ the 10-fold cross-validation (CV) approach (see Equation 2 in \cite{Yamaguchi2021}). This approach explores the optimal hyper-parameters by considering the errors between the model and observations. We first find a set of hyper-parameters $(\Lambda_{l,\min}, \Lambda_{tsv,\min})$ which give the minimum cross-validation error (CVE). In most cases, we judge that the image reconstructed with $(\Lambda_{l,\min}, \Lambda_{tsv,\min})$ is the one with the highest fidelity \citep{Akiyama2017ApJ}. However, in some cases, we find that the image with the minimum CVE exhibits artificial patchy structures. In such cases, we visually inspect a set of images obtained by hyper-parameters with $\Delta \log \Lambda_{l}\leq 1$ and $\Delta \log \Lambda_{tsv}\leq 1$, where $\Delta \log \Lambda$ stands for $|\log \Lambda - \log \Lambda_{\min}|$. For five objects (DM Tau, GO Tau, DS Tau, DN Tau, and RW Aur), we conservatively choose the image with $\Lambda_{tsv}$ one order of magnitude larger than $\Lambda_{tsv, \min}$ as optimal. The total fluxes of those SpM images are consistent with those of CLEAN images within $5-10\%$ errors. The final values for $(\Lambda_{l}, \Lambda_{tsv})$ are summarized in Table \ref{tab:fullsample}.

We assess the effective spatial resolution $\theta_{\rm eff}$ of the SpM image by the ``point-source injection'' method outlined in \cite{Yamaguchi2021}. We inject an artificial point source into the visibility data using the $\tt CASA$ task $\tt addcomponent$, $\tt ft$, and $\tt uvsub$. The artificial point source is placed in an emission-free area north of the central star but at the distance within MRS (Table \ref{tab:summary_obs}) and its flux is set to be $5\%$ of the total flux density of the target. Then, SpM imaging was performed for the point-source injected data using the same set of regularization parameters employed for generating the optimal image of non-injected data. The injected point source exhibits a Gaussian-like distribution in the reconstructed image. We fit it with an elliptical Gaussian function to obtain the Full Width at Half Maximum (FWHM), which is used as a measure of the effective spatial resolution $\theta_{\rm eff}$ of the SpM image. We have checked that the measured spatial resolution is altered only at the level of a few percent if we inject the point source in the east, west, or south of the central star and the total flux density of the point source is recovered within a $10\%$ error range. The effective spatial resolutions $\theta_{\rm eff}$ of SpM images are summarized in Table \ref{tab:fullsample}.

Here, we briefly describe our motivation for choosing SpM over the Maximum Entropy Method (MEM; \cite{Narayan1986}), which is another regularization method. MEM uses a relative entropy function that requires a ``prior image'' (e.g., a noise map derived from a dirty map or a circular Gaussian model; \cite{Carcamo2018, EHT2019}). However, incorporating prior information may introduce biases into the reconstructed image. In contrast to MEM, SpM does not impose such prior information. We also note that MEM simultaneously imposes a similar sparsity and smoothness constraint as the $\ell _1$ norm and TSV regularizations in a single regularization term (see Equation 27 in \cite{EHT2019}). SpM has the advantage of imposing them separately so that it is easier to control the balance between the sparsity and smoothness constraints.

\section{Reconstructed Images and SpM Imaging Performances}\label{sec:images}

\begin{figure*}[p]
\begin{center}
\includegraphics[width=0.98 \textwidth]{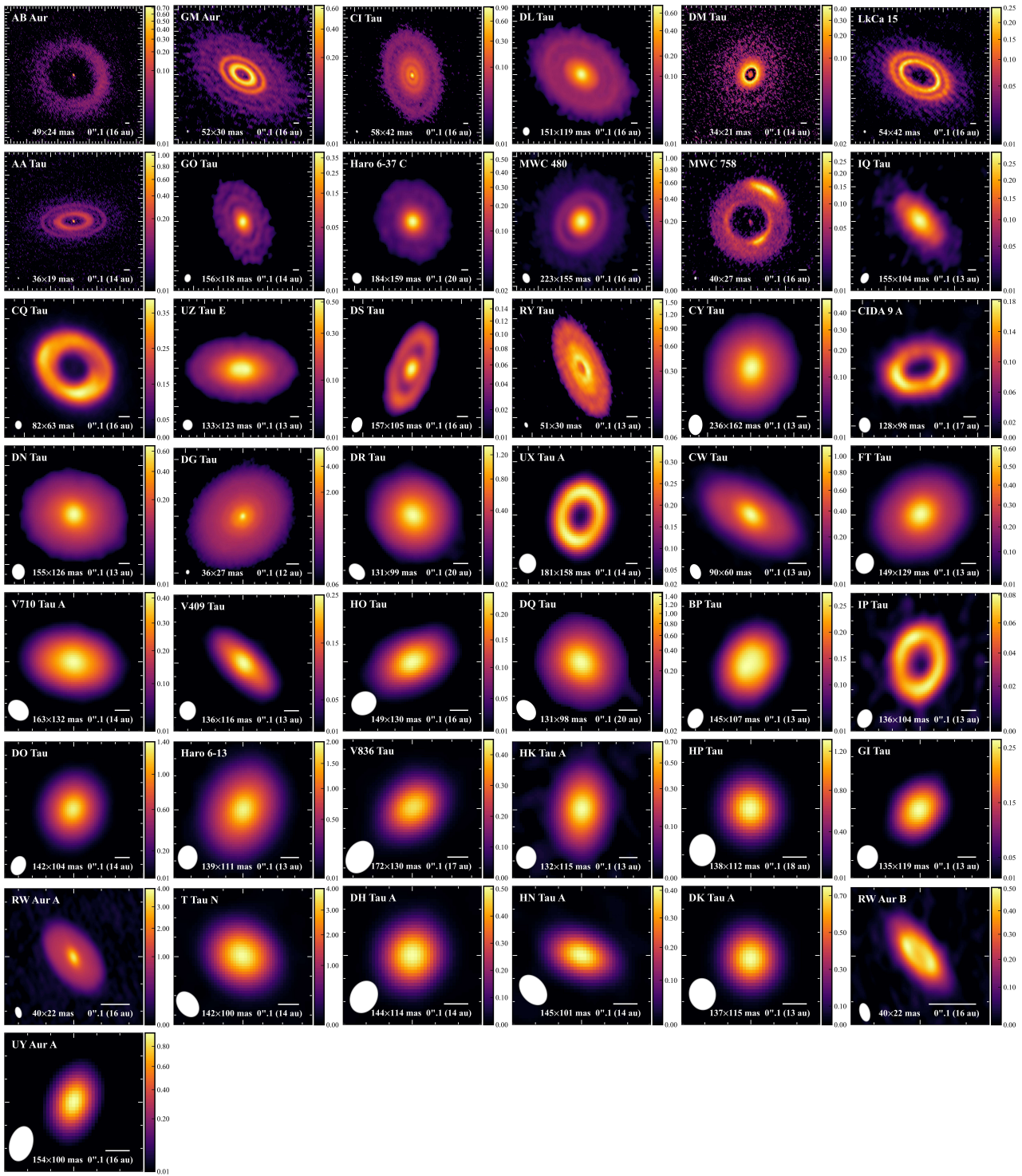}
\end{center}
\caption{Gallery of CLEAN images for 43 disks in our sample. Briggs weighting with a robust parameter of 0.5 is adopted for all images. These images are ordered by decreasing dust disk size with the astronomical unit from left to right and top to bottom. The color scale given by a power law is adopted. A white bar of $0.''1$ is provided for reference to the angular scales. The filled white ellipse denotes the synthesized beam $\theta_{\rm cl}$ in the bottom left corner. The unit of the CLEAN image is converted from $\rm Jy~beam^{-1}$ to $\rm Jy~arcsec^{-2}$.}
\label{fig:clean_image}
\end{figure*}

\begin{figure*}[p]
\begin{center}
\includegraphics[width=0.98 \textwidth]{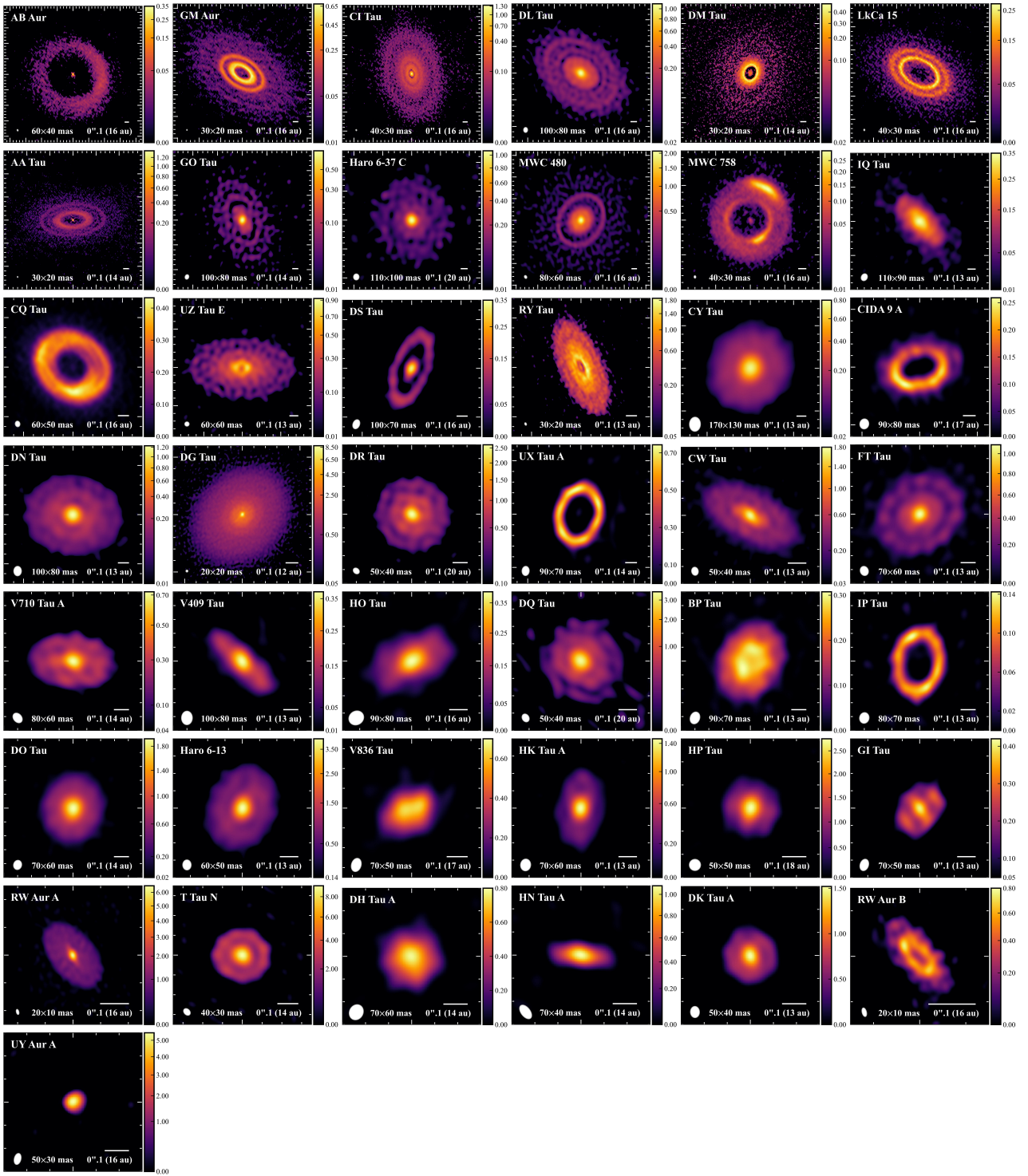}
\end{center}
\caption{Same as Figure~\ref{fig:clean_image} but for SpM images. Note that the SpM image is not convolved with the beam. The ellipse in the bottom left indicates the effective spatial resolution estimated as described in Section \ref{sec:imaging_spm}. The unit of the SpM image that is initially obtained from the imaging is $\rm Jy~pixel^{-1}$ and converted to $\rm Jy~arcsec^{-2}$.}
\label{fig:spm_image}
\end{figure*}

Figures \ref{fig:clean_image} and \ref{fig:spm_image} show the images of 43 disks of our targets using the CLEAN and SpM methods, respectively. Note that all images show the surface brightness distribution in units of $\rm Jy~asec^{-2}$ for consistency. We clearly see many more disk substructures in the SpM images. Distinct structures, such as gaps, rings, and crescents are seen in most of the disks.

Before going into details of analyses in Sections \ref{sec:images} to \ref{sec:corr_disksub}, we define the terms ``compact'' and ``large'' disks used in this study. Figure \ref{fig:diskradius} shows the distribution of dust disk radii $r_{\rm disk}$ taken from the SpM images (see Appendix \ref{appendix:dust_disk_size} for the definition of disk radius). The disk radii span a wide range from 8 to 238 au, with a median radius of 45 au. We set this median radius as a boundary of the ``compact'' and ``large'' disks. In the following subsections in Section \ref{sec:images}, we analyze the technical aspects of imaging methods: spatial resolution, goodness of visibility fit, and image fidelity.

\begin{figure*}[ht]
\begin{center}
\includegraphics[width=0.98 \textwidth]{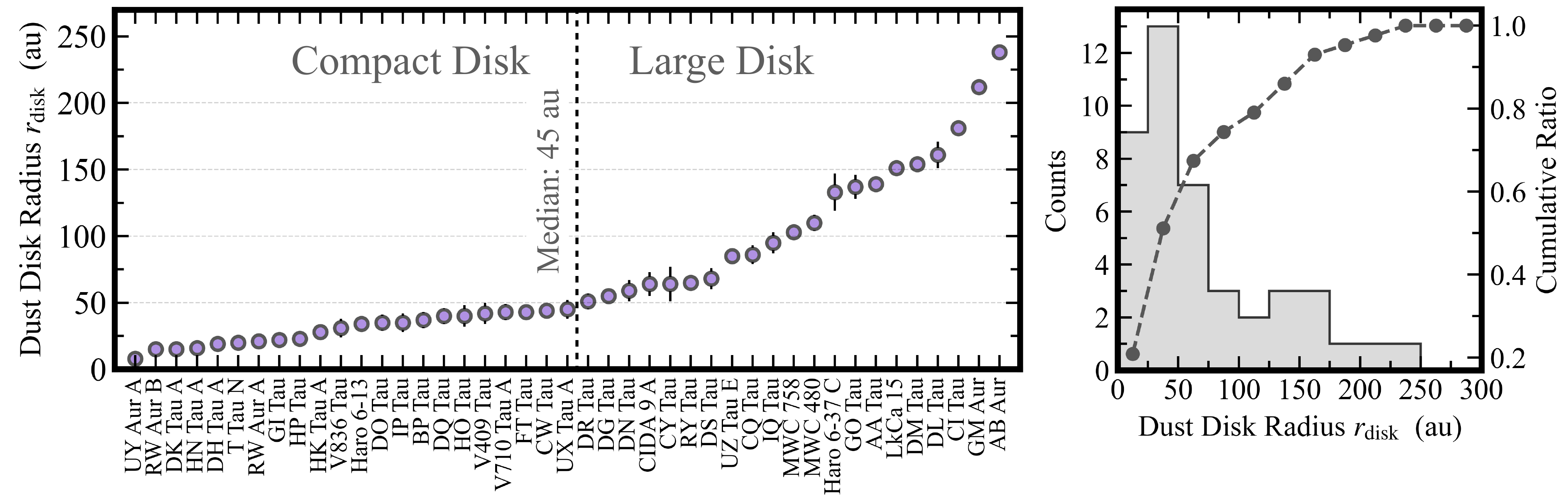}
\end{center}
\caption{Left: the dust disk radius ($r_{\rm disk}$) of each source in ascending order. The range of disk radii spans from 8 to 238 au with a median value of 45 au, which we use as a border of ``compact'' and ``large'' disks. Right: histogram and cumulative relative frequency of the dust disk radius with $\sim30$ au in radius are most common within the sample. Disks with $r_{\rm disk} \gtrsim 100$ au occupy only $30\% (11/43)$ of our sample.}
\label{fig:diskradius}
\end{figure*}


\subsection{Spatial Resolution}\label{sec:spatial_resolution}

\begin{figure*}[t]
\begin{center}
\includegraphics[width=0.98 \textwidth]{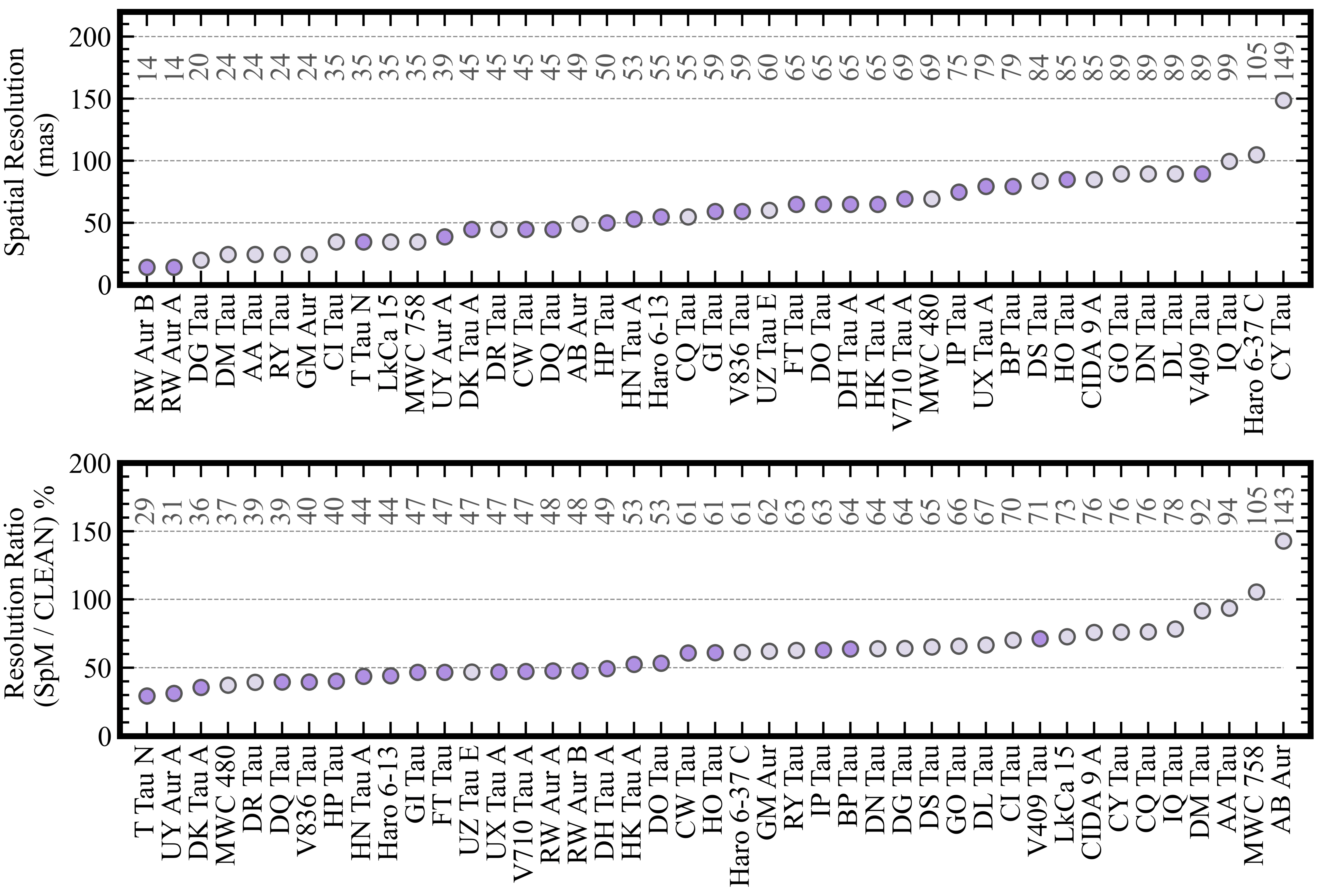}
\end{center}
\caption{Top: effective spatial resolutions of each SpM image. Bottom: the ratio of SpM image resolution to CLEAN beam size. Purple points are for compact disks and gray points are for large disks.}
\label{fig:resolution_ratio}
\end{figure*}

\begin{figure*}[ht!]
\begin{center}
\includegraphics[width=0.98 \textwidth]{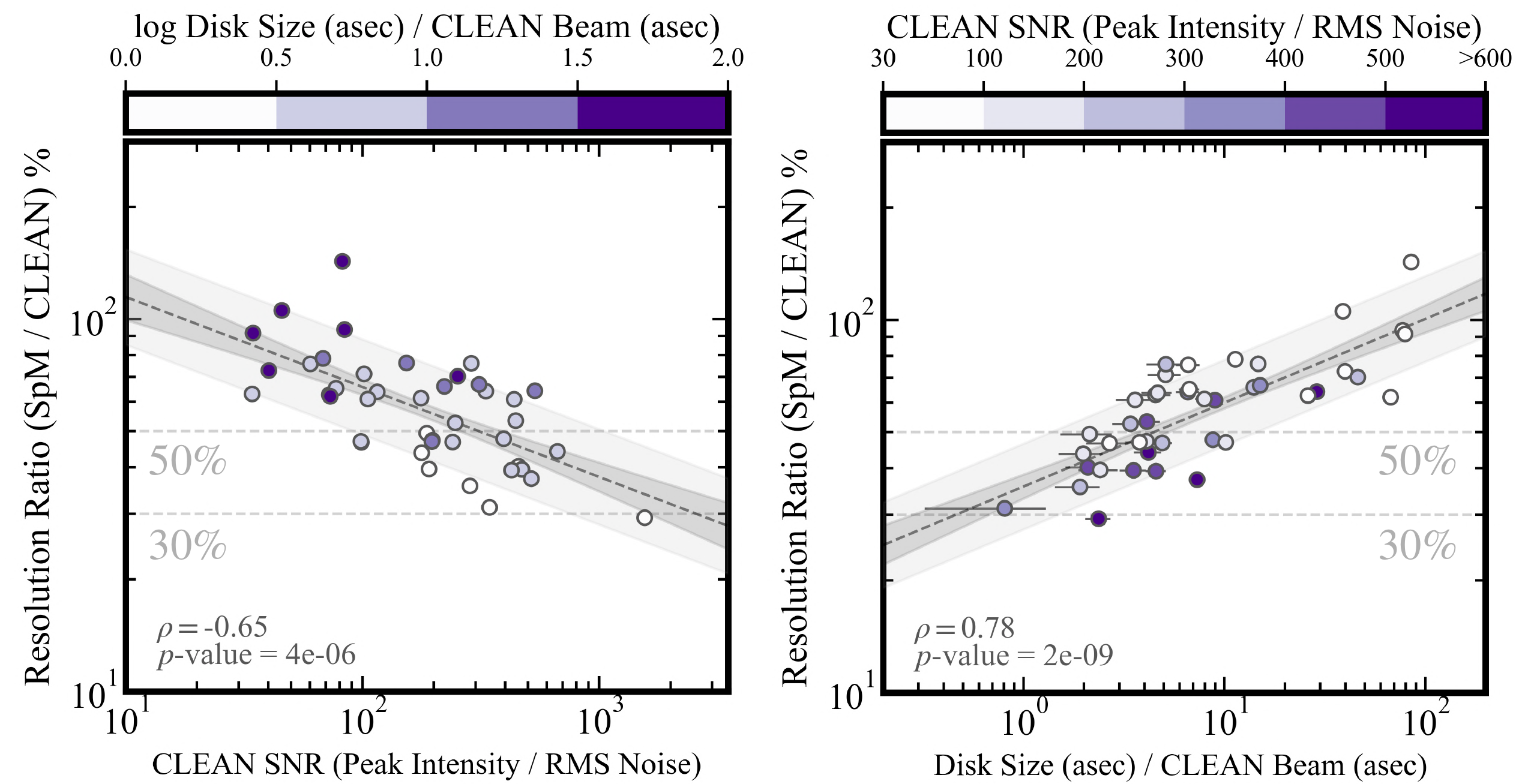}
\end{center}
\caption{Left: Relationship between resolution ratio (SpM/CLEAN) and CLEAN SNR (peak intensity to RMS noise) in the logarithmic plane. The sample is colored with the disk size ($=2 r_{\rm disk}$) normalized by CLEAN beam size on the logarithmic scale. The CLEAN beam size is the geometric mean of the major and minor axes. Right: Relationship between resolution ratio and the disk size normalized by the CLEAN beam size. The sample is colored with the CLEAN SNR. The dashed line is the median scaling relation from the Bayesian linear regression, and the dark gray area indicates the $68\%$ confidence interval for the regression. The light gray area corresponds to the inferred scatter. Pearson's correlation coefficient ($\rho$) and the $p$ value calculated from the sample distribution are shown in the lower left of each panel.}
\label{fig:resraatio_disksize}
\end{figure*}

Figure \ref{fig:resolution_ratio} shows the spatial resolution of SpM images and the ratio of SpM image resolution and CLEAN beam, that is, SpM/CLEAN = $\theta_{\rm eff}/\theta_{\rm cl}$ for each disk. 18 out of 43 targets show a factor of two to three improvement in spatial resolution compared to the conventional CLEAN method. All but three images show spatial resolution better than $0''.1$, with the highest achieved resolution reaching $0''.01 - 0''.02$.

In the left panel of Figure \ref{fig:resraatio_disksize}, we find a robust relationship between the improvement of spatial resolution and the SNR (peak/RMS away from emission) of the CLEAN image. The left panel of Figure \ref{fig:resraatio_disksize} shows a clear trend of decreasing resolution ratio as increasing SNR (Pearson correlation coefficient $\rho = -0.65$, $p$-value $< 0.01$). Using Bayesian linear regression in logarithmic space, this trend can be described as
\begin{equation}\label{eq:resolution_snr}
\log \left(\frac{\theta_{\rm eff}/\theta_{\rm cl}}{\%}\right)= (2.30\pm 0.11) - (0.24\pm0.05) \log \mathrm{SNR}
\end{equation}
\noindent
with a Gaussian scatter perpendicular to this regression with a standard deviation of $0.13\pm0.01$ dex. We have used $\tt Linmix$\footnote[4]{$\tt Linmix$ is a Bayesian approach to linear regression and is publicly available at \url{https://github.com/jmeyers314/linmix}. Hereafter we apply this method when employing the linear regression to data sets.} \citep{Kelly2007} for fitting. The resolution can be improved by a factor of two if the SNR reaches $\sim 100$ on the CLEAN image. It remains similar to that of the CLEAN image if the SNR is $\sim30$.

In the right panel of Figure \ref{fig:resraatio_disksize}, we also find a robust relationship ($\rho = 0.78$ and $p-$value $< 0.01$) between the improvement in spatial resolution and the disk size $\theta_{\rm disk}$ (= $2 \times r_{\rm disk}$) (arcsec) normalized by the CLEAN beam size $\theta_{\rm cl}$ (arcsec). We note that the sizes of these disks are measured on the SpM images, and all disks are spatially resolved. Using the Bayesian linear regression, we obtain
\begin{equation}\label{eq:resolution_disksize}
\log \left(\frac{\theta_{\rm eff}/\theta_{\rm cl}}{\%}\right)= (1.55\pm 0.03) + (0.22\pm0.03) \log \left( \frac{\theta_{\rm disk}}{\theta_{\rm cl}} \right)
\end{equation}
\noindent
with a Gaussian scatter of $0.11\pm0.01$ dex. The resolution can be improved by a factor of two when the disk size is $2-10$ times larger than the CLEAN beam size. It remains similar to that of the CLEAN image when the disk size is larger than the CLEAN beam size by more than 30 times.

From the two empirical relationships for SpM resolution improvement, we see that this improvement is more significant for higher SNR data or more compact disks close to the CLEAN beam size. This is because the visibility data of compact disks tend to have higher SNR at the longest baseline lengths. The large disk data are in many cases constructed with more than two antenna configurations, where the SNRs in the long baseline data are relatively low compared to those in the short baselines. With a better SNR of visibility at long baselines, it is possible to recover structures at smaller spatial scales and thus obtain better spatial resolution. The beam size of the CLEAN method, on the other hand, is determined by the array configuration, and the SNR is not taken into account (see ALMA Technical Handbook). Therefore, data with better SNR at long baselines tend to show more improvement in spatial resolution when SpM is applied. These two empirical relationships would provide an estimate of the required SNR or source size for SpM to be effective in improving spatial resolution.

In Appendix \ref{appendix:vis_profile_objects}, we elaborate more on how the resolution is improved with the disk size by using the data of two disks with different sizes. In Appendix \ref{appendix:noislevel_spm}, we evaluate how the noise level of images is related to the improvement of spatial resolution.

\subsection{Goodness of Visibility Fits}\label{sec:goodness_visfit}

\begin{figure*}[t]
\begin{center}
\includegraphics[width=0.98 \textwidth]{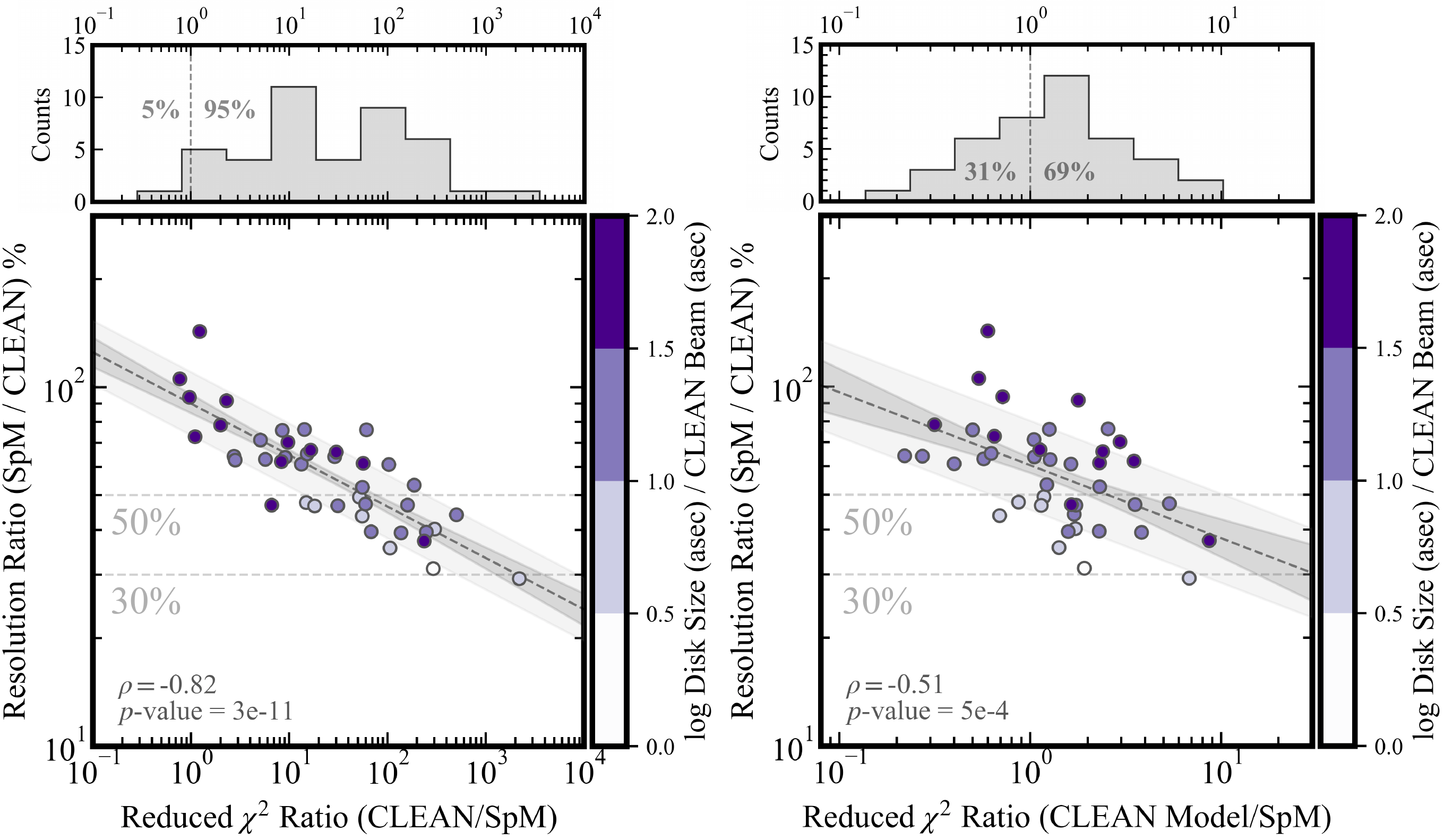}
\end{center}
\caption{Relationship between the resolution ratio (SpM/CLEAN) and the reduced $\chi^{2}$ ratio (CLEAN/SpM) in the logarithmic plane. Reduced $\chi^2$ values are computed for the CLEAN image after beam convolution (left panel) and the CLEAN model before the convolution (right panel). Each data sample is colored with the disk size normalized to the CLEAN beam size on the logarithmic scale. The dashed line, the dark gray area, and the light gray area have the same meaning as in Figure \ref{fig:resraatio_disksize}. Pearson's correlation coefficient ($\rho$) and the $p$ value calculated from the sample distribution are shown in the lower left of each panel. The upper histogram in each panel counts the number of the reduced $\chi^2$ ratio in each interval. The vertical dashed line indicates when the ratio is unity. The values shown in the histogram represent the fraction of the sample with the reduced $\chi^2$ ratio larger (smaller) than unity.}
\label{fig:resolutionratio_chisquare}
\end{figure*}


We assess the performance of SpM and CLEAN imaging by investigating how well the Fourier transform of the reconstructed images can fit the visibility. We use the reduced chi-square $\rm Red\chi^2$ as a measure of goodness-of-fit. We calculate $\rm Red\chi^2$ that compares the observed visibility and the model visibility that is obtained as the Fourier transform of the model image (see Appendix \ref{appendix:vis_profile}). For CLEAN images, we use ``CLEAN'' to indicate the image obtained after beam convolution while ``CLEAN model'' to indicate the one before the beam convolution, or a collection of clean components.

The top left panel of Figure \ref{fig:resolutionratio_chisquare} shows the histogram of the ratio of the reduced chi-square values obtained for CLEAN and SpM $(\rm Red\chi^{2}_{\rm cl}/ Red\chi^{2}_{\rm spm})$. SpM produces a better fit in $95\%$ of the cases (40/42). The remaining $5\%$ (2 cases) show comparable performance between SpM and CLEAN. The top right panel of Figure \ref{fig:resolutionratio_chisquare} shows the ratio of the reduced chi-square values for the CLEAN model and SpM $(\rm Red\chi^{2}{\rm clmodel}/ Red\chi^{2}{\rm spm})$. In this case, SpM shows better performance for $69\%$ (29/42), so we consider SpM and CLEAN models to be of comparable performance in terms of goodness-of-fit. In both CLEAN vs SpM and CLEAN model vs SpM comparisons, we find that the ratio of reduced chi-square values and the improvement of spatial resolution may be correlated. Further description is given in Appendix \ref{appendix:chisq_resolution}.

The difference in performance between the CLEAN and CLEAN model is due to beam convolution. In the CLEAN algorithm, clean components are placed in a way that the ``CLEAN model'' matches observed visibility. Then, the ``CLEAN model'' image is convolved with a restoring beam to obtain the ``CLEAN'' image. This process acts as a low-pass filter and causes the visibility amplitude to deviate from the original, especially in the long baselines.

Although the goodness-of-fit is comparable in the CLEAN model and SpM images, the quality of the actual 2D image is very different. The CLEAN model image is a collection of point sources (or multi-scale Gaussian distributions; \cite{Schwab1984, Rau2011}). The SpM image can better reproduce complex and smooth structures compared to the CLEAN model images. Therefore, we consider that the SpM images are more plausible for scientific analyses compared to the CLEAN model images. We note, however, that the CLEAN model can capture some of the features such as the gradient of the outer edge of a disk if the model is azimuthally averaged. Further discussion is given in Appendix \ref{appendix:utility_clmodel}.

\subsection{Image Fidelity}

The fidelity of SpM images is assessed with the ``cross-check method'' \citep{Yamaguchi2020}, wherein a comparison is made between the CLEAN image derived from the visibility data including the long baseline observations and the SpM image generated from those of the short baseline observations (refer to Appendix \ref{appendix:eval_spm} for a comprehensive description of this procedure). We have analyzed three bright and large disks that already have the necessary data: DG Tau, RY Tau, and CI Tau. We confirm that those images reconstructed from the shorter-baseline data using the SpM match those obtained with the longer-baseline data using CLEAN. Furthermore, azimuthally averaged radial intensity profiles of the SpM image and the CLEAN image with long baseline data in three disks agree within $10\%$ error. 

However, we note that SpM imaging can introduce artificial clumped/speckled features, especially for disks with the brightest emission in the central part (i.e., an inner disk) surrounded by the faint emission area (i.e., a ring skirt). We observe these features in several samples, including DQ Tau, DR Tau, FT Tau, GO Tau, GI Tau, Haro 63-C, UZ Tau E, and MWC 480 (see Figure \ref{fig:spm_image}). The possible cause is the intrinsic bias of the SpM imaging. This imaging preferentially fits the bright and compact part (e.g., inner disk) of the target, but this bias could introduce similar ``compact''  features such as clumps and speckles to the surrounding extended area (e.g., a ring or its surrounding).

These artifacts can be suppressed by taking the azimuthal average and extracting the radial intensity profile.  In this way, the image is ``smoothed'' in the azimuthal direction while the spatial resolution in the radial direction is unaffected (see Appendix \ref{appendix:eval_spm}). We thus employ the radial profile when extracting the physical properties of the disks.




\section{Total Flux and Disk Size}\label{sec:totalflux}

In this Section, we present the measurements of total flux and disk sizes. We compare the measurements with other telescopes and discuss the correlation between the flux and disk size. Methods of measuring the flux and size are outlined in Appendix \ref{appendix:dust_disk_size}. 

\subsection{Comparison with Other Telescopes }\label{sec:totalflux_comparison}

\begin{figure*}[h]
\begin{center}
\includegraphics[width=0.98 \textwidth]{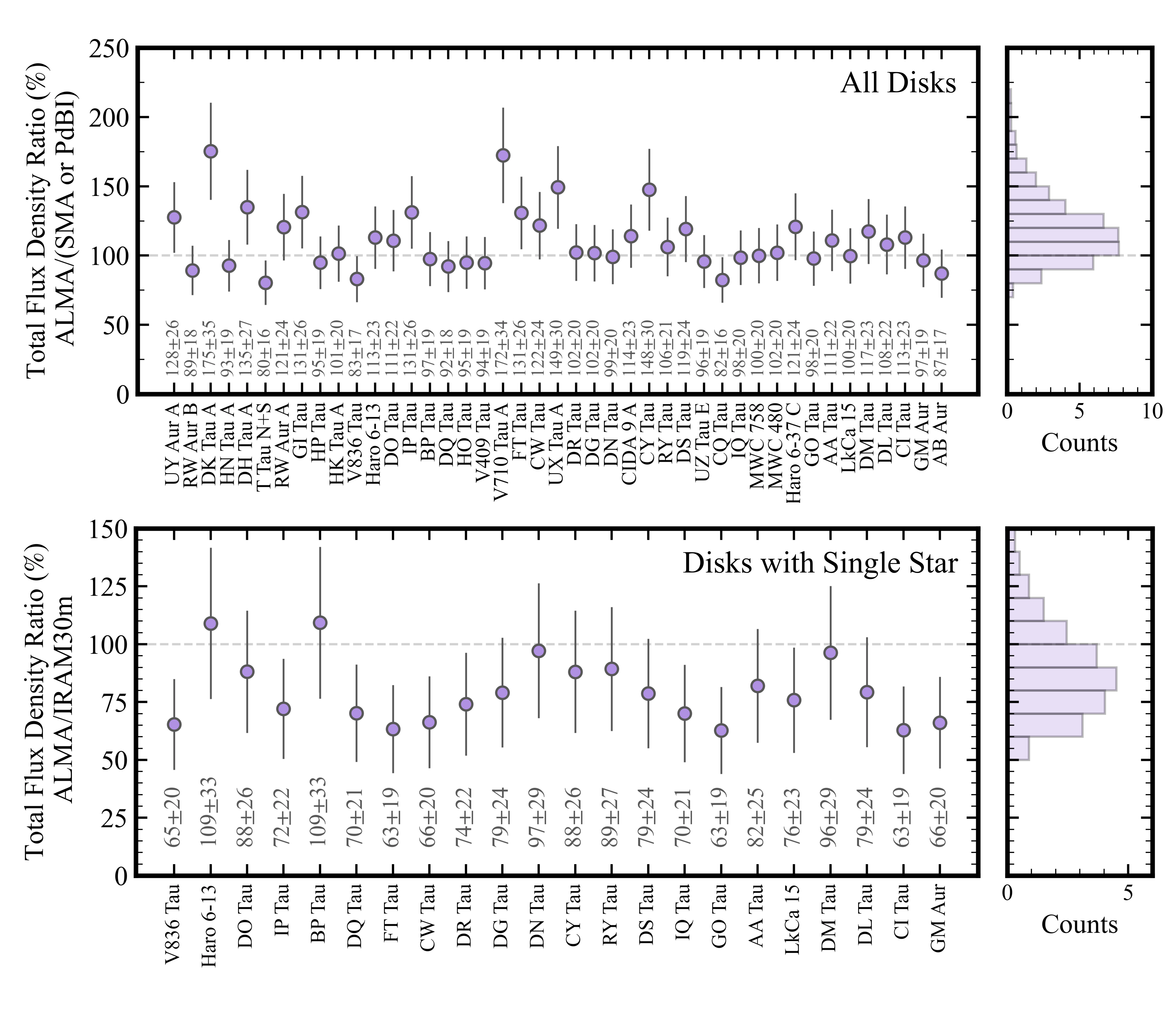}
\end{center}
\caption{Flux density ratio of each source between ALMA and other telescopes at 1.3 mm. The top panel shows the ratio between ALMA and compact interferometers (SMA and PdBI; \cite{Andrews2013, Chapillon2008}). The source names are listed in the ascending order of their dust disk radii. The bottom panel shows the ratio between ALMA and the IRAM 30-meter single-dish telescope \citep{Beckwith1990, Osterloh1995}. Only disks around a single star are shown. For both panels, error bars are calculated through error propagation using the measurement errors of the two data sets. Uncertainties of measurements include the absolute flux error, which is $10\%$ for interferometers and $20\%$ for the IRAM 30 m telescope. The right panels show probability histograms generated from the flux ratio distributions by using a Monte Carlo routine.}
\label{fig:totalfluxratio}
\end{figure*}

We compare our measurements of total flux with those taken by other telescopes. First, we consider data taken by other interferometers. We take the total fluxes taken by the SMA \citep{Andrews2013} for 41 Class II disks and those by the Plateau de Bure Interferometer \citep{Chapillon2008} for the two Herbig disks of CQ Tau and MWC 758. The total flux densities taken at different wavelengths (AA Tau at 1.2 mm and GM Aur at 1.1 mm) in ALMA data sets are modified for those at 1.3 mm, by applying spectral indices $\alpha_{\rm mm}$ of 1.6 for AA Tau \citep{Andrews2005} and 2.7 for GM Aur \citep{Huang2020}. These spectral indices are derived from the relationship in $F_{\nu}\propto \nu^{\alpha_{\rm mm}}$. These compact interferometers have larger maximum recoverable scales than the ALMA and have beam sizes ranging from $2''-5''$, corresponding to $130-350$ au from the central star at a distance of 140 pc. In the top left panel of Figure \ref{fig:totalfluxratio}, we show the ratio of total flux measured with ALMA and other interferometers. We see that most of the flux values measured with ALMA are consistent with those obtained from other compact interferometers. The top right panel shows the probability histogram of the flux ratios generated by a Monte Carlo routine, which uses random sampling of iterative 5000 calculations by incorporating the errors associated with each flux ratio measurement. The average of the histogram is $110\%\pm25\%$, where the error indicates the standard deviation. Therefore we consider that ALMA observations, in general, have detected dust emission signals all the way to the outer edge of the disk and do not resolve out large-scale structures. However, the total fluxes of DK Tau A and V710 Tau A are $\sim 50\%$ higher than those obtained with SMA. Since SMA has a shorter maximum baseline than ALMA, spatial filtering of the SMA data cannot explain this discrepancy. We actually find other ALMA observations at the same observing wavelength but 1.5 times higher sensitively and $\sim 2$ times lower resolution have obtained the total flux values that are consistent with our results within $10\%$ for the two disks \citep{Rota2022}. Therefore we consider that there may have been some calibration issues at the time of SMA observations or there is some time variability of mm flux for the two objects.

Next, we compare the total flux density with those obtained by the IRAM 30m single-dish telescope, which has a beam size of $11''$, corresponding to the area of 770 au in radius at the distance of 140 pc \citep{Beckwith1990, Osterloh1995}. Here, we use the data only for 21 disks with single-star systems to avoid contamination by dust emission from companions. The bottom panels of Figure \ref{fig:totalfluxratio} show the flux ratio and the probability histogram. The average flux ratio is $79\%\pm19\%$, suggesting that the ALMA observations drop $\sim 20\%$ of the total flux density taken by a single-dish telescope. This discrepancy may be accounted for as the contribution from the emission at the envelope scale surrounding the star and disk system. Further discussion on this interpretation is given in Section \ref{sec:envelope_remnants}.

\subsection{Relation of Flux and Disk Size}\label{{sec:disksize_mmlum_relation}}

\begin{figure*}[ht!]
\begin{center}
\includegraphics[width=0.98 \textwidth]{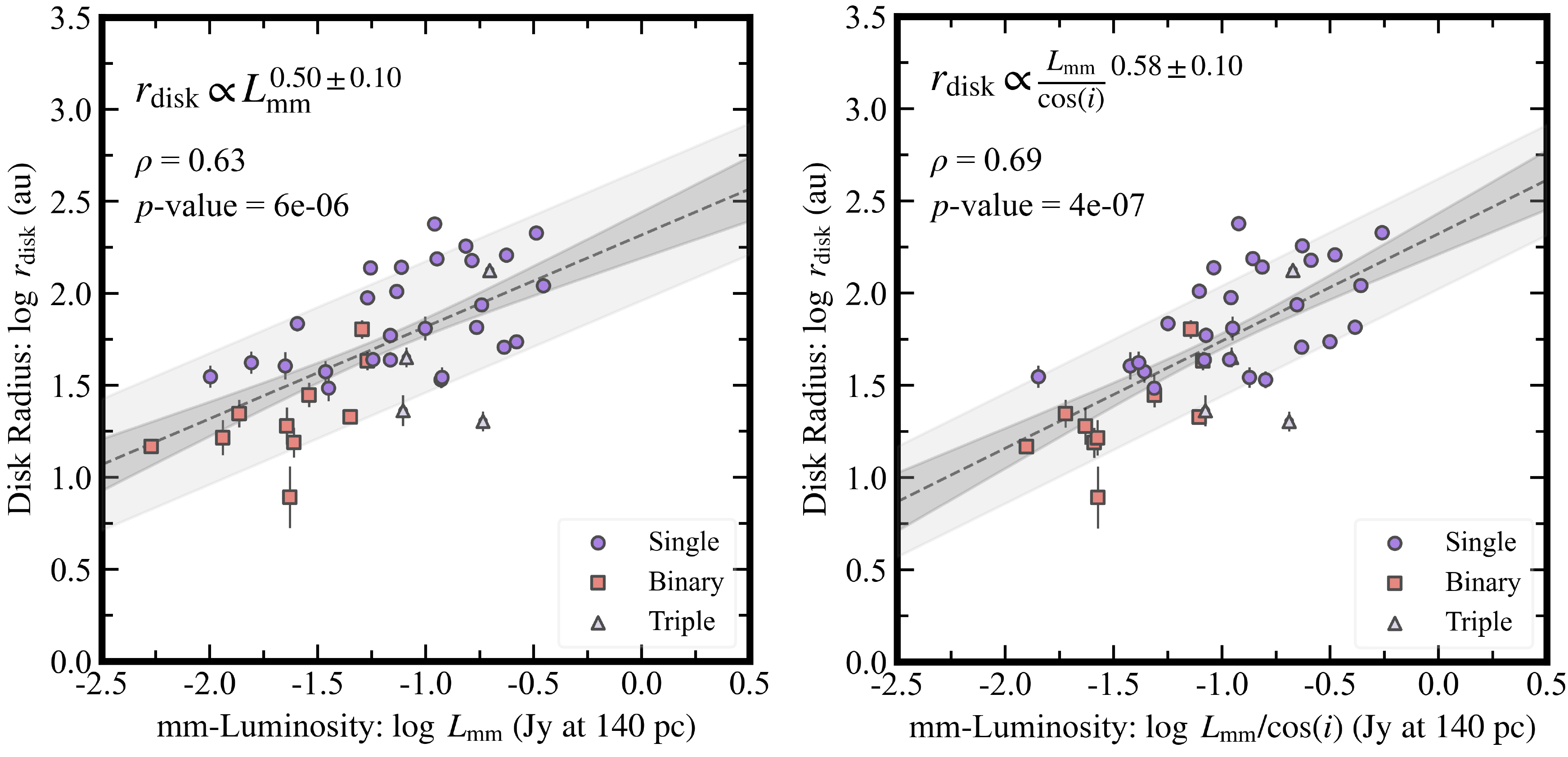}
\end{center}
\caption{Left: Relationship between millimeter-Luminosity ($L_{\rm mm}$; flux density scaled at the distance of 140~pc) and the disk radius. Circle, square, and triangle symbols are for single or spectroscopic binary (29), binary (10), and triple stars (4), respectively. Right: The same relation but with the effect of inclination angle ($i$) is taken into account for $L_{\rm mm}$. For each panel, the black dashed line denotes the median scaling relation obtained from Bayesian linear regression, and the dark gray area represents the $68\%$ confidence interval around the median. The light gray area corresponds to the inferred scatter. The best-fit linear regression model, Pearson's correlation coefficient ($\rho$), and the $p$-value calculated from the sample distribution are shown in the upper left of each panel.}
\label{fig:size_lumi_relation}
\end{figure*}

The correlation between millimeter luminosity (total flux density re-scaled to a common distance) and the disk size in a nearby star-forming region has been suggested using 0.9 mm data from SMA and medium-resolution ($ 0''.1-0''.2$) ALMA observations \citep{Tazzari2017, Tripathi2017, Barenfeld2017, Andrews2018b, Long2019, Hendler2020, Tazzari2021}. Here, we explore if there is a similar trend in our dataset. In previous studies, disk sizes were constrained by parametric modeling of disk shapes and fitting in the visibility domain. In our dataset, it is possible to measure the disk size directly in the image domain since all 43 disks are spatially resolved with SpM. Therefore, our approach provides a model-independent assessment of the disk size-luminosity relation. Figure \ref{fig:size_lumi_relation} presents the millimeter luminosity $L_{\rm mm}$ as a function of disk radii $r_{\rm dust}$. We see a strong correlation with $\rho = 0.63$ and a $ p-$value of $<0.01$. Linear regression on the logarithmic scale gives 
\begin{equation}
\log \left(\frac{r_{\rm disk}}{\mathrm{au}}\right)= (2.32_{-0.13}^{+0.13})+ (0.50_{-0.10}^{+0.10}) \log \left[F_{v}\left(\frac{d}{140 \mathrm{pc}}\right)^{2}\right]
\end{equation}
\noindent
with a scattering of $0.36 \pm 0.04$ dex. Also seen in Figure \ref{fig:size_lumi_relation} is that single-star disks primarily occupy the upper right region, or larger and brighter disks, while binary- and triple-star disks dominate the lower left region or smaller and fainter disks.

The correlation can be written as $L_{\rm {mm}} \propto r_{\rm {disk}}^2 $. The same correlation is found by 0.9~mm data of disks in Taurus-Auriga-Ophiuchus regions \citep{Tripathi2017, Hendler2020}. This relationship also suggests that the surface brightness intensity $\left\langle I_{\nu}\right\rangle$ (or temperature $\left\langle T_{\rm b} \right\rangle$) averaged over the area inside the dust disk is approximately constant \citep{Tripathi2017}. For our data in Band 6, the averaged surface brightness is $\left\langle I_{\nu}\right\rangle \simeq 0.14~\rm Jyasec^{-2}$ and the corresponding brightness temperature is $\left\langle T_{\rm b} \right\rangle \simeq 8$ K. This brightness temperature is similar to what is obtained by \citet{Tripathi2017} despite the difference of observing wavelengths (1.3~mm vs 0.9~mm).

\citet{Tazzari2021} discussed the influence of optical thickness on the size-luminosity relation. Following their approach, we checked the correlation between the disk size and the luminosity re-scaled by $\cos i$ where $i$ is the inclination angle. Employing the Bayesian linear regression, we have obtained
\begin{equation}
\log \left(\frac{r_{\rm disk}}{\mathrm{au}}\right) = (2.32_{-0.11}^{+0.11}) + (0.58_{-0.10}^{+0.10}) \log \left[\frac{F_{v}}{\cos i}\left(\frac{d}{140 \mathrm{pc}}\right)^{2}\right]
\end{equation}
\noindent
with the scattering of $0.30 \pm 0.02$ dex. We observe a slightly smaller scatter in the correlation when the $\cos i$ effect is considered. The correlation is improved by $\sim10\%$ ($\rho = 0.69$ with a $p-$value of $<0.01$) compared to the case without the $\cos i$ term, while the slope and intercept (normalization) remain statistically consistent within $1\sigma$ uncertainty.

\section{Categorization of Radial Structures}\label{sec:categorize_ra}

In this section, we investigate the radial intensity profile of disks. Gaps and rings have been identified by the analyses of radial intensity profiles of disks and the statistical trends of disk substructures have been discussed (e.g., \cite{Pinilla2018, Huang2018, Long2018}). Here, we perform analyses based on the newly obtained images with SpM that recover disk structures at smaller scales compared to CLEAN (see Section \ref{sec:images}). We define fundamental disk substructures such as gaps and rings and suggest categorizations of the disks based on the substructures they have. 

\subsection{Methods of Extracting Radial Profiles}\label{sec:radial_intensity_profile}

\begin{figure*}[ht]
\begin{center}
\includegraphics[width=0.98 \textwidth]{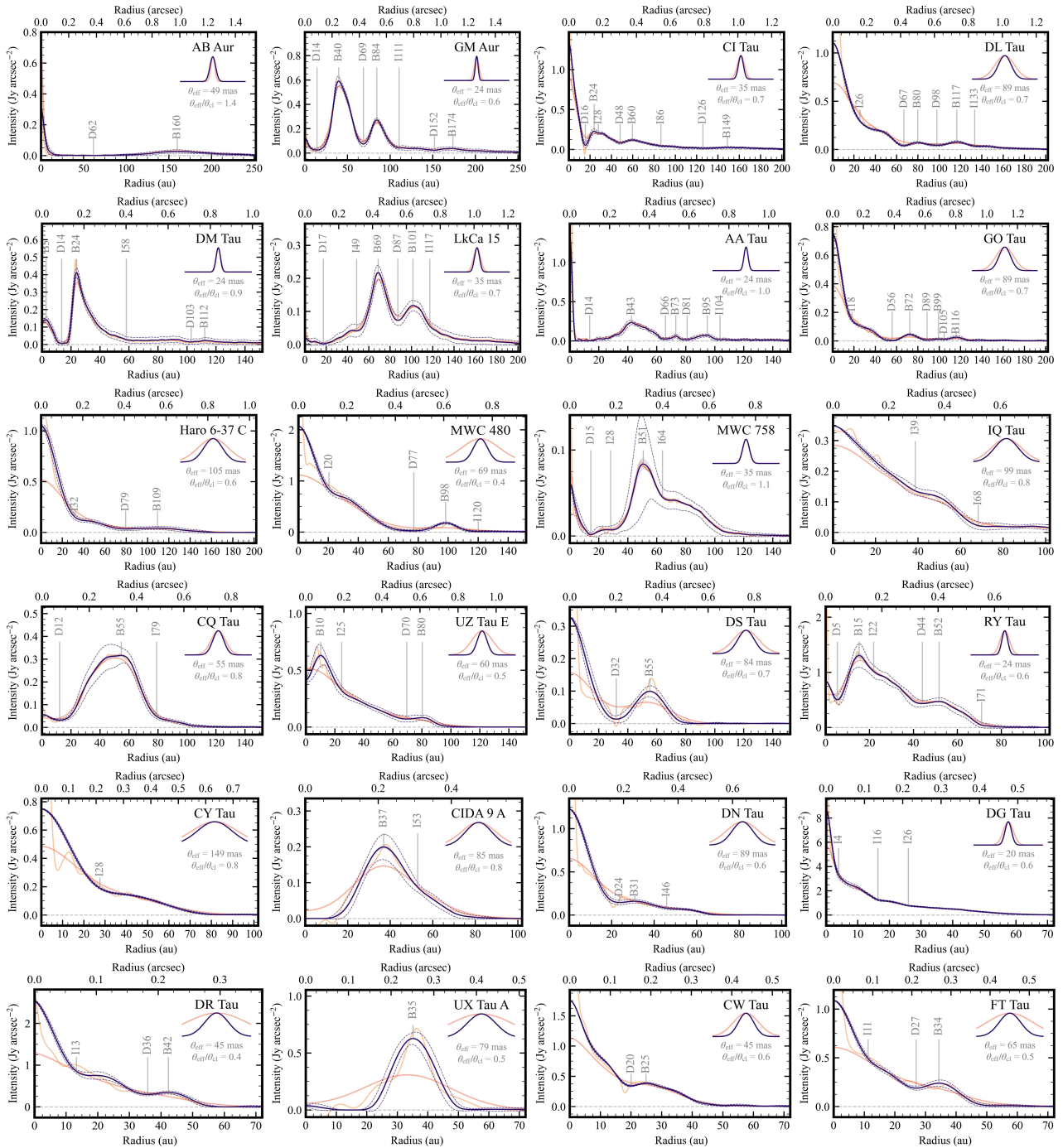}
\end{center}
\caption{Deprojected and azimuthally averaged radial intensity profiles obtained from SpM (purple), CLEAN (orange), and CLEAN model (yellow) images. The unit of the intensity is converted to $\rm mJy~arcsec^{-2}$. The profiles are shown in descending order of dust disk size from left to right and from top to bottom. Each profile is interpolated onto radial grid points spaced by 0.1 au with $\tt interpolate.interp1d$ in the $\tt SciPy$ module. For SpM data, the light purple ribbon shows the error of the mean at each radius ($\hat{\sigma}_{I}$), while its range is negligibly small for all of the data. The purple dashed lines show the standard deviation of the data in each azimuth ($\sigma_{I}$) for comparison. Vertical solid gray lines mark the gap ($D$), ring ($B$), and inflection ($I$) listed in Table \ref{tab:disk_substructure}. The geometric mean of the effective spatial resolutions of SpM (purple) and CLEAN (orange) are shown in the inset at the top right corner of each image.}
\label{fig:radialprofile_1}
\end{figure*}

\begin{figure*}[ht]
\begin{center}
\includegraphics[width=0.98 \textwidth]{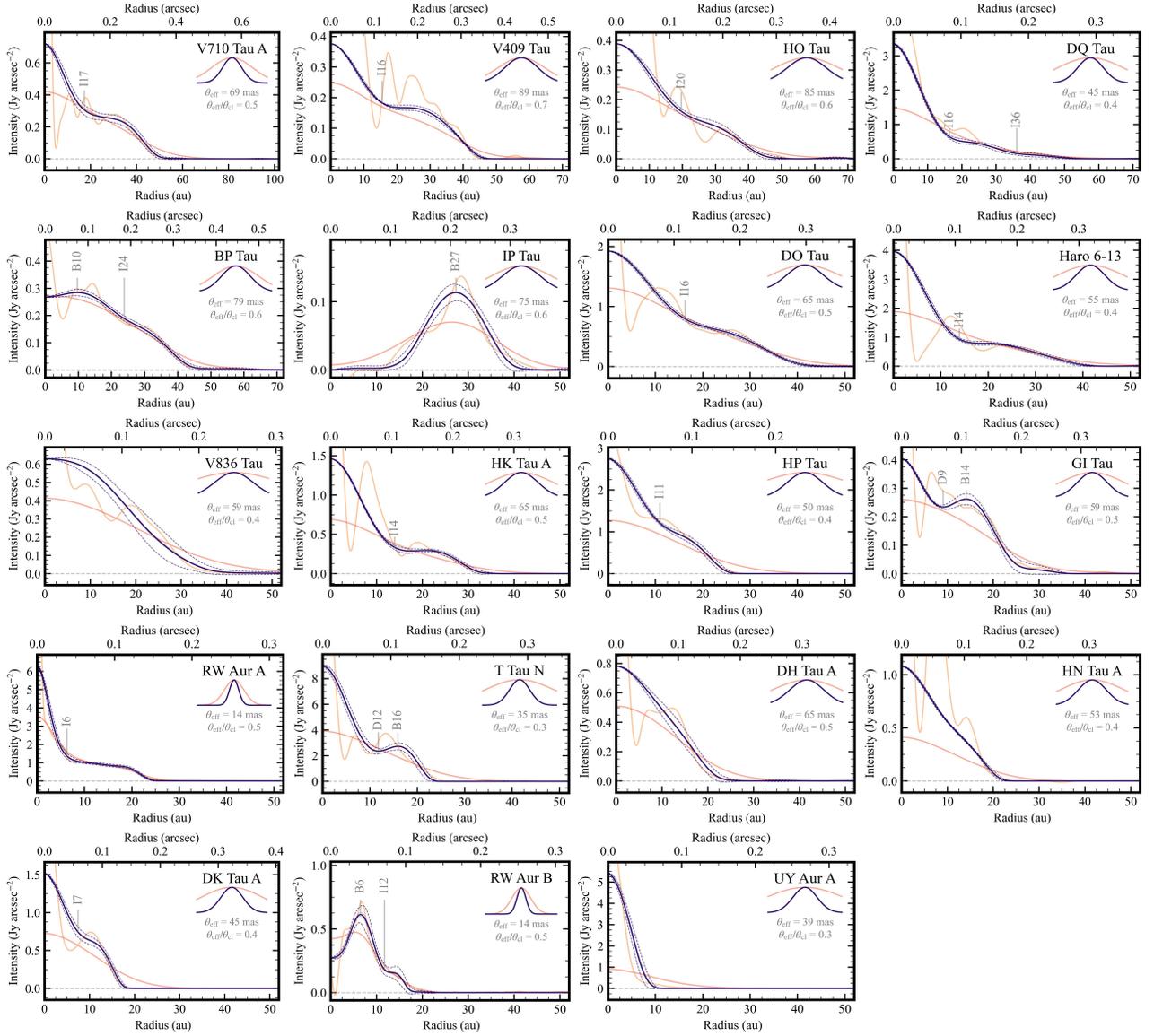}
\end{center}
\caption{Continues from Figure \ref{fig:radialprofile_1}.}
\label{fig:radialprofile_2}
\end{figure*}

Figure \ref{fig:radialprofile_1} and \ref{fig:radialprofile_2} show the azimuthally averaged radial intensity profiles $I_{\nu}(r)$. Here, we briefly describe how those profiles are obtained. We first deproject the SpM images shown in Figure \ref{fig:spm_image} to a face-on view. The methods for deriving the geometric parameters for deprojection (inclination and position angle) are outlined in Appendix \ref{appendix:dust_disk_size}.

The uncertainty $\hat{\sigma}_{I}$ of the radial intensity profiles is computed as the standard error of the mean at each radius $r_{i}$, considering one beam size as the smallest independent unit. It is given by $ \hat{\sigma}_{I}=\sigma_{I}/\sqrt{N_{B}}$, where $\sigma_{I}$ is the standard deviation of the brightness within the concentric ring at $r=r_{i}$ and $N_{B}= 2\pi r_{i}/ \left\langle\theta_{\mathrm{eff}}\right\rangle$ is the number of beams within the ring. Here, the beam size $\left\langle\theta_{\mathrm{eff}}\right\rangle$ is given by the geometric mean of the effective spatial resolution of the SpM image.

We determine the range for azimuthal averaging of the deprojected disks. For 24 disks not listed below, we average all over the azimuth as they show almost axisymmetric structures. For four highly inclined disks with $i \geq 60^{\circ}$ (AA Tau, CW Tau, IQ Tau, and V409 Tau), the azimuthal average is taken in the range from $20^{\circ}$ to $40^{\circ}$ relative to the semi-major axis since the substructures around the minor axis are not well spatially resolved. For disks with gaps and rings that are not identified over full azimuthal angles due to insufficient spatial resolution, low SNRs, or radial asymmetry (GM Aur, GO Tau, GI Tau, DG Tau, DL Tau, DN Tau, DQ Tau, DR Tau, HK Tau A, HO Tau, HP Tau, and RY Tau), we take average over the position angles where we can visually see the gaps and rings. For MWC758, we exclude the range of PA with $290-10^{\circ}$ due to the presence of a significant blob at $r=0''.4$. This blob produces an artificial ring-gap structure in the radial intensity profile. The ranges of the azimuthal angles for averaging for each disk are summarized in Table \ref{tab:disk_substructure}.

\onecolumn
\begin{landscape}
\begin{longtable}{l@{\hspace{0.02cm}}l@{\hspace{0.05cm}}l@{\hspace{0.05cm}}l@{\hspace{0.05cm}}l@{\hspace{0.05cm}}l@{\hspace{0.05cm}}l@{\hspace{0.05cm}}l@{\hspace{0.05cm}}l@{\hspace{0.05cm}}l@{\hspace{0.05cm}}c@{\hspace{0.01cm}}r}
\caption{Properties of Disk Structures.}\\
\hline
\label{tab:disk_substructure}
Name & Rings-Gaps & Inflection & Label & Gap            & Ring           & Inflection      & Gap Width              & Norm Width       & Gap Depth        & Avg Range & Method \\
      & Type      & Type       &       & $r_{\rm gap}$  & $r_{\rm ring}$ & $r_{\rm Inf}$   & $\Delta_{\rm I, unit}$ & $\Delta_{\rm I}$ & $\delta_{\rm I}$ &           &        \\
      &           &            &       & (au, mas)      & (au, mas)      & (au, mas)       & (au)                   &                  &                  & (degree)  &        \\
(1)   & (2)       & (3)        & (4)   & (5)            & (6)            & (7)             & (8)                    & (9)              & (10)             & (11)      &(12)    \\
\hline
\endfirsthead
\hline
Name & Rings-Gaps & Inflection & Label & Gap            & Ring           & Inflection      & Gap Width              & Norm Width       & Gap Depth        & Avg Range & Method \\
      & Type      & Type       &       & $r_{\rm gap}$  & $r_{\rm ring}$ & $r_{\rm Inf}$   & $\Delta_{\rm I, unit}$ & $\Delta_{\rm I}$ & $\delta_{\rm I}$ &           &        \\
      &           &            &       & (au, mas)      & (au, mas)      & (au, mas)      & (au)                     &                 &                  & (degree)  &        \\
(1)   & (2)       & (3)        & (4)   & (5)            & (6)            & (7)             & (8)                    & (9)              & (10)             & (11)      &(12)    \\
\hline
\endhead
\hline \multicolumn{12}{r}{\textit{Continued on next page}} \\
\endfoot
\hline 
\multicolumn{12}{l}{\hbox to 0pt{\parbox{230mm}{\footnotemark[]\textbf{Note.} Column description: (1) Name of the host star. (2) Rings-Gaps type specified in Section \ref{sec:ring-gap_feature}. (3) Inflection type specified in Section \ref{sec:inflection_feature}. (4) Substructure label specified in Section \ref{sec:ra_extraction}. (5) Radial gap location in astronomical unit (au) and millimeter-arcsecond (mas). (6) Radial ring location in au and mas. (7) Radial inflection location in au and mas. (8) Gap width in au. (9) Normalized gap width. (10) Gap depth. (11) Range of azimuthally averaging the radial profile in degree. (12) Method used to derive gap properties, specified in Section \ref{sec:gap_width_depth}. The uncertainties of the gap properties are $1\sigma$ and do not account for the uncertainty in the distance to the source. $^{*}$: The gap/ring structure identified by the improved spatial resolution of the SpM, whereas it was not previously identified in the conventional CLEAN image. $^{\dag}$: The spatially resolved gap whose width is larger than the effective spatial resolution (i.e., $\Delta_{\rm I, unit} > \theta_{\rm eff}$).}}}\\
\endlastfoot
\hline
AB Aur & Pre-Transition & $\cdots$ & D62/B160$^{\dag}$& 61.6(378) & 160.0(982) &$\cdots$& $116.10\pm2.01$ & $0.90\pm0.01$ & $<40$ & $0 \leq \theta \leq 359$ & case 1\\
\hline
GM Aur &Pre-Transition& $\cdots$ & D14/B40$^{\dag}$& 14.2(89) & 39.7(249) &$\cdots$& $19.63\pm1.28$ & $0.82\pm0.03$ & $23.94\pm2.56$ &$110 \leq \theta \leq 300$ & case 2\\
 & Ring-Gap  & $\cdots$ & D69/B84$^{\dag}$& 68.8(431) & 84.4(529) &$\cdots$& $17.72\pm0.16$ & $0.22\pm0.01$ & $3.67\pm0.09$ & $110 \leq \theta \leq 300$ & case 1\\
 & $\cdots$ &Outer Disk-Skirt& I111&$\cdots$&$\cdots$& 110.6(693) & $\cdots$ & $\cdots$ & $\cdots$ & $110 \leq \theta \leq 300$ & $\cdots$ \\
 & Ring-Gap & $\cdots$ & D152/B174& 152.1(953) & 173.6 (1088) &$\cdots$& $\sim 20$ & $\sim 0.1$ & $\sim 2$ & $110 \leq \theta \leq 300$ & case 1\\
\hline
CI Tau &Ring-Gap & $\cdots$ & D16/B24$^{\dag}$& 15.6(98) & 23.8(150) &$\cdots$& $6.51\pm0.48$ & $0.34\pm0.02$ & $4.22\pm0.52$ & $0 \leq \theta \leq 359$ & case 1\\
 & $\cdots$ &Outer Disk-Skirt& I28&$\cdots$&$\cdots$& 27.8(175) & $\cdots$ & $\cdots$ & $\cdots$ & $0 \leq \theta \leq 359$ & $\cdots$ \\
 & Ring-Gap & $\cdots$ & D48/B60$^{\dag}$& 48.4(305) & 59.8(377) &$\cdots$& $9.36\pm0.63$ & $0.18\pm0.01$ & $1.48\pm0.02$ & $0 \leq \theta \leq 359$ & case 1\\
 & $\cdots$ &Outer Disk-Skirt& I86&$\cdots$&$\cdots$& 86.5(545) & $\cdots$ & $\cdots$ & $\cdots$ & $0 \leq \theta \leq 359$ & $\cdots$ \\
 & Ring-Gap  & $\cdots$ & D126/B149& 125.8(793) & 148.9(938) &$\cdots$& $ \sim 20 $ & $\sim 0.2 $ & $\sim 2$ & $0 \leq \theta \leq 359$ & case 1\\
\hline
DL Tau &$\cdots$&Shoulder& I26&$\cdots$&$\cdots$& 25.6(161) & $\cdots$ & $\cdots$ & $\cdots$ & $\theta = 50-140, 320-359$ & $\cdots$ \\
 &Ring-Gap & $\cdots$ & D67/B80$^{*}$ & 66.9(420) & 79.8(501) &$\cdots$& $10.99\pm0.48$ & $0.15\pm0.0$ & $1.82\pm0.05$ & $\theta = 50-140, 320-359$ & case 1\\
 &Ring-Gap & $\cdots$ & D98/B117$^{* \dag}$& 97.8(614) & 116.6(732) &$\cdots$& $21.82\pm0.8$ & $0.2\pm0.01$ & $1.82\pm0.04$ & $\theta = 50-140, 320-359$ & case 1\\
 & $\cdots$ &Outer Disk-Skirt& I133&$\cdots$&$\cdots$& 133.3(837) & $\cdots$ & $\cdots$ & $\cdots$ &$\theta = 50-140, 320-359$ & $\cdots$ \\
\hline
DM Tau &Central Hole & $\cdots$ & B3&$\cdots$& 2.8(19) &$\cdots$& $\cdots$ & $\cdots$ & $\cdots$ & $0 \leq \theta \leq 359$ & $\cdots$ \\
 & Pre-Transition &$\cdots$ & D14/B24$^{\dag}$& 13.6(94) & 23.9(165) &$\cdots$& $11.17\pm0.15$ & $0.58\pm0.01$ & $51.79\pm15.15$ & $0 \leq \theta \leq 359$ & case 2\\
 & $\cdots$ &Outer Disk-Skirt& I58&$\cdots$&$\cdots$& 58.2(401) & $\cdots$ & $\cdots$ & $\cdots$ & $0 \leq \theta \leq 359$ & $\cdots$ \\
 & Ring-Gap & $\cdots$ & D103/B112& 102.9(709) & 112.3(774) & $\cdots$ & $\sim 10$ & $\sim 0.1$  & $\sim2$  & $0 \leq \theta \leq 359$ &  case 1\\
\hline
LkCa 15 &Pre-Transition& $\cdots$ & D17/B69$^{\dag}$& 17.3(109) & 68.8(433) &$\cdots$& $14.14\pm0.16$ & $0.56\pm0.01$ & $<110$ & $0 \leq \theta \leq 359$ & case 2 \\
 & $\cdots$ &Inner Disk-Skirt& I49&$\cdots$&$\cdots$& 48.6(306) & $\cdots$ & $\cdots$ & $\cdots$ & $0 \leq \theta \leq 359$ & $\cdots$ \\
 & Ring-Gap & $\cdots$ & D87/B101$^{\dag}$ & 87.4(550) & 101.4(639) &$\cdots$& $12.87\pm0.64$ & $0.14\pm0.01$ & $1.61\pm0.04$ & $0 \leq \theta \leq 359$ & case 1\\
 & $\cdots$ &Outer Disk-Skirt& I117&$\cdots$&$\cdots$& 117.3(738) & $\cdots$ & $\cdots$ & $\cdots$ & $0 \leq \theta \leq 359$ & $\cdots$ \\
\hline
AA Tau &Pre-Transition  & $\cdots$ & D14/B43$^{\dag}$ & 13.7(100) & 42.7(311) &$\cdots$& $29.5\pm0.14$ & $0.90\pm0.01$ & $<110$ & $|\theta - \mathrm{PA} |\leq30$  & case 1\\
 & Ring-Gap & $\cdots$ & D66/B73$^{\dag}$ & 66.0(481) & 73.1(533) &$\cdots$& $8.09\pm1.1$ & $0.12\pm0.02$ & $2.72\pm0.26$ & $|\theta - \mathrm{PA} |\leq30$ & case 1\\
 & Ring-Gap & $\cdots$ & D81/B95$^{\dag}$ & 80.9(590) & 94.8(691) &$\cdots$& $13.03\pm2.33$ & $0.15\pm0.02$ & $3.97\pm0.35$ & $|\theta - \mathrm{PA} |\leq30$ & case 1\\
 & $\cdots$ &Outer Disk-Skirt& I104&$\cdots$&$\cdots$& 103.6(755) & $\cdots$ & $\cdots$ & $\cdots$ & $|\theta - \mathrm{PA} |\leq30$  & $\cdots$ \\
\hline
GO Tau & $\cdots$  &Shoulder& I18&$\cdots$&$\cdots$& 18.1(125) & $\cdots$ & $\cdots$ & $\cdots$ & $|\theta - \mathrm{PA} |\leq20$ & $\cdots$ \\
 & Ring-Gap& $\cdots$ & D56/B72$^{\dag}$& 56.0(387) & 72.4(501) &$\cdots$& $19.67\pm0.43$ & $0.3\pm0.0$ & $14.49\pm1.4$ & $|\theta - \mathrm{PA} |\leq30$ & case 1\\
 &Ring-Gap & $\cdots$ & D89/B99$^{*}$& 88.8(614) & 99.3(687) &$\cdots$& $7.95\pm0.14$ & $0.09\pm0.01$ & $1.78\pm0.17$ & $|\theta - \mathrm{PA} |\leq20$ & case 1\\
 & Ring-Gap& $\cdots$ & D105/B116$^{*}$& 104.8(725) & 116.1(803) &$\cdots$& $5.35\pm3.47$ & $0.05\pm0.03$ & $2.23\pm0.12$ &$|\theta - \mathrm{PA} |\leq20$ & case 2\\
\hline
Haro 6-37 C & $\cdots$ &Shoulder& I32&$\cdots$&$\cdots$& 31.7(162) & $\cdots$ & $\cdots$ & $\cdots$ & $0 \leq \theta \leq 359$ & $\cdots$ \\
 &Ring-Gap & $\cdots$ & D79/B109$^{*}$& 79.4(406) & 109.3(559) &$\cdots$& $18.2\pm2.94$ & $0.2\pm0.03$ & $1.26\pm0.05$ & $0 \leq \theta \leq 359$ & case 1\\
\hline
MWC 480 & &Shoulder& I20&$\cdots$&$\cdots$& 20.4(126) & $\cdots$ & $\cdots$ & $\cdots$ & $0 \leq \theta \leq 359$ & $\cdots$ \\
 & Ring-Gap& $\cdots$ & D77/B98$^{\dag}$& 77.2(477) & 98.5(609) &$\cdots$& $29.93\pm0.16$ & $0.33\pm0.01$ & $7.08\pm0.3$ & $0 \leq \theta \leq 359$ & case 1\\
 & $\cdots$ &Outer Disk-Skirt& I120&$\cdots$&$\cdots$& 120.1(742) & $\cdots$ & $\cdots$ & $\cdots$ & $0 \leq \theta \leq 359$ & $\cdots$ \\
\hline
MWC 758 & Pre-Transition  & $\cdots$ & D15/B51$^{\dag}$& 14.6(91) & 50.8(317) &$\cdots$& $39.73\pm0.8$ & $0.93\pm0.01$ & $47.73\pm15.72$ & $20 \leq \theta \leq 290$ & case 1\\ 
& $\cdots$ & Inner Disk-Skirt& I28&$\cdots$&$\cdots$& 28.2(176) & $\cdots$ & $\cdots$ & $\cdots$ & $20 \leq \theta \leq 290$ & $\cdots$ \\
 & $\cdots$ &Outer Disk-Skirt& I64&$\cdots$&$\cdots$& 63.9(399) & $\cdots$ & $\cdots$ & $\cdots$ & $20 \leq \theta \leq 290$ & $\cdots$ \\
\hline
IQ Tau & $\cdots$ &Shoulder& I39&$\cdots$&$\cdots$& 38.7(295) & $\cdots$ & $\cdots$ & $\cdots$ & $|\theta - \mathrm{PA} |\leq40$ & $\cdots$ \\
& $\cdots$ &Outer Disk-Skirt& I68&$\cdots$&$\cdots$& 68.3(520) & $\cdots$ & $\cdots$ & $\cdots$ & $|\theta - \mathrm{PA} |\leq40$ & $\cdots$ \\
\hline
CQ Tau & Pre-Transition & $\cdots$ & D12/B55$^{* \dag}$& 12.2(75) & 54.8(336) &$\cdots$& $13.21\pm0.49$ & $0.72\pm0.02$ & $9.76\pm0.4$ & $0 \leq \theta \leq 359$ & case 2\\
 & $\cdots$ &Outer Disk-Skirt& I79&$\cdots$&$\cdots$& 79.1(485) & $\cdots$ & $\cdots$ & $\cdots$ & $0 \leq \theta \leq 359$ & $\cdots$ \\
\hline
UZ Tau E &Central Hole & $\cdots$ & B10&$\cdots$& 10.2(78) &$\cdots$& $\cdots$ & $\cdots$ & $\cdots$ & $0 \leq \theta \leq 359$ & $\cdots$ \\
 &$\cdots$ &Outer Disk-Skirt& I25&$\cdots$&$\cdots$& 24.8(189) & $\cdots$ & $\cdots$ & $\cdots$ & $0 \leq \theta \leq 359$ & $\cdots$ \\
 & Ring-Gap & $\cdots$ & D70/B80$^{* \dag}$& 69.5(530) & 80.3(612) &$\cdots$& $9.71\pm1.84$ & $0.13\pm0.02$ & $1.17\pm0.04$ & $0 \leq \theta \leq 359$ & case 1\\
\hline
DS Tau &Pre-Transition& $\cdots$ & D32/B55$^{\dag}$& 32.1(202) & 55.4(348) &$\cdots$& $22.75\pm0.48$ & $0.51\pm0.01$ & $7.72\pm0.71$ & $0 \leq \theta \leq 359$ & case 1\\
\hline
RY Tau & Pre-Transition & $\cdots$ & D5/B15$^{\dag}$& 5.4(42) & 15.3(119) &$\cdots$& $5.64\pm0.51$ & $0.68\pm0.05$ & $2.6\pm0.2$ & $-50 \leq\rm PA\leq220$ & case 2\\
 & $\cdots$  &Outer Disk-Skirt& I22&$\cdots$&$\cdots$& 21.9(171) & $\cdots$ & $\cdots$ & $\cdots$ & $-50 \leq\rm PA\leq220$ & $\cdots$ \\
 & Ring-Gap  & $\cdots$ & D44/B52$^{\dag}$& 44.1(344) & 51.7(403) &$\cdots$& $6.15\pm2.05$ & $0.13\pm0.04$ & $1.09\pm0.03$ & $-50 \leq\rm PA\leq220$ & case 1\\
 & $\cdots$ &Outer Disk-Skirt& I71&$\cdots$&$\cdots$& 71.0(554) & $\cdots$ & $\cdots$ & $\cdots$ & $-50 \leq\rm PA\leq220$ & $\cdots$ \\
\hline
CY Tau & $\cdots$ &Shoulder& I28&$\cdots$&$\cdots$& 27.6(214) & $\cdots$ & $\cdots$ & $\cdots$ & $0 \leq \theta \leq 359$ & $\cdots$ \\
\hline
CIDA 9 A &Transition& $\cdots$ & B37&$\cdots$& 37.0(215) &$\cdots$& $\cdots$ & $\cdots$ & $\cdots$ & $0 \leq \theta \leq 359$ & $\cdots$ \\
         & $\cdots$ &Outer Disk-Skirt& I53&$\cdots$&$\cdots$& 52.9(308) & $\cdots$ & $\cdots$ & $\cdots$ & $0 \leq \theta \leq 359$ & $\cdots$ \\
\hline
DN Tau &Ring-Gap & $\cdots$ & D24/B31$^{*}$& 23.7(185) & 30.9(241) &$\cdots$& $5.38\pm1.41$ & $0.2\pm0.04$ & $1.1\pm0.04$ &$230 \leq \rm PA\leq 310$ & case 1\\
 &$\cdots$ &Outer Disk-Skirt& I46&$\cdots$&$\cdots$& 45.9(358) & $\cdots$ & $\cdots$ & $\cdots$ & $230 \leq \rm PA\leq 310$ & $\cdots$ \\
\hline
DG Tau & $\cdots$ &Shoulder& I4&$\cdots$&$\cdots$& 4.0(33) & $\cdots$ & $\cdots$ & $\cdots$ & $60 \leq \theta\leq 320$  & $\cdots$ \\
 & $\cdots$&Shoulder& I16&$\cdots$&$\cdots$& 16.4(135) & $\cdots$ & $\cdots$ & $\cdots$ & $60 \leq \theta\leq 320$ & $\cdots$ \\
 & $\cdots$ &Outer Disk-Skirt& I26&$\cdots$&$\cdots$& 26.1(215) & $\cdots$ & $\cdots$ & $\cdots$ & $60 \leq \theta\leq 320$ & $\cdots$ \\
\hline
DR Tau & $\cdots$ &Shoulder& I13&$\cdots$&$\cdots$& 13.3(68) & $\cdots$ & $\cdots$ & $\cdots$ & $60 \leq\theta \leq160$ & $\cdots$ \\
 & Ring-Gap & $\cdots$ & D36/B42& 35.8(183) & 42.3(216) &$\cdots$& $5.68\pm0.20$ & $0.15\pm0.01$ & $1.16\pm0.02$ & $60 \leq\theta \leq160$ & case 1\\
\hline
UX Tau A &Transition& $\cdots$ & B35&$\cdots$& 35.1(251) &$\cdots$& $\cdots$ & $\cdots$ & $\cdots$ & $0 \leq \theta \leq 359$ & $\cdots$ \\
\hline
CW Tau &Ring-Gap& $\cdots$ & D20/B25$^{*}$& 20.0(151) & 24.9(188) &$\cdots$& $3.84\pm0.13$ & $0.17\pm0.01$ & $1.13\pm0.01$ & $|\theta - \mathrm{PA} | \leq40$ & case 1\\
\hline
FT Tau & $\cdots$ &Shoulder& I11&$\cdots$&$\cdots$& 11.4(89) & $\cdots$ & $\cdots$ & $\cdots$ & $0 \leq \theta \leq 359$ & $\cdots$ \\
 & Ring-Gap & $\cdots$ & D27/B34$^{*}$& 27.1(212) & 34.4(269) &$\cdots$& $7.41\pm1.15$ & $0.24\pm0.03$ & $1.26\pm0.03$ & $0 \leq \theta \leq 359$ & case 1\\
\hline
V710 Tau A & $\cdots$ &Shoulder& I17&$\cdots$&$\cdots$& 17.4(122) & $\cdots$ & $\cdots$ & $\cdots$ & $0 \leq \theta \leq 359$ & $\cdots$ \\
\hline
V409 Tau & $\cdots$ &Shoulder& I16&$\cdots$&$\cdots$& 15.5(118) & $\cdots$ & $\cdots$ & $\cdots$ & $|\theta - \mathrm{PA} |\leq20$ & $\cdots$ \\
\hline
HO Tau & $\cdots$ &Shoulder& I20&$\cdots$&$\cdots$& 19.7(122) & $\cdots$ & $\cdots$ & $\cdots$ & $|\theta - \mathrm{PA} |\leq20$ & $\cdots$ \\
\hline
DQ Tau  & $\cdots$ &Shoulder& I16&$\cdots$&$\cdots$& 16.2(82) & $\cdots$ & $\cdots$ & $\cdots$ & $ 200 \leq\theta \leq 360$ & $\cdots$ \\
 & $\cdots$ &Shoulder& I36&$\cdots$&$\cdots$& 36.0(182) & $\cdots$ & $\cdots$ & $\cdots$ & $200 \leq\theta \leq 360$ & $\cdots$ \\
\hline
BP Tau &Central Hole& $\cdots$ & B10&$\cdots$& 9.7(75) &$\cdots$& $\cdots$ & $\cdots$ & $\cdots$ & $0 \leq \theta \leq 359$ & $\cdots$ \\
 & $\cdots$ &Outer Disk-Skirt& I24&$\cdots$&$\cdots$& 23.9(185) & $\cdots$ & $\cdots$ & $\cdots$ & $0 \leq \theta \leq 359$ & $\cdots$ \\
\hline
IP Tau &Transition& $\cdots$ & B27&$\cdots$& 27.4(210) &$\cdots$& $\cdots$ & $\cdots$ & $\cdots$ & $0 \leq \theta \leq 359$ & $\cdots$ \\
\hline
DO Tau & $\cdots$ &Shoulder& I16&$\cdots$&$\cdots$& 16.3(117) & $\cdots$ & $\cdots$ & $\cdots$ & $0 \leq \theta \leq 359$ & $\cdots$ \\
\hline
Haro 6-13 & $\cdots$ &Shoulder& I14&$\cdots$&$\cdots$& 14.0(107) & $\cdots$ & $\cdots$ & $\cdots$ & $0 \leq \theta \leq 359$ & $\cdots$ \\
\hline
V836 Tau & $\cdots$ & $\cdots$ & $\cdots$ &$\cdots$&$\cdots$&$\cdots$& $\cdots$ & $\cdots$ & $\cdots$ & $0 \leq \theta \leq359$ & $\cdots$ \\
\hline
HK Tau A & $\cdots$ &Shoulder& I14&$\cdots$&$\cdots$& 14.0(105) & $\cdots$ & $\cdots$ & $\cdots$ & $|\theta - \mathrm{PA} | \leq20$ & $\cdots$ \\
\hline
HP Tau & $\cdots$ &Shoulder& I11&$\cdots$&$\cdots$& 11.0(62) & $\cdots$ & $\cdots$ & $\cdots$ & $\theta = 30-110, 210-290$ & $\cdots$ \\
\hline
GI Tau &Ring-Gap& $\cdots$ & D9/B14$^{*}$& 9.1(70) & 14.1(108) &$\cdots$& $4.44\pm0.39$ & $0.38\pm0.03$ & $1.12\pm0.01$ & $|\theta - \mathrm{PA} |\leq20$ & case 1\\
\hline
RW Aur A & $\cdots$ &Shoulder& I6&$\cdots$&$\cdots$& 6.4(39) & $\cdots$ & $\cdots$ & $\cdots$ & $0 \leq \theta \leq 359$ & $\cdots$ \\
\hline
T Tau N &Ring-Gap& $\cdots$ & D12/B16$^{*}$& 11.6(81) & 15.7(109) &$\cdots$& $3.88\pm0.43$ & $0.28\pm0.02$ & $1.22\pm0.06$ & $0 \leq 	\theta \leq 359$ & case 1\\
\hline
DH Tau A & $\cdots$ & $\cdots$ & $\cdots$ &$\cdots$&$\cdots$&$\cdots$& $\cdots$ & $\cdots$ & $\cdots$ & $0 \leq \theta \leq 359$ & $\cdots$ \\
\hline
HN Tau A & $\cdots$ & $\cdots$ & $\cdots$ &$\cdots$&$\cdots$&$\cdots$& $\cdots$ & $\cdots$ & $\cdots$ & $|\theta - \mathrm{PA} |\leq20$ & $\cdots$ \\
\hline
DK Tau A & $\cdots$ &Shoulder& I7&$\cdots$&$\cdots$& 7.4(58) & $\cdots$ & $\cdots$ & $\cdots$ & $0 \leq \theta \leq 359$ & $\cdots$ \\
\hline
RW Aur B &Central Hole & $\cdots$ & B6&$\cdots$& 6.4(39) &$\cdots$& $\cdots$ & $\cdots$ & $\cdots$ &  $|\theta - \mathrm{PA} | \leq 40$ & $\cdots$ \\
 & $\cdots$ &Outer Disk-Skirt& I12&$\cdots$&$\cdots$& 11.8(72) & $\cdots$ & $\cdots$ & $\cdots$ &  $|\theta - \mathrm{PA} | \leq 40$ & $\cdots$ \\
\hline
UY Aur A & $\cdots$ & $\cdots$ & $\cdots$ &$\cdots$&$\cdots$&$\cdots$& $\cdots$ & $\cdots$ & $\cdots$ & $0 \leq \theta \leq 359$ & $\cdots$ \\
\end{longtable}
\end{landscape}
\twocolumn

\subsection{Definition of Rings, Gaps, and Inflections}\label{sec:ra_extraction}

\begin{figure}[t]
\begin{center}
\includegraphics[width=0.48 \textwidth]{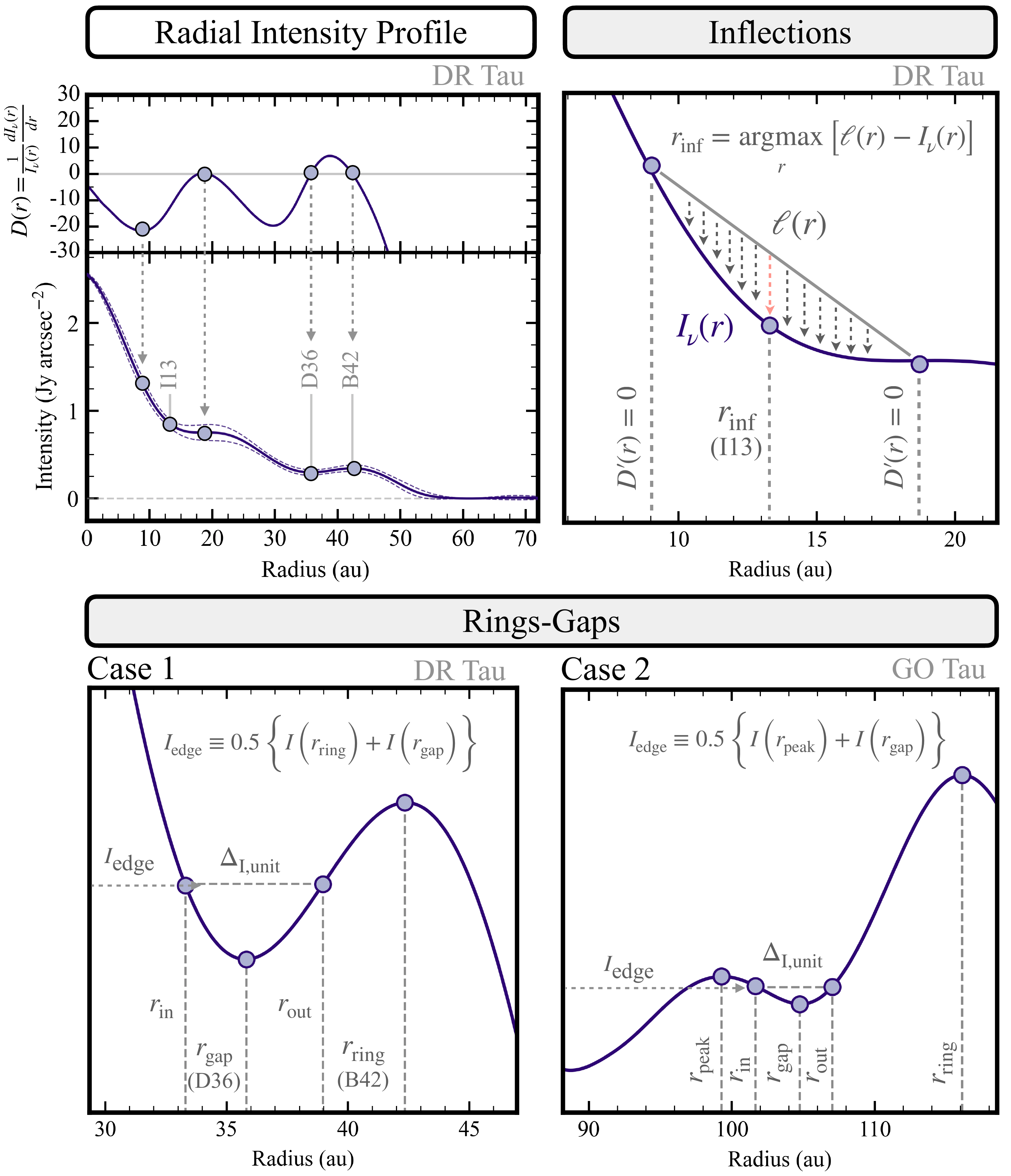}
\end{center}
\caption{Definition of disk substructures. Upper left: the identification of the inflection point ($I$, $r_{\rm inf}$), gap ($D$, $r_{\rm gap}$), and ring ($B$, $r_{\rm ring}$) from the slope of the azimuthally averaged radial intensity profile using the case of DR Tau as an example. Upper right: the method for determining the inflection point ($I$, $r_{\rm inf}$) using the case of DR Tau. Lower left and right: the methods for determining the gap width ($\Delta_{\rm I, unit}$) and depth ($\delta_{\rm I}$) using the cases of DR Tau and GO Tau.}
\label{fig:definition_substructre}
\end{figure}

From the radial profiles obtained in the previous subsection, we identify characteristic substructures in the profile. We define rings, gaps, and inflections based on the derivative of the radial intensity profile with respect to the radial coordinate. We define this derivative $\mathcal{D}(r)$ as 
\begin{equation}
\mathcal{D}(r) = \frac{1}{I_{\nu}(r)}\frac{d I_{\nu}(r)}{dr}.
\end{equation}
\noindent
The following describes the definition of the disk substructures we consider in this study.

\noindent
\textbf{Ring ($r_{\rm ring}$).} Rings are defined as the local maximum of the radial intensity profile, i.e., $\mathcal{D}(r_{\rm ring}) = 0$ and $\mathcal{D^{\prime}}(r_{\rm ring}) < 0$, where $\prime$ denotes the derivative with respect to $r$. Following the notation of \cite{ALMAPartnership2015}, we label the ring features with ``B'' (Bright) followed by a number indicating their location in astronomical units.

\noindent
\textbf{Gap ($r_{\rm gap}$).} Gaps are defined as the local minimum of the radial intensity profile, i.e., $\mathcal{D}(r_{\rm gap}) = 0$ and $\mathcal{D^{\prime}}(r_{\rm gap}) > 0$. We label the gap features with “D” (Dark) followed by a number indicating the location in astronomical units.

\noindent
\textbf{Inflection ($r_{\rm inf}$).} Inflections are defined as $\mathcal{D}^{\prime\prime}(r_{\rm inf})=0$, $\mathcal{D}^{\prime}(r_{\rm inf})>0$, and $\mathcal{D}(r_{\rm inf}) < 0$. We label the inflections with ``I'' followed by a number indicating the location in astronomical units. Intuitively, this is a small ``dip'' in a decreasing profile of radial intensity, as shown in the left panel of Figure \ref{fig:definition_substructre}. Such features may be caused by beam smearing of a narrow gap-like structure in a decreasing intensity profile. This feature has been visually identified in previous studies due to numerical difficulties in specifying the higher-order derivative of intensity profiles (e.g., \cite{Cieza2021}). In our analysis, this feature is more rigorously identified using the consecutive local minima and maxima of $\mathcal{D}(r)$ that exist in the region $\mathcal{D}(r)<0$. As shown in the left panel of Figure \ref{fig:definition_substructre}, the radial location of the inflection $r_{\rm inf}$ can be determined as the point at which the maximum deviation occurs between the linear line $\ell(r)$ connecting the local minimum and maximum of $\mathcal{D}(r)$ and the intensity profile $I_{\nu}(r)$,
\begin{equation}\label{eq:inflection}
r_{\rm inf} = \argmax_{r} [\ell(r) - I_{\nu}(r)] ~ \mathrm{subject~to} ~ \ell(r) \geq I_{\nu}(r)
\end{equation}
\noindent
To identify the location of $r_{\rm inf}$, we partially use the $\tt Kneed$\footnote[5]{The $\tt Kneed$ approach detects the knee point on a radial profile and is publicly available at \url{https://github.com/arvkevi/kneed}}, a Python package that identifies the inflection point to fit the data based on Equation \ref{eq:inflection}.

\subsubsection{Gap Width and Depth}\label{sec:gap_width_depth}

At the location of gap features, we measure the gap width $\Delta_{\rm I, unit}$, normalized gap width $\Delta_{\rm I}$, and gap depth $\delta_{\rm I}$. To determine the width and the depth of the gap, we first need to define the inner and the outer edges, $r_{\rm in}$ and $r_{\rm out}$, of the gap. In our samples, we find that any gap at $r=r_{\rm gap}$ is associated with a ring at $r=r_{\rm ring,1}$ just outside the gap.

In previous studies (e.g., \cite{Huang2018, Zhang2018}), the location of the outer gap edge $r_{\rm out}$ is defined as the largest value satisfying the criteria $I_{\rm edge}= I_{\nu}(r_{\rm out}) $ and $ r_{\rm gap} < r < r_{\rm ring}$ and that of the inner gap edge $r_{\rm in}$ is defined as the smallest value satisfying the criteria $I_{\rm edge}= I_{\nu}(r_{\rm in}) $ and $ r < r_{\rm gap}$. Here, $I_{\rm edge}$ is defined as
\begin{equation}
I_{\rm edge} \equiv 0.5 \left\{ I_{\nu}\left(r_{\rm ring}\right)+I_{\nu}\left(r_{\rm gap}\right) \right\}.
\end{equation}
\noindent
This definition is illustrated as Case 1 in the middle panel of Figure \ref{fig:definition_substructre}.

In some cases, we find that this definition of gap edge locations fails because the brightness inside the gap location is too weak for $r_{\rm in}$ to be well-defined. We consider that there is still a ``gap'' if there is a local peak at $r=r_{\rm peak} < r_{\rm gap}$. This situation is illustrated in the right panel of Figure \ref{fig:definition_substructre} as Case 2. We define $r_{\rm in}$ and $r_{\rm out}$ as in Case 1 but with
\begin{equation}
I_{\rm edge} \equiv 0.5 \left\{ I_{\nu}\left(r_{\rm peak}\right)+I_{\nu}\left(r_{\rm gap}\right) \right\}.
\end{equation}
\noindent
We also apply Case 2 when the system has a localized emission at the position of the central star (i.e., inner disk) surrounded by a gap structure. In this case, we set $r_{\rm peak} = 0$ to define the inner and the outer edges of the gap. The targets that have gaps around a central emission are GM Aur (D14), DM Tau (D14), CQ Tau (D12), and RY Tau (D5).

Once we have obtained the location of the gap edges, it is possible to define the width and the depth of the gap. We define the gap width in units of length $\Delta_{\rm I, unit}$ and normalized gap width $\Delta_{\rm I}$ as
\begin{equation}
\Delta_{\rm I, unit}= r_{\rm out}-r_{\rm in}, \rm ~ and ~\Delta_{\rm I}=\frac{\left(r_{\rm out}-r_{\rm in}\right)}{r_{\rm out}},
\end{equation}
\noindent
respectively.
For the gap depth $\delta_{\rm I}$, we define
\begin{equation}
\delta_{\rm I} = \frac{I_{\nu}\left(r_{\rm ring}\right)}{ I_{\nu}\left(r_{\rm gap}\right)}.
\end{equation}
\noindent
It should be noted that the measurements of the gap depth can be significantly affected by the faint emission at the gap location. Specifically, when the width of the gap is larger than the spatial resolution by a factor of several, the emission at the gap location $I_{\nu}\left(r_{\rm gap}\right)$ can easily reach the noise level. This effect is prominent in three disks: AB Aur (D62), LkCa 15 (D17), and AA Tau (D14). Therefore, we consider these to be upper limits of depth and exclude them from our analyses when we use $\delta_{\mathrm{I}}$.

We have detected 33 gap–ring pairs in 21 disks and have been able to measure the widths, depths, and uncertainties for $90\%$ (30/33) of all gaps. We have failed to accurately measure the properties of three gaps located at very large distances ($r_{\rm gap}>100$~au; GM Aur (D152), CI Tau (D126), and DM Tau (D103)) since the emission at these locations are too faint. While presenting their approximations, we regard them as spatially unresolved gaps and exclude them from our analyses involving depths or widths of the gaps.

Among the gaps with measurements of widths and depths, Case 1 accounts for $80\%$ (27/33), while Case 2 represents $20\%$ (6/33). Out of all the gaps studied, $60\%$ (20/33) show $\Delta_{\rm I, unit}$ being larger than the geometric mean of the effective spatial resolution so they are considered to be spatially resolved. The remaining $40\%$ (13/33) of the gaps have not been fully spatially resolved. Their widths and contrasts appear narrower and lower than the spatially resolved ones, although we should be careful in drawing more quantitative conclusions. The measurements of gap properties are summarized in Table \ref{tab:disk_substructure}.

\subsection{Categorization of Disk Structures}\label{sec:categorize_radialdisk}

In the previous Section, we have defined three characteristic axisymmetric structures seen in the radial intensity profiles: rings, gaps, and inflections. The combination(s) of these structures as well as some asymmetric structures appear in actual observations. From the investigations of the morphology of the 43 disks, we have found that the disk substructures can be grouped into nine categories: eight types of axisymmetric structures and asymmetry.

\begin{figure*}[t]
\begin{center}
\includegraphics[width=0.98 \textwidth]{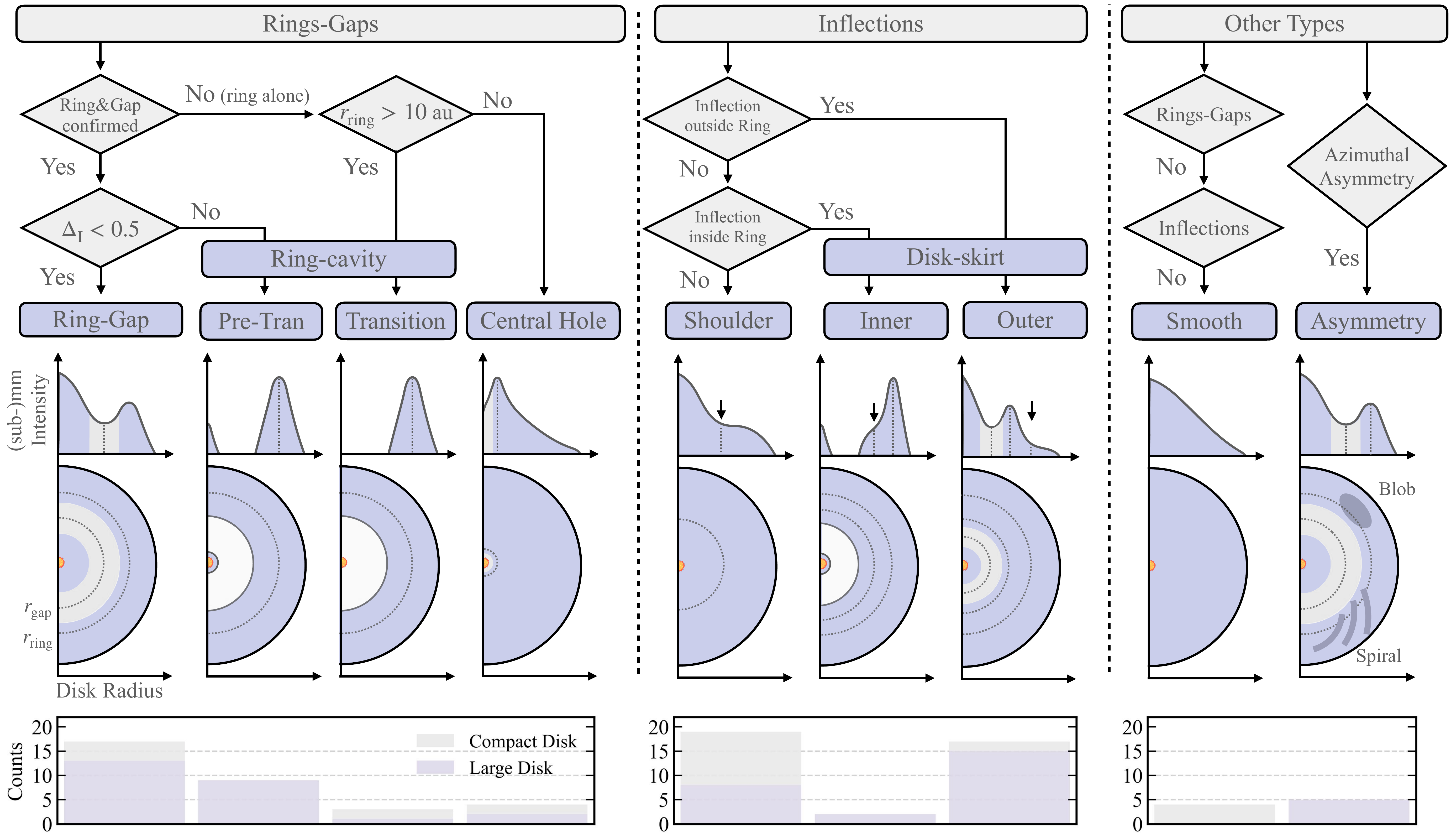}
\end{center}
\caption{Flow chart to determine (sub-)mm disk morphologies. We define nine morphologies starting from three primary features: ``Rings-Gaps'', ``Inflections'', and ``Other Types''. Some disks contain a combination of different features. The endpoints of the flowcharts show schematic representations of radial intensity profiles for each morphology. The bottom panels show the histogram counting the number of disk morphologies. When there are the same morphologies within a disk, we count them as one unit. The bar color differs between compact disks (gray) and large disks (purple).}
\label{fig:disk_clasification}
\end{figure*}

Figure \ref{fig:disk_clasification} presents the flowchart of categorization. We start with three large categories: ``Rings-Gaps'', ``Inflections'', and ``Others''. Then, each of the large categories is subdivided into small groups. Some disks have two or more substructures. The morphological category for each disk is summarized in Table \ref{tab:disk_substructure}.

\subsubsection{Rings-Gaps}\label{sec:ring-gap_feature}

As mentioned in Section \ref{sec:ra_extraction}, there is always a ring outside a gap. We denote this pair of a ring and a gap as the ``Rings-Gaps'' feature. This is subdivided into ``Ring Gap'', ``Ring-Cavity'', and ``Central Hole'' depending on the gap width and location. The ``Ring-Cavity'' is further divided into ``Transition Disk'' and ``Pre-transition Disk''.

\noindent
\textbf{Ring-Gap.} We denote a narrow gap with the width of $\Delta_{\rm I} < 0.5$ as ``Ring-Gap''. This is a structure where we observe ``a concentric, axisymmetric pattern of alternating intensity enhancements (rings) and depletions (gaps)'' as noted by \cite{Andrews2020}. This pattern appears most frequently in our sample. We find a total of 24 narrow gap/ring structures in 17 disks. The improvement of spatial resolution by SpM has enabled us to detect 12 gap/ring structures that were not previously identified in the conventional CLEAN images (see Table \ref{tab:disk_substructure}). It is worth noting that the substructure within the T Tau N disk had been reported in our earlier studies \citep{Yamaguchi2021} and ring-gap features in several other disks were also identified using a non-parametric one-dimensional fitting approach with $\tt frank$ \citep{Jennings2022taurus}. The consistency or difference in the obtained disk substructure between $\tt frank$ and the SpM approach is described in Appendix \ref{appendix:comparison_other_imaging}.

\noindent
\textbf{Ring-cavity.} We use ``Ring-Cavity'' to denote either a wide gap with $\Delta_{\rm I} > 0.5$ or a ring at $r_{\rm ring} > 10$ au. We further sub-categorize the former as ``Transition Disk'' (TD) and the latter as ``Pre-transition Disk'' (PTD). The Ring-cavity structure is present in 12 disks, comprising three TDs and nine PTDs (see Table \ref{tab:disk_substructure}).

All TDs in our sample (CIDA 9A, IP Tau, and UX Tau) show only one ring structure and no additional ring-gap features are found. To further investigate the ring structures in TDs, we measure the ring width by fitting the radial intensity profile with a Gaussian function. We use $\tt optimize.leastsq$ within $\tt Scipy$ that employs the Levenberg-Marquardt algorithm for a nonlinear least-squares problem. We determine the ring width $\Delta_{\rm ring}$ as the FWHM of the best-fit Gaussian function. The widths are $0''.13$ (22 au), $0''.11$ (14 au), and $0''.11$ (15 au) for CIDA 9A, IP Tau, and UX Tau, respectively and the error is at the negligible level with $\pm 0''.01$. All of the ring widths are found to be spatially resolved compared to the effective spatial resolutions (i.e., $ \Delta_{\rm ring} > \theta_{\rm eff}$) and, interestingly, the ratio of the ring location to the ring width is $\sim 2$ for all the three TDs.

TD and PTD are conventionally categorized based on the excess emissions in the near-infrared (e.g., \cite{Espaillat2014, Pessah2017}), which is an indicator of the close to the central star. However, our TDs/PTDs do not exactly match with those determined by near-infrared excess. For example, GM Aur and AA Tau are classified as PTD in our study while those are classified as TD from infrared, and vice versa for IP Tau and UX Tau. Here, we have used the TD/PTD classification based on infrared presented in \citet{Francis2020}. Such discrepancies may be attributed to different distributions between $\mu$m-size grains (traced by near-infrared observations) and mm-size grains (traced by sub-mm observations), as noted by \cite{Marel2023}.

\noindent
\textbf{Central Hole.} We denote disks with a local minimum at the central star and with a small inner ring of $r_{\rm ring} < 10$ au as ``Central Hole''. This structure is seen in four disks: DM Tau, UZ Tau E, BP Tau, and RW Aur B. There could be a small-scale inner disk at the central star (e.g., \cite{Perez2018, Hashimoto2021dmtau}), but the interpretation of this structure is difficult due to limited spatial resolutions. We note that the SpM image indicates that the inner ring of UZ Tau E is potentially asymmetric. The emission in the west is $\sim 10\%$ brighter than in the east.

\subsubsection{Inflections}\label{sec:inflection_feature}

The Inflection feature is subdivided into two groups, ``Shoulder'' and ``Disk Skirt'', based on the presence of adjacent ring features.

\noindent
\textbf{Disk-skirt.} We denote the structure of a combination of an inflection point and a ring that are separated by 40 au or less as ``Disk-skirt''. If the inflection point is located inside the ring, we use ``Inner Disk-skirt'' while if it is outside the ring, we use ``Outer Disk-skirt''. The radial intensity profiles with this ``Disk-Skirt'' feature tend to zero slower than in the case of just a ``Ring'' as the distance from the ring increases.

We have observed Disk-skirt features in 15 of our targets. Two of them (LkCa 15 and MWC 758) show both the inner and outer Disk-skirt features while the other disks show only outer Disk-skirt features. In addition, visual inspection of the radial intensity profiles at the outer radii of DG Tau (I26) and IQ Tau (I68) show very similar behavior to the outer Disk-skirt although they do not show a clear ring structure. We add these two disks as exceptional cases having outer Disk-skirt. In total, we have observed the Disk-skirt features in $40\%$ (17/43) of all disks, with two disks having inner Disk-Skirt and all 17 disks having outer Disk-skirt. Interestingly, Disk-skirt is found in $70\%(15/21)$ of the large disks while only $10\%(2/22)$ of the small disks have Disk-skirt. The large difference in the number of inner and outer Disk-skirt might indicate that they are actually of different origins. The outer Disk-skirt feature may be created by drifting dust particles that will be trapped in a pressure bump at an inner radius (e.g., \cite{Cieza2021, Leiendecker2022}). The inner Disk-skirt, on the other hand, may just be caused by limited spatial resolution. Two narrow rings that are not spatially resolved could produce a structure that resembles the inner Disk-skirt.

\noindent
\textbf{Shoulder.} We denote the inflections that are not associated with any ring as ``Shoulder'', except for DG~Tau(I26) and IQ~Tau(I68), which are included in the outer Disk-skirt. The number of disks showing Shoulder features amounts to 19, which is $40\%$ (19/43) of our sample. The shoulder features are typically found at $5-40$ au from the central star. Most of the compact disks harbor shoulder features and do not show other disk features in radial intensity profiles. The shoulder feature can be caused by insufficient spatial resolution for narrow and shallow gaps. Therefore, more sensitive and higher spatial resolution observations are needed to judge if the actual structure is gap-like or shoulder-like.

\subsubsection{Other Types}\label{sec:other_feature}

As disk structures that do not fall into either ``Rings-Gaps'' or ``Inflections'', we define ``Smooth'' and ``Asymmetry''.

\noindent
\textbf{Smooth Disk.} We denote disks without rings, gaps, or inflections as ``Smooth''. Four ($=9\%$) disks of our 43 samples, V836 Tau, DH Tau A, HN Tau A, and UY Aur A, are categorized into this group. Three out of four of these disks are associated with the primary stars of binary systems and all of them have disk radii less than 40 au. The disks are spatially resolved only by $2-3$ beams even with SpM imaging. Therefore, disk structures, if exist, may have been smeared out.

In the disk around V836~Tau, we see a hint of some substructures. A dual blob-like feature at the center of the disk is seen in the SpM image, whose spatial resolution is improved by a factor of $\sim 3$ compared to the CLEAN image. The peaks of the blobs are at $\sim 0''.03$ from the central star and are only $\sim 2\%$ brighter than the center of the image. This feature could indicate the presence of a small cavity but we conservatively categorize the disk around V833~Tau as ``Smooth''.

\noindent
\textbf{Asymmetry.} Finally, we denote ``Asymmetry'' for disks with asymmetric structures on 2D SpM images. In this paper, we only identify the asymmetric structures with visual inspection since we mainly focus on the structures that appear in azimuthally averaged radial profiles. We identify five disks as having obvious asymmetric structures: AB Aur, CQ Tau, RY Tau, MWC 758, and CIDA 9A. Except for CIDA 9A in the M-type star, they are intermediate-mass stars of the F-A type ($M_{*} = 1.5-2.2~M_{\odot}$). Their disks are of Ring-cavity type in common. We will explore the implications of these substructures in the asymmetric disks in more detail in future investigations by treating each case as an individual study.

\section{Distribution of Gap Radii}\label{sec:radialdistribution_gap}

\begin{figure*}[!htbp]
\begin{center}
\includegraphics[width = 0.85 \textwidth]{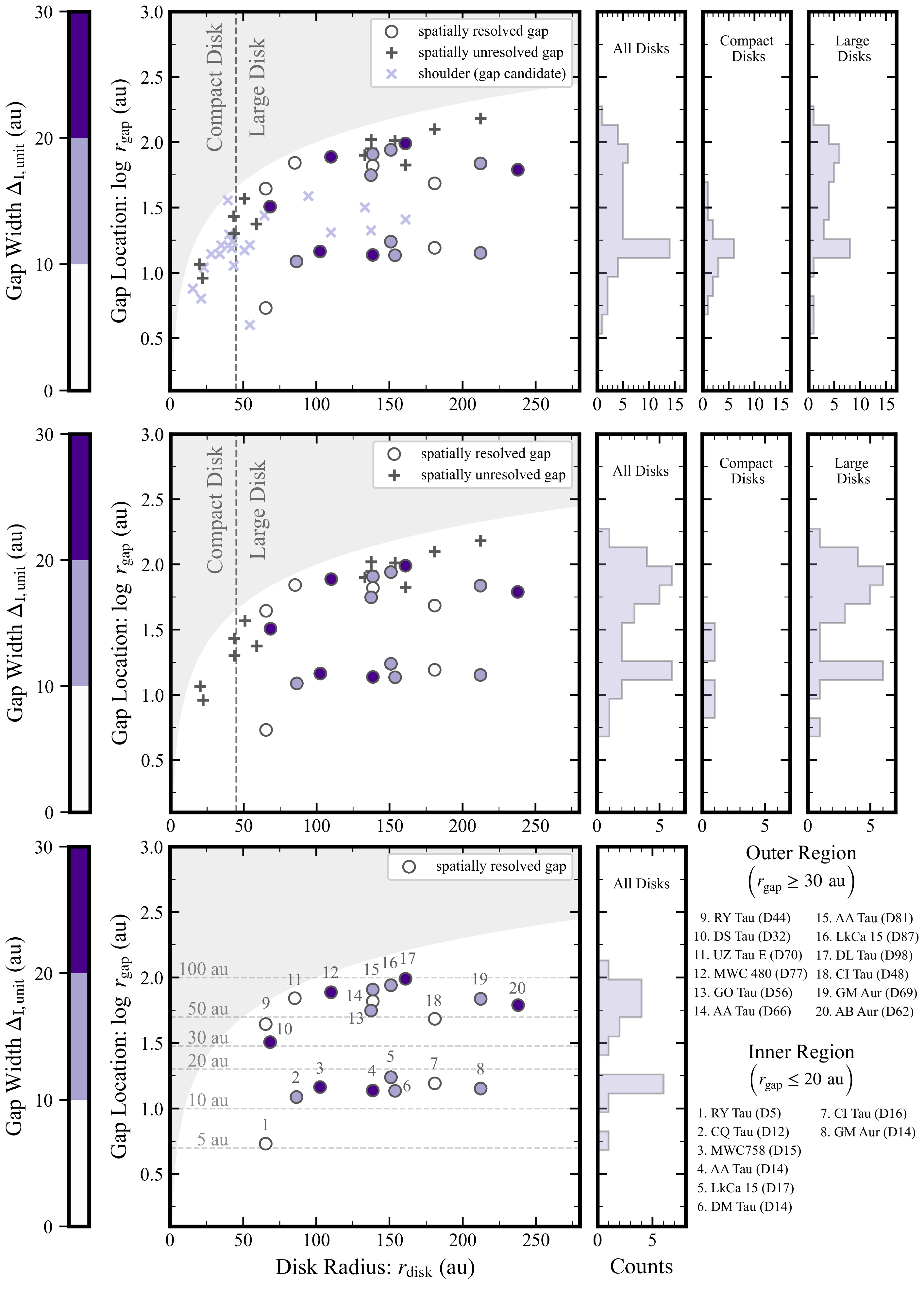}
\end{center}
\caption{
The distribution of gap locations ($r_{\rm gap}$) of our targets as a function of disk radii ($r_{\rm disk}$). The upper panel is for all identified gaps and shoulders, where circles, plus symbols and cross symbols indicate spatially resolved gaps, spatially unresolved gaps, and shoulders, respectively. The colors of spatially resolved gap symbols indicate the gap width in units of au. Gray regions indicate the gap location larger than the disk radius so that no gap exists in this region. The middle and bottom panels indicate the distributions without shoulders and with spatially resolved gaps only, respectively. In the bottom panel, each spatially resolved gap is numbered and the corresponding source names are indicated. Histograms on the right illustrate the distribution of gap locations. The histograms are produced for all the disks, for compact disks ($r_{\rm disk}<45$~au), and for large disks ($r_{\rm disk}>45$~au) in the top and middle panels. We define the inner region gaps ($r_{\rm gap}\leq20$~au) and outer region gaps ($r_{\rm gap}\geq30$~au) based on the bimodality of the distribution of spatially resolved gaps.}
\label{fig:radialgap_location}
\end{figure*}

In this Section, we investigate the distribution of the radial location of gaps. Figure \ref{fig:radialgap_location} shows the location of gaps and gap candidates, which are identified as Shoulder, as a function of disk radius $r_{\rm disk}$. We use different symbols for spatially resolved gaps (gap width larger than spatial resolution; 20 gaps), spatially unresolved gaps (gap width smaller than spatial resolution; 13 gaps), and the structures identified as Shoulder (19 in total). The histograms of the locations of all the gaps and candidates are shown in the top right panel of Figure \ref{fig:radialgap_location}. We find a broad distribution of gap locations, ranging from 5 au to 100 au from the central host stars, with a notable concentration around 10-20 au. The absence of gaps located at 10 au or less may be due to observational difficulty. Spatial resolution may not be enough to identify small-scale structures at the innermost radii. Moreover, the inner region of the disk is likely to be optically thick so it is hard to observe surface density structures.

The gaps in compact disks (size less than 45~au) are predominantly located at 10-20~au from the central star, but they are all either spatially unresolved or gap candidates. Therefore, again, the statistical properties of gaps in compact disks may be affected by insufficient spatial resolution. The gaps and gap candidates around large disks, on the other hand, show qualitatively different distributions. They are distributed at all the range of 5-100~au from the central star. Interestingly, however, the gap candidates (Shoulders) and the gaps are distributed differently. The gap candidates are mainly at 20-30 au from the central star while the gaps are located either inside the outside of the region dominated by the candidates.

In the middle and bottom panels of Figure \ref{fig:radialgap_location}, we dropped the gap candidates (Shoulders) from the plot. We now clearly see the bimodality of gap distribution. The gaps are located either in the inner ($r\lesssim 20$~au) or the outer ($r\gtrsim 30$~au) part of the disk, and the region in between may be considered a ``gap desert''. We find that the classification of the disks (discussed in Section \ref{sec:categorize_radialdisk}) is different for disks having gaps in the inner region and in the outer region. Disks bearing gaps in the outer region include both Ring-Gap (nine disks) and PTDs (two disks), while $90\%(=7/8)$ of disks having gaps in the inner region are PTDs. The difference in disk types may be connected to the bimodality of clear gaps that can be observed. Weak gaps that may be observed as ``gap candidates'' may exist all over the disk.

\section{Correlation of Disk Substructures}\label{sec:corr_disksub}

In this Section, we explore the relationships of ring and gap properties. We consider 20 spatially resolved gaps listed in Figure \ref{fig:radialgap_location} and investigate the correlation among their locations of the gaps and the rings associated with them, widths, and depths. We note that it is not possible to obtain reliable measurements of depths for three of the wide gaps (i.e., AB Aur (D62), LkCa 15 (D17), and AA Tau (D14)) so we exclude them when calculating the correlation between the depth and other properties.

\subsection{Overview of Correlations}

\begin{figure*}[!ht]
\begin{center}
\includegraphics[width = 0.98 \textwidth]{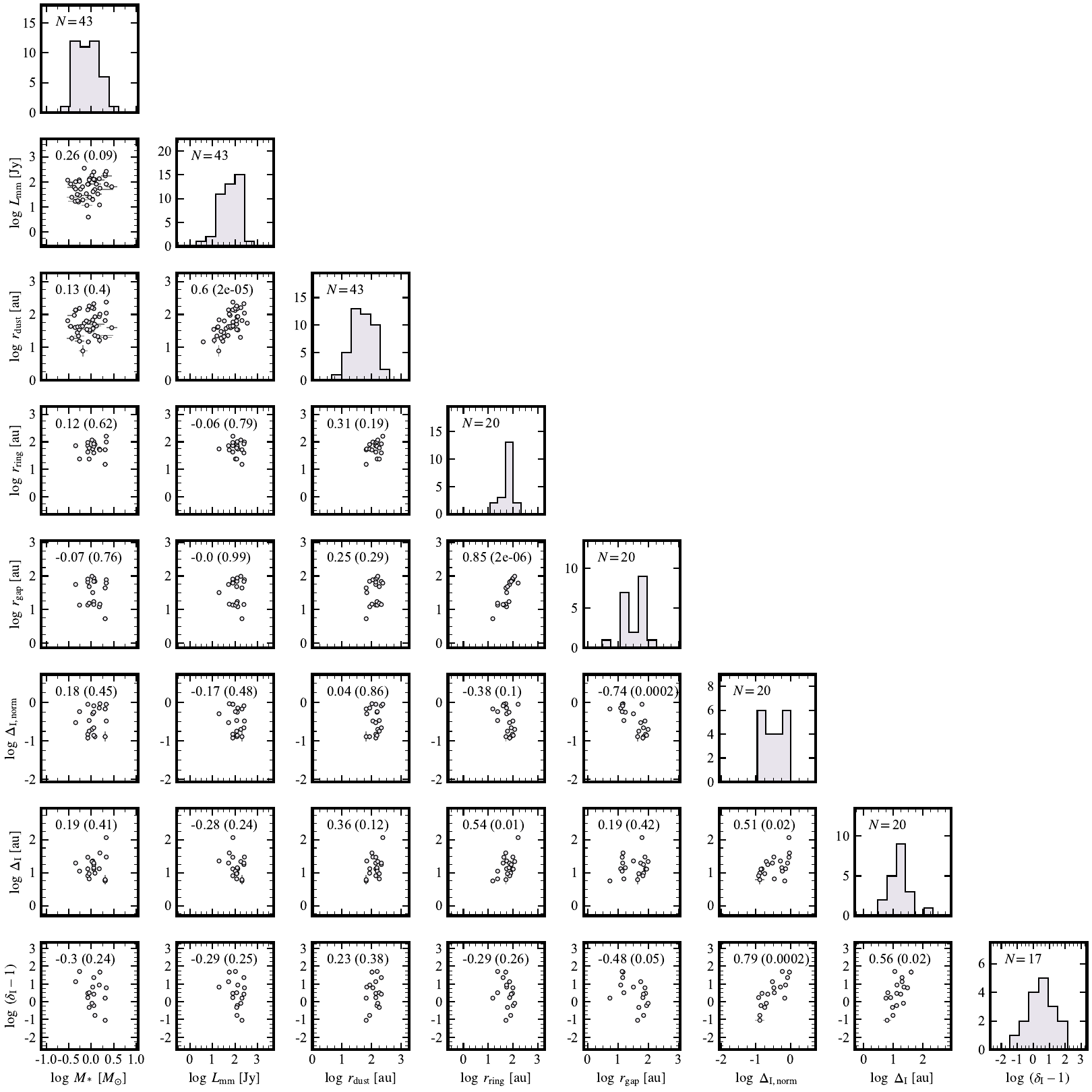}
\end{center}
\caption{Correlation matrix of the morphological quantities of the disks, stellar mass, and millimeter luminosity (flux density scaled at the distance of 140~pc). We have explored the correlation on a logarithmic scale. The values in each figure indicate the Pearson correlation coefficients along with the $p$-values in parenthesis. For $(r_{\rm gap}, r_{\rm ring}, \Delta_{\rm I}, \Delta_{\rm I, unit}, \rm ~and~ \delta_{\rm I})$, we only use spatially resolved cases (i.e., $\Delta_{\rm I, unit} > \theta_{\rm eff}$). The histograms show the number of samples for each quantity.}
\label{fig:correlation_matrix}
\end{figure*}

Figure \ref{fig:correlation_matrix} shows the correlations among gap and ring quantities. We also calculate the correlation between quantities of disk substructures and stellar properties or disk size. In this analysis, we have found correlations in (1) the radial positions of gaps and rings, (2) the gap locations and their widths, and (3) the gap widths and depths. Meanwhile, we do not observe a robust correlation ($\rho < 0.3$) between stellar mass and disk substructure properties. We discuss each of the correlations in subsequent sections in more detail.

\subsection{Ring and Gap Radii}\label{sec:ring_gap_relation}

\begin{figure*}[ht]
\begin{center}
\includegraphics[width=0.98 \textwidth]{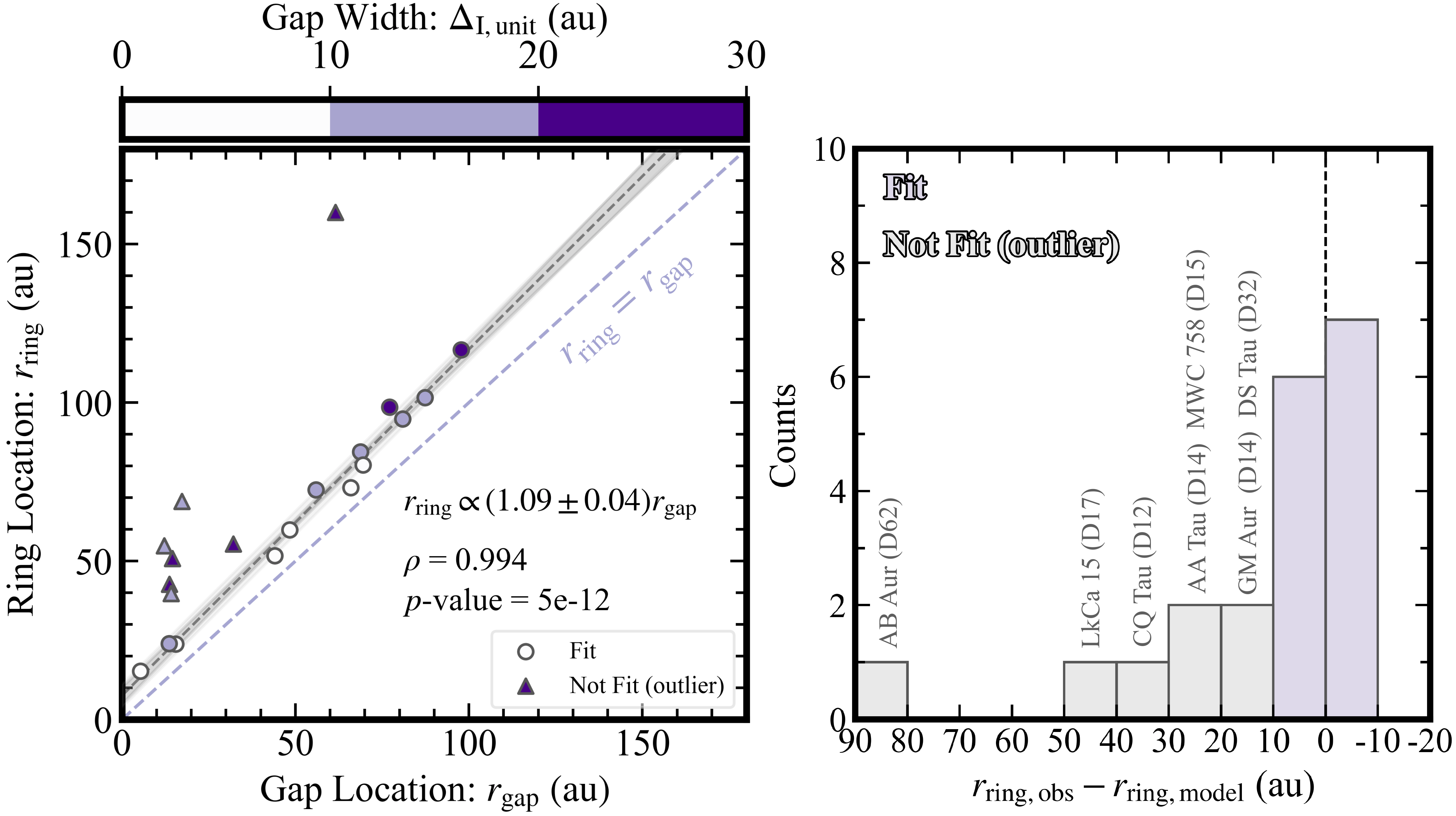}
\end{center}
\caption{Left: the scaling relation between the locations of gaps and the corresponding rings (the closest ring outside the gap). The black dashed line indicates the median scaling relation obtained from Bayesian linear regression and the dark gray area represents the $68\%$ confidence interval around the median. The light gray area corresponds to the inferred scatter. Circles are data points used for the regression while triangles are outliers that are excluded from the regression. The colors of each symbol indicate the gap width $\Delta_{\rm I, unit}$. The light purple dashed line represents $r_{\rm ring} = r_{\rm gap}$ for reference. The model equation obtained by the linear regression, Pearson's correlation coefficient ($\rho$), and $p$-value calculated from the sample distribution are shown in the lower left corner. Right: the histogram of residuals from the linear regression model for the observed samples. The purple bars are for the gaps used for deriving the linear regression model and the gray bars are outliers that are not used to derive the model. The source names and the locations of gaps of the outliers are indicated for reference.}
\label{fig:ring-gap_relation}
\end{figure*}

We investigate the relationship between the radial location of gaps and rings that are associated with the gaps. The left panel of Figure \ref{fig:ring-gap_relation} shows that these two quantities are tightly correlated. There are, however, a few outliers. We find all of the outliers are in the Ring-cavity type (see Section \ref{sec:ring-gap_feature}). The Pearson correlation coefficient after removing these outliers is $\rho = 0.994$ with the $p$-value falling below the significance threshold ($p$-value $ < 0.01$). Bayesian linear regression in a linear-space results in the relation:
\begin{equation}
\left( \frac{r_{\rm ring}}{\mathrm{au}} \right)= 7.58^{+2.59}_{-2.62} + 1.09^{+0.04}_{-0.04} \left( \frac{r_{\rm gap}}{\mathrm{au}} \right),
\end{equation}
\noindent
with a scatter of $3.65\pm 1.02$ au. For our samples, most of the gaps are located at $>10$~au ($\gg 1.09$~au) from the central star, so we conclude that the rings associated with narrow gaps (not a cavity-like structure) are located at $\sim 10\%$ larger distance from the central star compared to the gap. We need higher spatial resolution data to investigate if this holds for gaps and rings at a few au scales from the central star. 

In the right panel of Figure \ref{fig:ring-gap_relation}, we show a histogram of the residual of the linear regression model. All the samples used for fitting show residuals less than 10~au. The ring locations of the outliers (not used for fitting), in contrast, exhibit substantial deviations ranging from 10 to 70 au from the model.

\subsection{Gap Width and Gap location}\label{sec:gapwidth_gaploc_relation}

\begin{figure*}[htbp]
\begin{center}
\includegraphics[width=0.98 \textwidth]{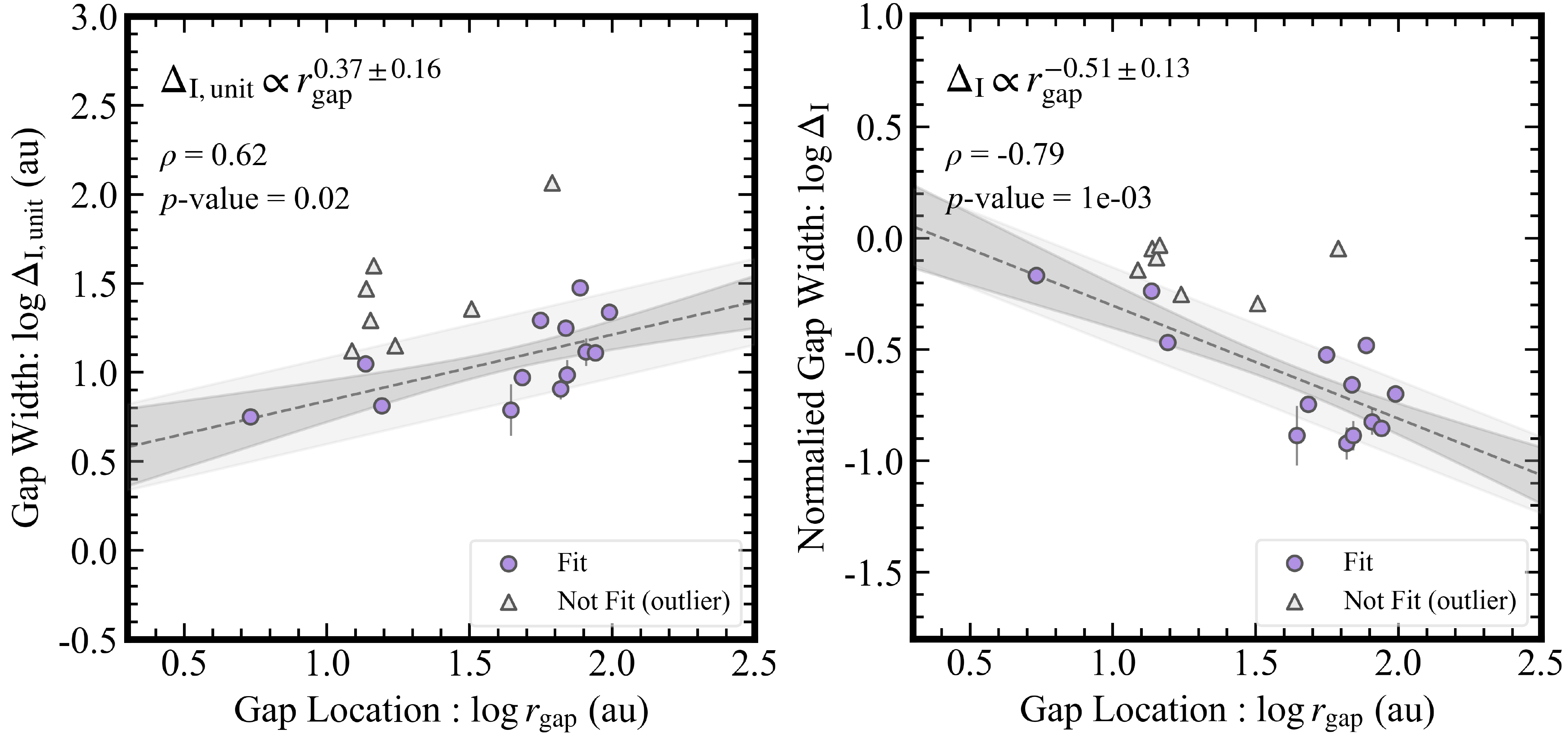}
\end{center}
\caption{Relationship between gap width and gap location on the logarithmic scale. The left panel uses the gap width in units of au ($\Delta_{\rm I, unit}$) and the right panel uses the normalized gap width ($\Delta_{\rm I}$). The black dashed line is the median scaling relation obtained from Bayesian linear regression, while the dark gray area indicates the $68\%$ confidence interval around the median. The light gray area corresponds to the inferred scatter. Circles are data points used in the regression, while triangles are the outliers in Figure \ref{fig:ring-gap_relation}. The model obtained by the linear regression, Pearson's correlation coefficient ($\rho$), and $p$-value calculated from the sample distribution are shown in the upper left corner.}
\label{fig:gapwidth_rgap_relation}
\end{figure*}

Figure \ref{fig:gapwidth_rgap_relation} shows the correlation between the gap location and widths. We investigate both the gap width in units ($\Delta_{\rm I,unit}$, left panel) and the normalized one ($\Delta_{\rm I}$, right panel). We exclude the ``outliers'' in the $r_{\rm ring}-r_{\rm gap}$ correlation and we obtain positive correlation between $\log \Delta_{\rm I,unit}$ and $\log r_{\rm gap}$ ($\rho = 0.62$, $p$-value $= 0.02$) while negative correlation between $\log \Delta_{\rm I}$ and $\log r_{\rm gap}$ ($\rho = -0.79$, $p$-value $>0.01$). Bayesian linear regression in the logarithmic plane results in the relationship
\begin{equation}
\log\left(\frac{\Delta_{\rm I, unit}}{\mathrm{au}} \right) = 0.47^{+0.27}_{-0.27}+0.37^{+0.16}_{-0.16}\log \left(\frac{r_{\rm gap}}{\mathrm{au}} \right) ,
\end{equation}
\noindent
with a scatter of $0.24\pm 0.12$ dex for the width with units while 
\begin{equation}
\log \Delta_{\rm I} = 0.21^{+0.22}_{-0.23}-0.51^{+0.13}_{-0.13}\log \left(\frac{r_{\rm gap}}{\mathrm{au}} \right),
\end{equation}
with a scatter of $0.17\pm 0.05$ dex for the normalized width. If a gap is created by a planet, $\Delta_{\rm I}$ is larger if the mass of the planet in the gap is larger (e.g., \cite{Kanagawa2015, Zhang2018}). The negative correlation between $\Delta_{\rm I}$ and $r_{\rm gap}$ may suggest that the planet mass is low at the outer radii. We discuss further the implication of planet formation in Section \ref{sec:inferred_planetmass}.

\subsection{Gap Width and Gap Depth}\label{sec:gap_widthdepth_relation}
\begin{figure*}[!htbp]
\begin{center}
\includegraphics[width=0.91 \textwidth]{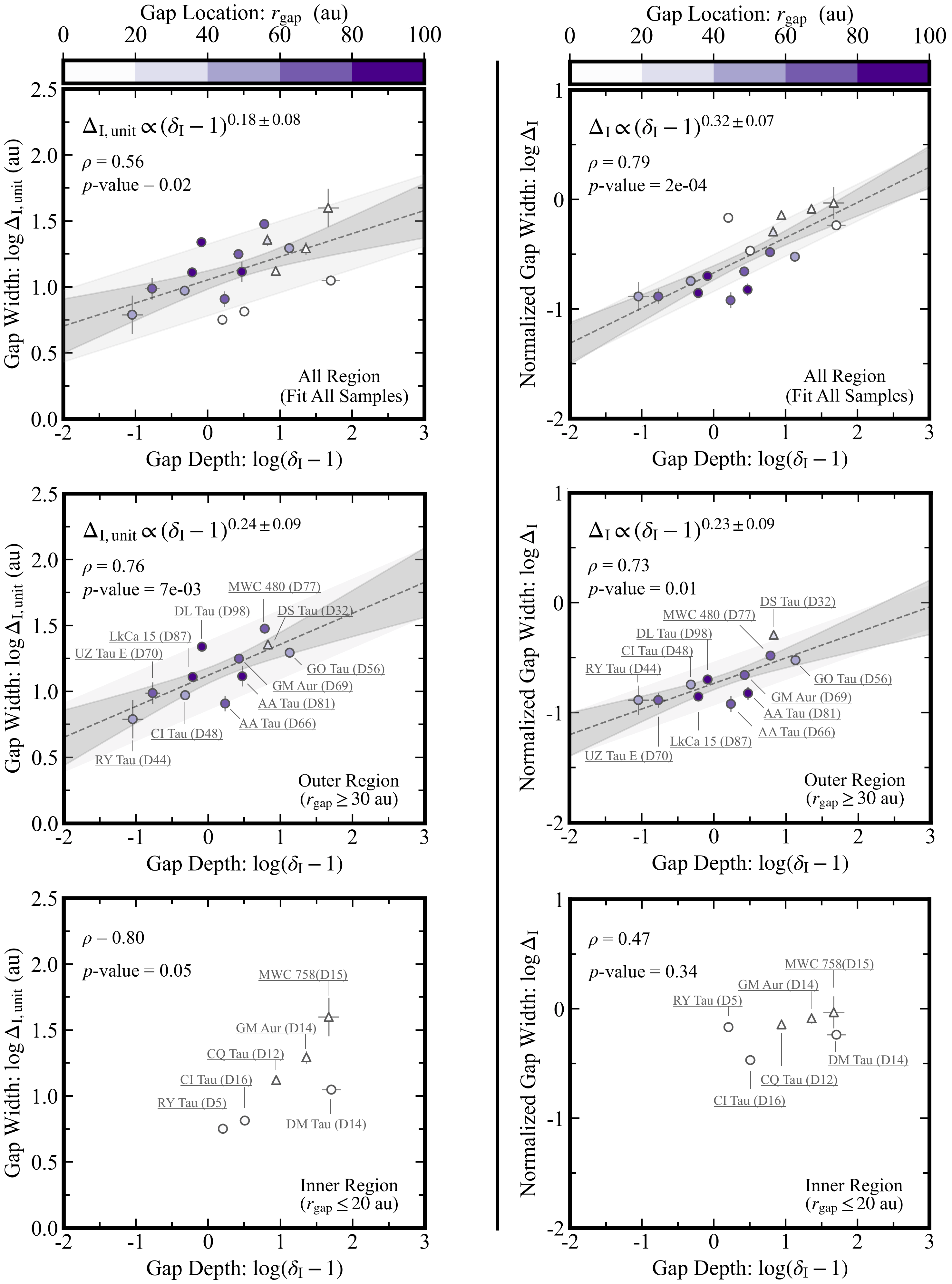}
\end{center}
\caption{Relationship between gap width and depth of the disks. The left and right panels are for gap width in units of au $\Delta_{\rm I, unit}$, and normalized gap width $\Delta_{\rm I}$, respectively. The top, middle, and bottom rows show the distributions using all samples, the samples from the outer region ($r_{\rm gap}>30$ au), and the samples from the inner region ($r_{\rm gap}<20$ au), respectively. The spatially resolved gaps ($\Delta_{\rm I, unit}>\theta_{\rm eff}$) are only used. The black dashed line indicates the median scaling relation obtained from Bayesian linear regression and the dark gray area is the $68\%$ confidence interval around the median. The light gray area corresponds to the inferred scatter. The circle and triangle symbols are the same as in Figure \ref{fig:ring-gap_relation} and their colors indicate the radial gap location $r_{\rm gap}$.}
\label{fig:gapwidthdepth_relation}
\end{figure*}

Figure \ref{fig:gapwidthdepth_relation} presents the distribution of gap widths and depths on a logarithmic scale. We use $\delta_{\rm I}-1$ rather than $\delta_{\rm I}$ for gap depth so that the value is zero when there is no gap as in the case of $\Delta_{\rm I}$, and we consider two measurements of gap width: one with units $\Delta_{\rm I, unit}$ and the normalized one $\Delta_{\rm I}$. We find a significant positive correlation ($\rho\ge 0.5$, $p-$value $\le 0.02$) between gap width and depth. Interestingly enough, the correlations are even stronger ($\rho>0.7$, $p-$value $<0.01$) if we restrict the gaps in the outer region with $r_{\rm ring}>30$~au. The inner region samples ($r_{\rm ring}<20$~au) also indicate some correlation. However, the $p-$value of $\geq 0.05$ is not small enough due to the small sample size and we do not further explore a trend for the gaps at the inner region. The Bayesian linear regression in the logarithmic plane is performed for the full sample and the samples at outer radii. The linear model is given by 
\begin{equation}
\label{eq:SL_gapwidth_depth}
\log \Delta = \mathcal{A} + \mathcal{B} \log (\delta_{\rm I}-1) + \varepsilon,
\end{equation}
\noindent
where $\Delta$ is either $\Delta_{\rm I, unit}$ or $\Delta_{\rm I}$, $\mathcal{A}$ is the intercept, $\mathcal{B}$ is the slope, and $\varepsilon$ is the Gaussian scatter along the vertical axis. The regression parameters and correlation coefficients are summarized in Table \ref{table:gapwidthdepth_relation}.

\begin{table}
\tbl{Results of the linear regressions for the gap relation.}{%
\renewcommand{\arraystretch}{1.4}\begin{tabular}{p{15mm}p{15mm}p{15mm}p{15mm}p{13mm}r}
\hline
Region & $\mathcal{A}$ & $\mathcal{B}$ & $\sigma$ & $\rho$ & $p-$val\\
\hline
$\Delta_{\rm I, unit}$ &&&&&\\
\hline
Full   & $1.05_{-0.07}^{+0.07}$  & $0.18_{-0.08}^{+0.08}$ & $0.27_{-0.11}^{+0.11}$ & 0.56 & 0.02\\
Outer & $1.12_{-0.06}^{+0.06}$  & $0.24_{-0.09}^{+0.09}$ & $0.26_{-0.21}^{+0.21}$ & 0.76 & $<0.01$\\
Inner & $\cdots$  & $\cdots$ & $\cdots$ & 0.80 & 0.05 \\
$\Delta_{\rm I}$ &&&&& \\
\hline
Full &  $-0.67_{-0.07}^{+0.06}$  & $0.32_{-0.07}^{+0.07}$ & $0.17_{-0.03}^{+0.03}$ & 0.79 & $<0.01$\\
Outer & $-0.74_{-0.06}^{+0.06}$  & $0.23_{-0.09}^{+0.09}$ & $0.19_{-0.10}^{+0.10}$ & 0.73 & $0.01$\\
Inner & $\cdots$  & $\cdots$ & $\cdots$ & 0.47 & 0.54 \\
\hline
\end{tabular}}
\begin{tabnote}
\textbf{Note}. Linear regression model is given by Equation \ref{eq:SL_gapwidth_depth}. The values quoted for $\mathcal{A}$ (intercept) and $\mathcal{B}$ (slope) are the medians of their posterior distribution, and the uncertainties are the $68\%$ confidence interval. $\sigma$ represents the standard deviation of Gaussian scatter around the linear regression. $\rho$ and $p-$val are the Pearson correlation coefficient and the corresponding $p-$value, respectively.
\end{tabnote}
\label{table:gapwidthdepth_relation}
\end{table}

\section{Discussion}\label{sec:discussion}

\subsection{Remnants of Dust Envelope}\label{sec:envelope_remnants}

In Section \ref{sec:totalflux_comparison}, we find that the flux density observed with single-dish telescopes may be $\sim 20\%$ larger than that with ALMA. This excess may be due to the emission at the envelope scale, which is resolved out in interferometric observations. \cite{Federman2023} studied the 0.9 mm continuum flux density of Orion protostars using ALMA (disk-scale) and Atacama Compact Array (ACA; envelope-scale). The survey revealed that the ALMA/ACA flux ratio shows an evolutionary trend. The ratio is below 0.5 for Class 0 protostars, which are predominantly envelope-dominated, whereas it ranges from 0.5 to 1.0 for Class I stars. This indicates that there is a transition from envelope-dominated to disk-dominated phase at the Class 0/I stage, and our derived flux ratio of $\sim 0.8$ for Class II disks is also in line with this evolutionary sequence.

We estimate the envelope mass $M_{\rm env}$ (gas + dust) from the envelope flux density $F_{\nu,\rm env}$, which is the flux density of IRAM~30m subtracted by that of ALMA. Under the assumption that the envelope is isothermal and optically thin, the total mass is given by:

\begin{equation}
M_{\rm env} = \frac{F_{\nu,\rm env} d^{2}}{\kappa_{\nu} B_{\nu}(T)},
\end{equation}

\noindent
where $d$ is the distance toward the source, $B_{\nu}(T)$ is the Planck function at the dust temperature $T_{\rm dust}$ and $\kappa_{\nu}$ is the opacity per unit gas mass at the observed frequency. Assuming that the gas and dust are at the same temperature of $T = 15$ K \citep{Federman2023} and the opacity of $\kappa_{\nu} = 0.023$ cm$^{2}$ g$^{-1}$ \citep{Beckwith1991}, average envelope masses over the 21 single-star targets is derived to be $M_{\rm env} = 10 \pm 8 $ $M_{\rm Jup}$. This estimate has an uncertainty of a factor of several due to the assumed opacity values (e.g., $\kappa_{\nu} = 0.009$ cm$^{2}$ g$^{-1}$; \cite{Ossenkopf1994}) but is an order of magnitude smaller than the envelope masses of Class I YSOs \citep{Federman2023}. Yet, our results indicate that there is still enough material to form Jupiter-mass planets if all the envelope material accretes onto the disk.

The presence of envelope remnants around the disks with typical ages of $1-5$ Myr \citep{Long2019} poses a question on conventional models that predict that most envelope materials dissipate within $0.5-1.0$ Myr since the onset of envelope collapse \citep{Young2005, Dunham2010}. One possible explanation for the remaining envelope is the ashfall phenomenon \citep{Wong2016, Tsukamoto2021, Tsukamoto2023}. In this scenario, an active outflow driven by a magnetic field entrains the dust particles grown to $\gtrsim 0.1$ mm that reside in the inner region of the disk. The dust particles are scattered all over at the envelope scale of $\sim 1000$ au and then fall back to the disk to replenish it with large grains. The grown dust within the envelope remnants can still be present during the Class II stage. The presence of grown dust in envelopes of Class 0/I stages has been suggested by several studies \citep{Kwon2009, Chiang2012, Miotello2014, Galametz2019}. To identify the grown dust even in the envelope of the Class II stage, the multi-band ACA observations, along with spectral energy distribution fitting \citep{Li2017}, would be desired.

\subsection{Possible Origin of Gap Formation}\label{sec:origin_gapformation}

\begin{figure*}[!htbp]
\begin{center}
\includegraphics[width=0.98 \textwidth]{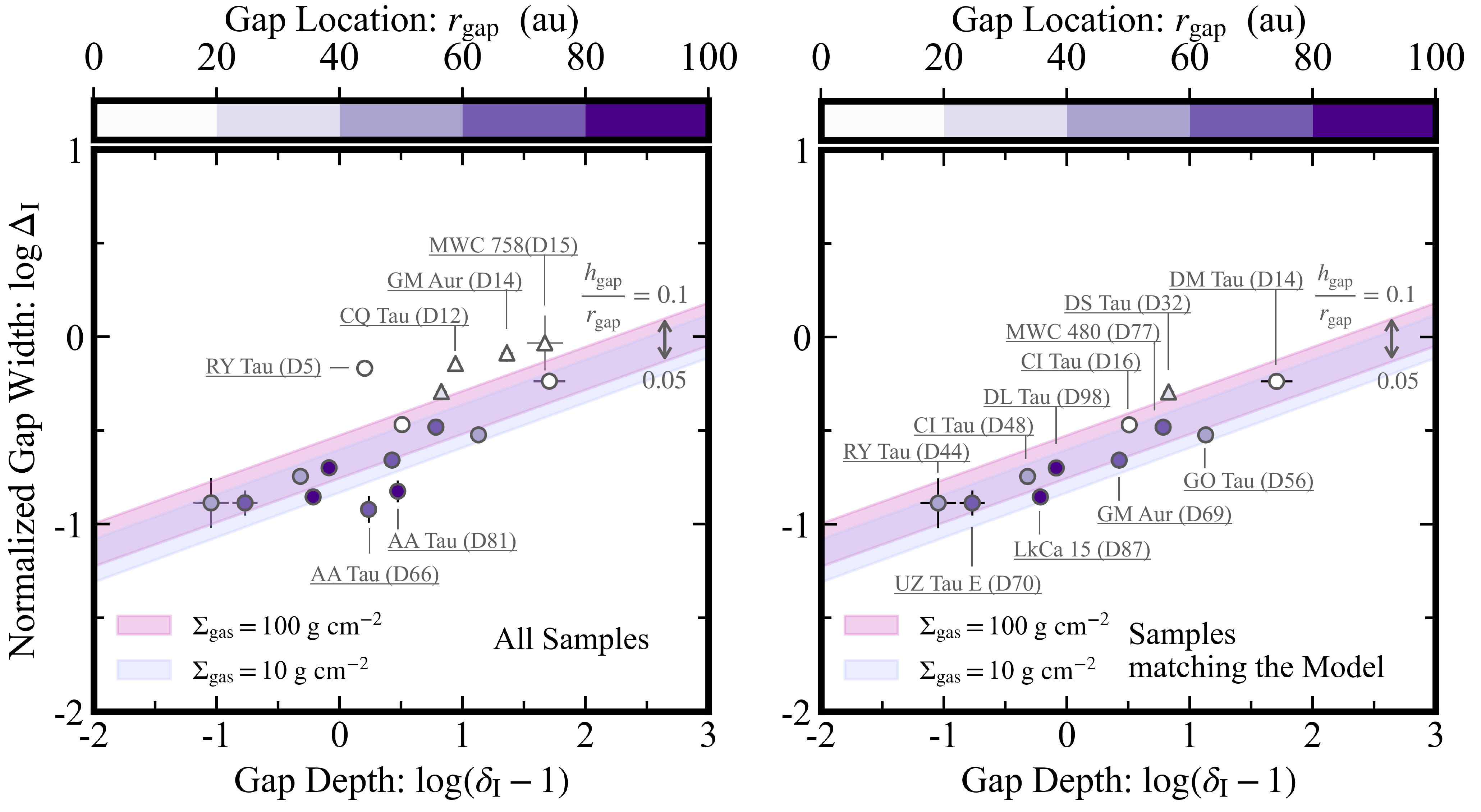}
\end{center}
\caption{The relationship of gap width and depth from the model by \citet{Zhang2018} (shaded area) overplotted by spatially resolved gaps (symbols). We fix the maximum dust particle size to be 0.1 mm and the viscous parameter to be $\alpha_{\rm vis} = 10^{-3}$, and show the models of the gas surface density $\Sigma_{\rm g}$ of $100~\rm g~cm^{-2}$ (purple) and $10~\rm g~cm^{-2}$ (blue).  The range of the model corresponds to varying scale heights from 0.05 to 0.1. The left panel plots all the samples of spatially resolved gaps (same as Figure \ref{fig:gapwidthdepth_relation}) and the right panel shows only the gaps that are consistent with the gap model.}
\label{fig:compare_planetdiskmodel}
\end{figure*}


\begin{figure}[ht]
\begin{center}
\includegraphics[width=0.48 \textwidth]{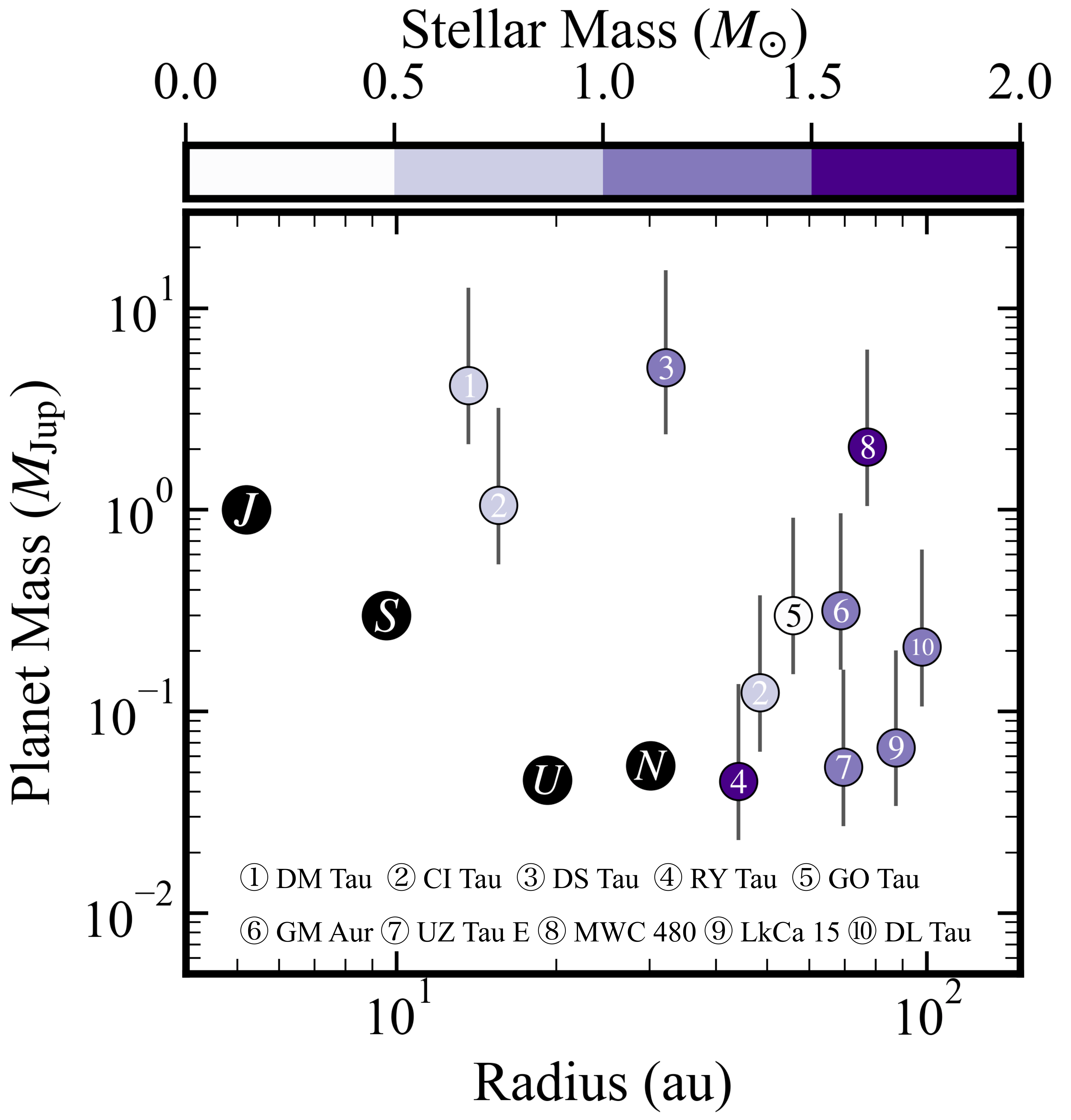}
\end{center}
\caption{Relationship between the inferred planet mass and the orbital radius. The circles indicate planet masses estimated assuming the viscous parameter of $\alpha_{\rm vis}= 10^{-3}$ and the error. The bars indicate the range of mass with different viscous parameters, ranging from $10^{-4}$ to $10^{-2}$. The colors of each symbol correspond to the host stellar mass. Black circles indicate planets of our Solar System, with labels indicating Jupiter, Saturn, Uranus, and Neptune.}
\label{fig:estimate_planetmass}
\end{figure}

In Section \ref{sec:gap_widthdepth_relation}, we have found that the gap width is correlated with the gap location and the depth. Intriguingly, we find that the derived power-law index of $\mathcal{B}={0.23 \pm 0.09}$ in the $\delta_{\rm I}-\Delta_{\rm I}$ relationship (Equation \ref{eq:SL_gapwidth_depth}) for the outer region sample (gaps at $30-100$ au from the central star) is consistent with the power-law indices predicted by planet-disk interaction models, which are 0.25 in \cite{Kanagawa2015, Kanagawa2016} and $\sim 0.23$ in \cite{Zhang2018}. 

In comparison with other gap-opening mechanisms, there is a theoretical prediction given by secular gravitational instability (SGI; \cite{Takahashi2014, Tominaga2019}). SGI generates annular ring structures for producing planetary embryos even at large orbital separations ($r>10$ au). \cite{Tominaga2019} proposed that the narrowness of observable dust ring systems can be explained by SGI. When the distance between the peaks of adjacent rings $\Delta r_{\rm ring}$ normalized by the gap scale height $h_{\rm gap}$ at the gap location falls below a threshold of 3.6 (i.e., $\Delta r_{\rm ring} / h_{\rm gap} < 3.6$), SGI becomes a plausible mechanism for shaping these rings. We compute $h_{\rm gap}$ using the dust temperature $T_{\rm d}(r)$ assumed by
\begin{equation}\label{eq:disk_temperature}
T_{\rm d}(r)=\left(\frac{\phi L_{*}}{8 \pi r^{2} \sigma_{\rm SB}}\right)^{1 / 4},
\end{equation}
\noindent
where $\sigma_{\rm SB}$ is the Stefan-Boltzmann constant, $L_{*}$ is the stellar luminosity (given in Table \ref{tab:stellar_properties}), $\phi$ and is the flaring angle \citep{Chiang1997, Dullemond2001}. The flaring angle is assumed to be constant at 0.02 except for DM Tau and GM Aur, for which we use 0.05 since low $\phi$ results in the dust temperature below the observed brightness temperature. Examining the double (or multiple) gaps/rings in eight disks (see Table \ref{tab:disk_substructure}), the sources except for GO Tau do not appear to be induced by the SGI model, as the derived ratios range from 4.0 to 14.9 with an average of 8.6. The GO Tau disk has three narrow gaps/rings (D56/B72, D89/B99, and D105/B116), and their derived ratios (3.6 for B72-B99 and 1.8 for B99-B116) are either equal to or below the threshold. It should be noted that these gap structures of D89 and D105 (and perhaps nearby areas) in GO Tau are not fully spatially resolved ($\Delta_{\rm I, unit} \leq \theta_{\rm eff}$), and it is too early to conclude that they are caused by SGI. Observations with higher spatial resolutions ($\theta < 0''.1$) are needed to determine the origin of these narrow gaps in the GO Tau disk.

The comparison with other gap-opening mechanisms, such as dust processes including ice lines and streaming instability, remains unclear because their theoretical predictions on gap morphology have not yet been established. Therefore, we focus our attention on the comparison of our results with the predictions of the planet-disk interaction model proposed by \cite{Zhang2018}.

The planet-disk interaction model outlined in \cite{Zhang2018} predicts the relationship between the gap depth $\delta_{\rm I}$ and normalized width $\Delta_{\rm I}$ in the surface brightness profile $I_{\nu}(r)$. The relationship can be derived from Equations (22)-(24) of \citet{Zhang2018} and reads
\begin{eqnarray}\label{eq:zhang_theory}
\Delta_{\mathrm{I}}=\mathrm{A}
\left[
0.635
\left(\frac{h_{\mathrm{gap}}}{r_{\mathrm{gap}}}\right)^{2.63}
\left(\frac{\alpha_{\mathrm{vis}}}{10^{-3}}\right)^{0.07} \left(\frac{\delta_{\mathrm{I}}-1}{\mathrm{C}}\right)^{\frac{1}{D}} 
\right]^{\mathrm{B}},
\end{eqnarray}
\noindent
where $h_{\mathrm{gap}} / r_{\mathrm{gap}}$ and $\alpha_{\rm vis}$ indicate aspect ratio at a planet location and viscous parameter, respectively. The constants $A$, $B$, $C$, and $D$ depend on the gas surface density $\Sigma_{\rm g}$ and the maximum grain size $s_{\rm max}$.

Figure \ref{fig:compare_planetdiskmodel} shows the comparison between the observations (for the spatially resolved gap samples) and the model prediction on the $\Delta_{\rm I}- \delta_{\rm I}$ relation. We consider the cases with gas surface density $\Sigma_{\rm g}$ of 10 and 100 $\rm g~cm^{-2}$ (see also Section \ref{sec:inferred_planetmass}) and various disk aspect ratios at the location of gap $h_{\rm gap}/r_{\rm gap}$ from 0.05 to 0.1. We fix the maximum grain size $s_{\max}$ to be 0.1~mm (DSD1 model for \cite{Zhang2018}) based on ALMA observations by \cite{Bacciotti2018} and the viscous parameter $\alpha_{\rm vis}$ to be $10^{-3}$. In this case, $\Delta_{\rm I}$ scales with $(\delta_{\rm I}-1)^{0.25}$.

We find that $65\%$ (11 out of 17) of the total sample falls in the range predicted by the model. Nine of them are gaps located in the outer region ($r_{\rm gap}>30$~au), while only two (DM Tau (D14) and CI Tau (D16)) are located in the inner region. Four of the six gaps that deviate from the model are wide ($\Delta_{\rm I} > 0.6$; MWC 758 (D15), GM Aur (D14), CQ Tau (D12), and RY Tau (D5)). A possible origin of these gaps may be a giant planet or a brown dwarf, which is out of the range of the model by \citet{Zhang2018}, in a disk \citep{Ubeira2019, Calcino2020}.

The model actually has large uncertainty in the parameter space when a Jupiter-mass planet opens a very wide gap ($\Delta_{\rm I} \sim 1$; see \citet{Zhang2018}) and therefore more refined model of planet-disk interaction is desired. When a planet is sufficiently massive ($>3~M_{\rm jup}$) and opens a wide gap, it may induce eccentricity in the gap edge \citep{Kley2006, Tanaka2022}. Such effects may need to be taken into account to construct models that are applicable to a wider range of planet mass and disk parameters.

Two adjacent gaps in the AA Tau disk (D66 and D81) show a narrower gap width than the range of model prediction. This disk has a large cavity feature (PTD) and narrow rings similar to the other disk around HD 169142 \citep{Perez2019}. One possible interpretation of this particular structure is the presence of a super-Earth planet ($\sim 10~M_{\rm \oplus}$) that induces multiple gaps in a low-viscosity ($\alpha_{\rm vis}<10^{-4}$) environment \citep{Dong2018_multigap}. The widths of the gaps are narrower than those models predicted in \citet{Zhang2018}, which is qualitatively consistent with observations. We note that such a low-viscous model is outside the scope of the models in \citet{Zhang2018}.

\subsection{Inferred Planet Mass in Gap Structure} \label{sec:inferred_planetmass}

For the 11 gaps that lie within the range of the planet-disk interaction model, we derive the planet mass. We follow the methods outlined in \cite{Zhang2018}, and use the following empirical relationship between the planet mass and the normalized gap widths $\Delta_{\mathrm{I}}$:
\begin{eqnarray}\label{eq:zhang_theory_gapwidth}
\frac{M_{\mathrm{p}}}{M_{*}} = 0.115 \left(\frac{\Delta_{\mathrm{I}}}{A}\right)^{1 / B} \left(\frac{h_{\mathrm{gap}}}{r_{\mathrm{gap}}}\right)^{0.18} \left(\frac{\alpha_{\mathrm{vis}}}{10^{-3}}\right)^{0.31},
\end{eqnarray}
\noindent
where the parameters $A$ and, $B$ are the parameters that appear in Equation \ref{eq:zhang_theory} and tabulated in \cite{Zhang2018}. The three physical parameters, the averaged gas surface density $\Sigma_{\mathrm{g}}$, the aspect ratio at the gap location $h_{\mathrm{gap}}/r_{\mathrm{gap}}$, and the maximum grain size $s_{\mathrm{max}}$ are needed to determine $A$ and $B$. We assume $s_{\mathrm{max}}$ to be 0.1~mm as stated in Section \ref{sec:origin_gapformation}.

To estimate $\Sigma_{\mathrm{g}}$, we use the relationship between $\Sigma_{\mathrm{g}}$ and $\Sigma_{\mathrm{d}}$ presented in Figure 18 of \citet{Zhang2018}, where $\Sigma_{\mathrm{d}}$ is the averaged dust surface density taken from 1.1 to 2.0 $r_{\rm gap}$. To obtain $\Sigma_{\mathrm{d}}$, we derive the radial dust surface density profile $\Sigma_{\mathrm{d}}(r)$ using the simple radiative transfer equation $\Sigma_{\mathrm{d}}(r)=-\ln \left\{ 1-I_{v}(r) / B_{v}[T_{d}(r)] \right \} / \kappa_{v}$, where $B_{\nu}$, $T_{\rm d}(r)$, and $\kappa_{\nu}$ denote the Planck function, the dust temperature in Equation \ref{eq:disk_temperature}, and the dust absorption opacity, respectively. The opacity $\kappa_{\nu}$ is assumed to be $0.43~\mathrm{cm}^{2}~\mathrm{g}^{-1}$ which is the value of DSHARP opacity model \citep{Birnstiel2018} with $s_{\mathrm{max}} = 0.1$ mm at $\lambda=1.3$~mm. With $T_{\mathrm{d}}(r)$, we can compute the disk aspect ratio at the gap location $h_{\mathrm{gap}}/r_{\mathrm{gap}}$ as well as the dust surface density. The values of planet mass and disk physical parameters ($h_{\mathrm{gap}}/r_{\mathrm{gap}}$, $\Sigma_{\mathrm{d}}$, $\Sigma_{\mathrm{g}}$, and $M_{\mathrm{p}}$) are summarized in Table \ref{tab:estimate_planetmass}.

All of our estimates of planet mass for the eleven gaps are in the range of $\sim$Neptune to $\sim$Jupiter mass. The derived planet mass does not vary much if we change the viscosity parameter $\alpha_{\rm vis}$ from $10^{-4}$ to $10^{-2}$. Figure \ref{fig:estimate_planetmass} shows the distribution in the plane of planet mass versus orbital radius. Interestingly, Saturn- to Jupiter-mass planets seem to be expected in the inner region ($r\sim 10$~au) and the Neptune-mass planets are in the outer region ($r\sim 40-100$~au). This trend is consistent with the results obtained for DSHARP samples that consist of bright and large disks \citep{Zhang2018}.

We note that \cite{Zhang2023} reanalyzed the continuum datasets of \cite{Long2019} and estimated planet masses for single-star systems within the Taurus-Auriga region. They used parametric fitting of azimuthally averaged visibilities to identify gaps and rings and estimated the planet masses using the same methods as ours. The measurements of $\Delta_{\rm I}$ for eight large disks in their analysis are consistent with ours within the uncertainties of 0.1 to 0.2, while the number of the gap/ring in GO Tau and DL Tau disks are not consistent with our results. This is probably due to the difference in the methods of ring/gap identification. We do not assume any prescribed function for gap profiles while we work in the image plane rather than the visibility plane. The qualitative results of planet mass distribution, that is, larger mass for smaller orbital distance, are consistent in both works.


\begin{table}
\centering %
\tbl{Inferred Planet Masses from Spatially Resolved Gap Structures and Associated Physical Quantities.}{
\renewcommand{\arraystretch}{1.5}
\begin{tabular}{p{15mm}p{13mm}p{13mm}p{13mm}p{13mm}p{13mm}} 
\hline
Name & $r_{\rm gap}$ & $h_{\rm gap}/r_{\rm gap}$ & $\Sigma_{\rm d}$ & $\Sigma_{\rm g}$ & $M_{\rm p}$\\
     & (au) &    & $\rm (g~cm^{-2})$ & $\rm (g~cm^{-2})$ & $(M_{\rm Jup})$\\
(1) & (2) & (3) & (4) & (5) & (6) \\
\hline
DM Tau & 13.6 & 0.05 & 1.85 & 100 & $4.15^{+4.31}_{-2.12}$\\
CI Tau & 15.6 & 0.05 & 0.75 & 100 & $1.06^{+1.09}_{-0.54}$\\
CI Tau & 48.4 & 0.06 & 0.54 & 100 & $0.12^{+0.13}_{-0.06}$\\
DS Tau & 32.1 & 0.06 & 0.7 & 100 & $5.07^{+5.28}_{-2.37}$\\
RY Tau & 44.1 & 0.03 & 0.78 & 30 & $0.04^{+0.05}_{-0.02}$\\
GO Tau & 56.0 & 0.08 & 0.29 & 30 & $0.30^{+0.31}_{-0.15}$\\
GM Aur & 68.8 & 0.05 & 0.59 & 100 & $0.32^{+0.33}_{-0.16}$\\
UZ Tau E & 69.5 & 0.05 & 0.62 & 100 & $0.05^{+0.06}_{-0.02}$\\
MWC 480 & 77.2 & 0.07 & 0.26 & 30 & $2.05^{+2.14}_{-1.05}$\\
LkCa 15 & 87.4 & 0.05 & 0.49 & 100 & $0.07^{+0.07}_{-0.04}$\\
DL Tau & 97.8 & 0.07 & 0.41 & 100 & $0.21^{+0.22}_{-0.11}$\\
\hline
\end{tabular}}
\tabnote{
\textbf{Note}. Column descriptions: (1) Host star name with spatially resolved gaps compatible with the planet-disk interaction model. (2) Radial position of the gap. (3) Disk aspect ratio at the gap's location. (4) Average dust surface density within the range from 1.1 $r_{\rm gap}$ to 2.0 $r_{\rm gap}$ on the radial dust surface density profile. (5) Gas density estimated from Figure 18 in \cite{Zhang2018} (6) Inferred planet mass assuming the viscosity parameter of $\alpha_{\rm vis} = 10^{-3}$. The errors represent masses with different viscous parameters of $10^{-2}$ and $10^{-4}$.
}
\label{tab:estimate_planetmass}
\end{table}


\subsection{Implication of planet formation}

The inferred planet mass for the eleven gaps is in the range of $\sim$Neptune to $\sim$Jupiter mass. The Saturn- to Jupiter-mass planets seem to be located in the inner region ($r\sim 10$~au), while the Neptune-mass planets are in the outer region ($r\sim 40-100$~au). As shown in Figure \ref{fig:estimate_planetmass}, this distribution is similar to the architecture of our Solar System, while the orbital radii of the inferred planets in protoplanetary disks are larger than those of the planets in our Solar System. The formation of giant planets at large orbital radii suffers from the problem of timescale in the traditional core accretion scenario. This is a bottom-up framework in which a planetary embryo of a few Earth masses accretes surrounding gas to form a gas giant (e.g., \cite{Mizuno1980, Pollack1996, Ida2013}). The planetary embryo formed at $5-10$ au from the central star can be massive enough to form a gas giant (e.g., \cite{Helled2014}). However, this process takes longer than the disk lifetime when planets are formed at larger orbital radii (e.g., \cite{Rafikov2004, Levison2010}). The pebble accretion scenario has been proposed as a way to solve the issue (e.g., \cite{Lambrechts2012}) but numerical simulations still show that this mechanism does not really produce the cold giant planets (Neptune-Jupiter masses) at larger orbital radii \citep{Bitsch2015}.

The gravitational instability scenario is a top-down framework that can potentially form gas giants at large radii from a massive disk (e.g., \cite{Vorobyov2013}). However, we do not see conspicuous asymmetric structures that indicate the presence of turbulence due to instability. Another possible scenario is the combination of the ashfall phenomenon that replenishes grown dust into the large radii (see Section \ref{sec:envelope_remnants}) and subsequent SGI (e.g., \cite{Takahashi2016}). This may qualitatively explain the annular ring/gap formation by the cold giant planet at the large radii \citep{Tsukamoto2021}. However, the distribution of rings and gaps is not consistent with the prediction of those predicted by SGI (see Section \ref{sec:gap_widthdepth_relation}). The distribution of rings during the formation of planets and that of already formed planets may be different due to, for example, planet-planet interaction. The prediction of this scenario should be explored in more detail.

We note that the inferred population of planets may actually be different when we interpret that the shoulder feature, which is a candidate for a weak gap structure is formed by a lower-mass planet. The shoulder features appear at many places in the disks; this might indicate the presence of an additional population of planets distributed all over the orbital radii from 5 to 40 au (see Figure \ref{fig:radialgap_location}).

The upper limit of the planet mass for the 21 shoulder features can be estimated by the empirical relationship between planet mass and gap width (Equation 5 in \cite{Kanagawa2016}),
\begin{equation}\label{eq:upper_planetmass}
\frac{M_{p}}{M_{*}}=0.19\left(\frac{\Delta_{\rm I, unit }}{r_{\rm{p }}}\right)^{2}\left(\frac{h_{\rm p}}{r_{\rm p}} \right)^{1.5} \left(\frac{\alpha_{\mathrm{vis}}}{10^{-3}}\right)^{0.5}.
\end{equation}
\noindent
Here, we assume that the dust is well coupled with the gas. This gives a large upper limit since dust particles tend to move to the pressure maximum at the edges of the gap. We estimate the upper limit of the planet mass by assuming that this gap width to be smaller than the effective spatial resolution, with $\Delta_{\rm I, unit}\leq\theta_{\rm eff}$ and $r_{\rm p} = r_{\rm inf}$. We fix the disk scale height $h_{\rm p}/r_{\rm p}$ and viscous parameter $\alpha_{\mathrm{vis}}$ to be 0.05 and $10^{-3}$, respectively, and apply this formula to the 21 shoulder features. We have obtained the average upper limit of $M_{p} = 0.6\pm0.5~M_{\rm Jup}$ at the spatial resolutions of $66\pm32$ mas, where the errors represent standard deviations. This suggests that Neptune (or lower) mass planets may exist anywhere in the disk but the gaps in the inner part of the disks are not clearly detected.

It may be possible to form low-mass planets through core accretion or pebble accretion without suffering timescale problems. The prevalence of planets with masses lower than Jupiter might indicate that these are the main mechanisms of planet formation. The giant planets at larger orbital radii may be formed in exceptionally massive disks and/or there might be other (relatively minor) mechanisms to accelerate massive planet formation.

To verify this argument, it is crucial to conduct observations that can measure the shape of the candidates of the weak gap structures more accurately. From Equation \ref{eq:upper_planetmass}, we need an extremely high resolution of $24 \pm 17$ mas to resolve the gaps carved by Neptune mass planets, respectively. Long baseline observations at higher frequencies than the current Band 6 observations (e.g., \cite{Asaki2020}) may provide new insights, and it is necessary to have better gap models that take into account optical depth effects that are crucial in analyzing high-resolution observations.

\section{Conclusion}\label{sec:conclusion}
We present ALMA 2D Super-resolution imaging of Taurus-Auriga protoplanetary disks for probing statistical properties of disk substructures. We use archival ALMA Band 6 continuum data from 43 disks, comprising 39 disks around Class II objects and four disks around Herbig Ae stars. To enhance the fidelity and spatial resolution of the images, we employ a novel 2D super-resolution imaging technique based on sparse modeling (SpM). Our main findings with the SpM images are summarized as follows:

\begin{enumerate}
\item{All dust disks are successfully spatially resolved. By applying super-resolution imaging, 18 out of 43 targets show the improvement of spatial resolution by a factor of two to three compared to the conventional CLEAN method. All but three images show spatial resolution better than $0''.1$, with the highest achieved resolution reaching $0''.02$. The radii of the disks range from 8 to 238 au with a median radius of 45 au.}

\item{We find two empirical relationships on spatial resolution improvement. One is the relationship with the image SNR and the other is with the disk size normalized by the synthesized beam size. The improvement of spatial resolution is more significant for higher SNR data or more compact disks whose size is close to the beam size.}

\item{We assess the performance of SpM and CLEAN imaging by investigating how well the Fourier transform of the reconstructed images can fit the observed visibility. SpM produces a better fit in $95\%$ of cases (40/42). The remaining $5\%$ (2 cases) show comparable performance between SpM and CLEAN.}

\item{The fidelity of SpM images is assessed with the ``cross-check method'' \citep{Yamaguchi2020}, wherein a comparison is made between the CLEAN image derived from long baseline observations and the SpM image generated from short baseline observations. Employing three bright and large disks for the evaluation, we confirm that the image reconstructed from the shorter-baseline data using the SpM matches well with that obtained by the longer-baseline data using CLEAN. However, we note that SpM imaging can produce artificial features such as clumps and speckles on an extended faint emission, especially for disks with the brightest emission in the central part surrounded by the extended area of faint emission.}

\item{The flux densities observed with the IRAM 30m single-dish telescope are $\sim 20\%$ larger than that with ALMA. This discrepancy may be accounted for as the contribution from the emission at the envelope scale surrounding the star and disk system. It is indicated that there are still enough materials to form Jupiter-mass planets if all the envelope material accretes onto the disk eventually.}

\item{We confirm the correlation between millimeter luminosity and the disk size $L_{\rm {mm}} \propto r_{\rm {disk}}^2 $ on the analysis of image domain, which is suggested based on the 0.9~mm data of disks in Taurus-Auriga-Ophiuchus regions \citep{Tripathi2017, Hendler2020}. The correlation is improved by $\sim10\%$ if the luminosity is re-scaled by the inclination angle.}

\item{We identify three characteristic structures in the azimuthally averaged radial intensity profiles: rings, gaps, and inflections. The combination(s) of these structures as well as some asymmetric structures appear in actual observations. From the investigations of the morphology of the 43 disks, we suggest nine categories of disk morphology (Figure \ref{fig:disk_clasification}): eight types of axisymmetric structures and asymmetry.}

\item{We find the bimodality in the distribution of gaps whose width and depth are measurable. The gaps are located either in the inner ($r\lesssim 20$~au) or the outer ($r\gtrsim 30$~au) part of the disk. We find that the classification of the disks is different for disks having gaps in the inner region and in the outer region. The difference in disk types may be connected to the bimodality of the gap distribution.}

\item{All the gaps have a ring outside of them. We find correlations in the properties of gaps and rings that are associated with gaps. The correlations are in (1) the radial positions of gaps and rings, (2) the gap locations and their widths, and (3) the gap widths and depths. Meanwhile, we do not observe a robust correlation between stellar mass and disk substructure properties.}

\item{The power-law index in the correlation between the gap depth $\delta_{\rm I}$ and normalized gap width $\Delta_{\rm I}$ is of $0.23 \pm 0.09$ for the outer region sample ($r\gtrsim 30$~au). This is consistent with the power-law indices predicted by planet-disk interaction models \citep{Kanagawa2015, Kanagawa2016, Zhang2018}.}

\item{We estimate the planet mass for each gap with clear measurements of depths and widths, assuming that these gaps are formed by planet-disk interaction. The messes of putative planets are in the range of $\sim$Neptune to $\sim$Jupiter mass. Saturn- to Jupiter-mass planets seem to be expected in the inner region ($r\sim 10$~au) and the Neptune-mass planets are in the outer region ($r\sim 40-100$~au). Questions remain open about the formation of such cold giant planets.}

\item{We note that the inferred population of planets may actually be different when we interpret that the shoulder features, which are a candidate for weak gap structures, are formed by lower-mass planets (Neptune mass or lower). The shoulder features appear at many places in the disks and therefore, our results indicate that there is an additional population of planets distributed all over the orbital radii from 5 to 40 au.}

\end{enumerate}

\begin{ack}
The authors thank the anonymous referee for all of the comments and advice that helped improve the manuscript and the contents of this study. The authors thank Sai Jinshi for valuable discussions, and Takahiro Ueda for giving us the ALMA data of CW Tau. This work was financially supported by JSPS KAKENHI Grant Numbers 18H05441,18H05442, 20H01951, and 23K03463. T.M. is supported by Yamada Science Foundation Overseas Research Support Program. N.H. acknowledges support from the National Science and Technology Council (NSTC) of Taiwan with grant NSTC 111-2112-M-001-060. This study uses the following ALMA data: 
2015.1.00889.S,
2019.1.00579.S,
2017.1.01151.S,
2018.1.01230.S,
2016.1.01164.S,
2016.1.01370.S,
2018.1.01631.S,
2015.1.01207.S,
2013.1.00498.S,
2018.1.01755.S,
2018.1.01255.S,
2018.1.00945.S,
2013.1.01070.S,
2016.1.01205.S,
2018.1.01829.S,
2013.1.00105.S,
2016.1.00724.S,
2017.1.00940.S,
$2016.\mathrm{A}.00026.\mathrm{S}$
2017.1.01404.S,
2017.1.01460.S,
2015.1.01268.S,
2016.1.00846.S,
2019.1.01108.S. ALMA is a partnership of ESO (representing its member states), NSF (USA) and NINS (Japan), together with NRC (Canada), MOST and ASIAA (Taiwan), and KASI (Republic of Korea), in cooperation with the Republic of Chile. The Joint ALMA Observatory is operated by ESO, AUI/NRAO, and NAOJ. Data analysis was in part carried out on the multi-wavelength data analysis system operated by the Astronomy Data Center (ADC), National Astronomical Observatory of Japan. This study used data from the European Space Agency (ESA) mission $\it Gaia$ (\url{https://www.cosmos.esa.int/gaia}), processed by the {\it Gaia} Data Processing and Analysis Consortium (DPAC, \url{https://www.cosmos.esa.int/web/gaia/dpac/consortium}). Funding for the DPAC has been provided by national institutions, in particular the institutions participating in the $\it Gaia$ Multilateral Agreement.
\end{ack}

\section*{Software}
AnalysisUtilities (\url{https://casaguides.nrao.edu/index.php?title=Analysis_Utilities}), Astropy \citep{Astropy2022}, CASA \citep{CASA2022}, kneed (\url{https://github.com/arvkevi/kneed}), Linmix \citep{Kelly2007}, matplotlib \citep{Hunter2007}, PRIISM \citep{Nakazato2020}, SciPy \citep{Virtanen2020}, NumPy \citep{Harris2020}


\appendix

\section{Disk Size and Total Flux Density}\label{appendix:dust_disk_size}

We use the curve of growth method (e.g., \cite{Ansdell2016}) to measure the dust disk radius $r_{\rm disk}$ on each SpM image. We first derive the inclination and position angle of each disk on the image domain. For disks showing clear ring structures in the outer region, we fit the shape of the ring with an ellipse.  For disks that do not show clear ring structures or with low SNR for ellipse fitting, we fit the surface brightness distribution around the central star with an elliptical Gaussian function. The measurements of the inclination and position angle for each source, as well as the methods to derive them listed in Table \ref{tab:stellar_properties}. The average of measured disk inclination is $42^{\circ} \pm 15^{\circ}$. We deproject the images to produce the face-on view using the inclination and the position angle.  Then, we define the incremental flux density $F_{\nu}(r)$ from the radial surface brightness distribution $I_{\nu}(r)$:
\begin{equation}
  F_{\nu}(r)= 2 \pi \int_{0}^{r} I_{v}\left(r^{\prime}\right)  r^{\prime} \mathrm{d} r^{\prime},
\end{equation}
\noindent
where the total flux density $F_{\nu}$ is the limiting value of $F_{\nu}(r)$ at $r\to\infty$. In practice, we use successively larger photometric apertures to obtain $F_{\nu}(r)$ and find the values where $F_{\nu}(r)$ becomes constant.
The disk radius $r_{\rm disk}$ is then measured by $0.95~F_{\nu} =  F_{\nu}(r)$. We estimate the uncertainty $\sigma_r$ of the disk radius using the effective spatial resolution. We assume $\sigma_{\mathrm{r}}=\left\langle\theta_{\mathrm{eff}}\right\rangle /2 \sqrt{2 \ln 2}$, where $\left\langle\theta_{\mathrm{eff}}\right\rangle$ indicates the geometric mean of the spatial resolution. The derived disk radii are summarized in Table \ref{tab:stellar_properties}.

\section{Improvement of Spatial Resolution by Sparse Modeling}\label{appendix:vis_profile_objects}

\begin{figure*}[t]
\begin{center}
\includegraphics[width=0.98 \textwidth]{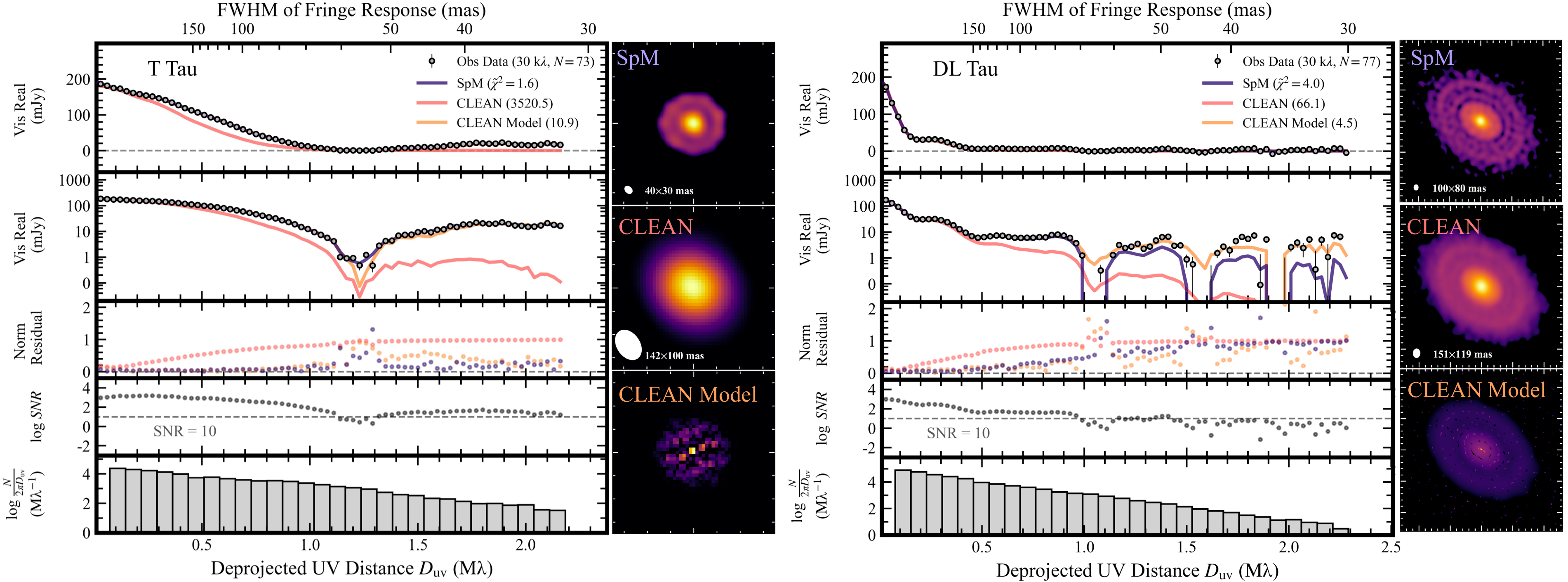}
\end{center}
\caption{The azimuthally averaged radial visibility profiles of the compact disk around T Tau N (left) and the large disk around DL Tau (right). The observed visibility data are shown by dots, and the visibility models with SpM, CLEAN, and CLEAN models are represented by black, red, and orange lines, respectively. The data are binned every $30k\lambda$. The reduced-$\chi^{2}$ values calculated from the observed data and the models are shown in the labels of the top panels. The left panels of each target display, from top to bottom, the amplitude of the real part of the visibility, its logarithmic scale, the normalized residual between the observation and the model, the SNR of visibility within each bin, and the data density of each bin in $uv$-space. The SNR is the ratio of the real part amplitude to its noise. See Appendix \ref{appendix:vis_profile} for details. The images in the right panels of each target are SpM images, beam-convolved CLEAN image, and the CLEAN model from top to bottom.}
\label{fig:radialvisprofile}
\end{figure*}


\begin{figure}[!p]
\begin{center}
\includegraphics[width=0.48 \textwidth]{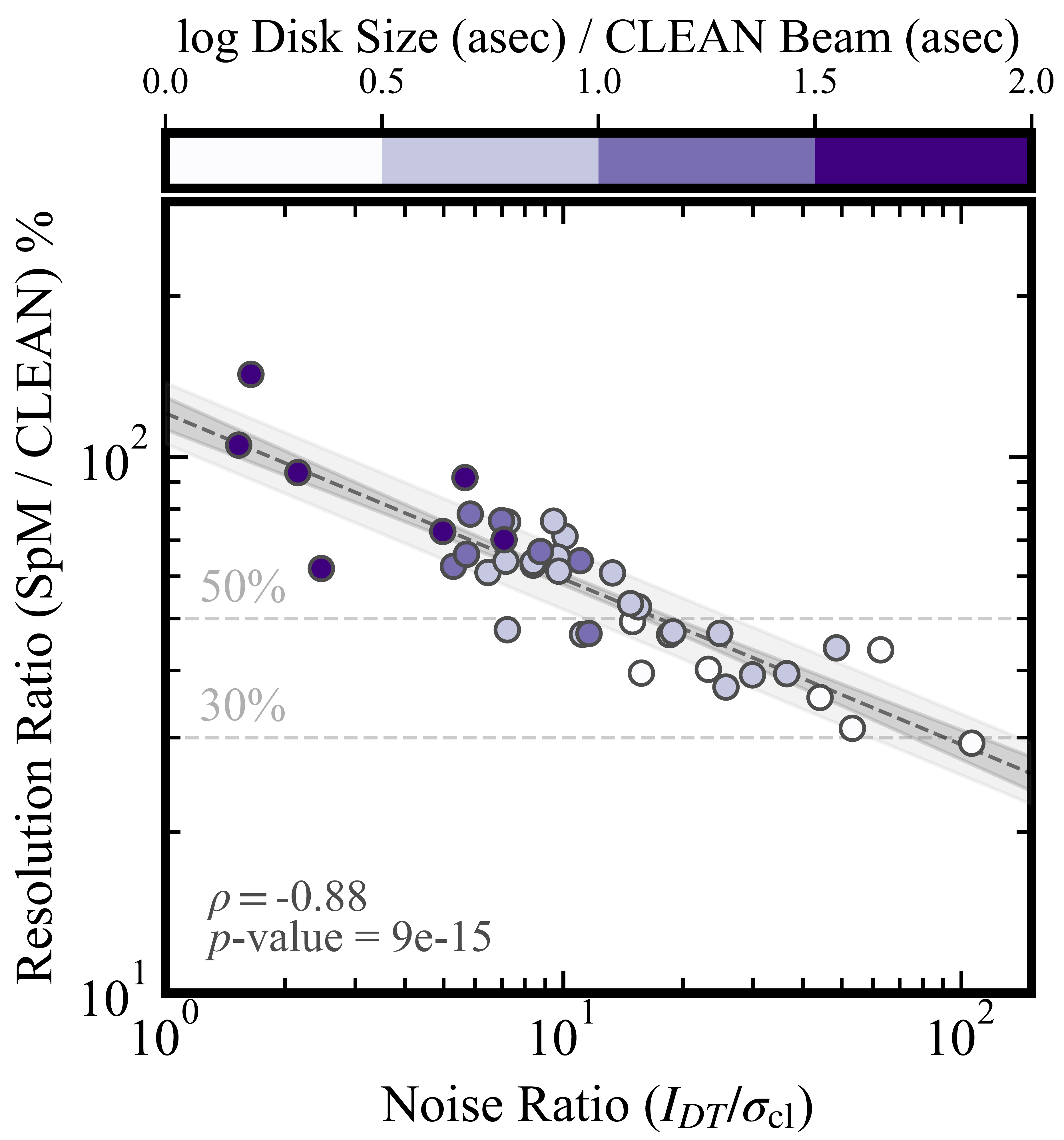}
\end{center}
\caption{Relationship between resolution ratio (SpM/CLEAN) and noise ratio ($I_{\rm DT}/\sigma_{\rm cl}$). Each data sample is colored with the disk size normalized to the CLEAN beam size on the logarithmic scale. The dashed line, the dark gray area, and the light gray area have the same meaning as in Figure \ref{fig:resraatio_disksize}. Pearson's correlation coefficient ($\rho$) and the $p$-value calculated from the sample distribution are shown in the lower left. Note that the definitions of these noise levels are different; $I_{\rm DT}$ denotes the maximum intensity on the emission-free area of a SpM image, while $\sigma_{\rm cl}$ denotes the RMS value on the emission-free area of a CLEAN image. The units of both $I_{\rm DT}$ and $\sigma_{\rm cl}$ are converted into $\rm Jy~arcsec^{-2}$ for consistency.}
\label{fig:resolutionratio_noiseratio}
\end{figure}

Figure \ref{fig:radialvisprofile} shows azimuthally averaged radial visibility profiles of two disks of different sizes: T Tau N and DL Tau. The data of the compact disk T Tau N show consistently high SNRs of $\gtrsim10$ up to the longest baseline. SpM is particularly effective in improving spatial resolution in such cases, and indeed the effective spatial resolution is $30\%$ less than the CLEAN beam size in the case of T Tau N observations. On the other hand, the data of the large disk DL Tau show low SNRs of less than 10 at baselines longer than $\gtrsim \rm 1~M\lambda$, mainly due to the low visibility amplitudes ($\lesssim1$ mJy). In this case, the SpM model data tend to deviate from the observed data at long baselines, and we obtain a moderate improvement in spatial resolution. The effective spatial resolution is $70\%$ of the CLEAN beam size for DL Tau. This behavior can be attributed to the TSV term in the SpM equation \citep{Yamaguchi2020}. This suppresses small-scale rapid spatial variations in the image and thus avoids over-fitting the visibility data, while the spatial structure is smeared by the TSV term. Therefore, the improvement in spatial resolution is moderate for low SNR data at large $ uv$ distances.

\section{Evaluation of Noise Levels in Images}\label{appendix:noislevel_spm}

We investigate the relationship between the noise level of the images and the improvement of the spatial resolution. As a measure of noise levels, we use the detection threshold $I_{\rm DT}$ for SpM images and the RMS noise level $\sigma_{\rm cl}$ for CLEAN images. We note that the two measurements have different definitions: $I_{\rm DT}$ is the maximum intensity on the emission-free area, while $\sigma_{\rm cl}$ is the RMS value in the emission-free area
 (see Table \ref{tab:fullsample} for actual values for each target). The reason why we do not use the RMS noise in the SpM images is the non-negative constraints in the imaging processes. The SpM image has only positive intensity in the off-source area, and its noise distribution is biased towards the positive side with a longer tail than that of a Gaussian distribution (see Figure 6 in \cite{Yamaguchi2020}).

Figure \ref{fig:resolutionratio_noiseratio} shows the relationship between the resolution ratio (SpM/CLEAN) and its noise ratio ($I_{\rm DT}/\sigma_{\rm cl}$). We find a clear trend of decreasing resolution ratio as increasing noise ratio in logarithmic space (Pearson correlation coefficient $\rho = -0.88$, $p$-value $< 0.01$). Using Bayesian linear regression, this trend can be described as
\begin{equation}\label{eq:resolution_snr}
\log \left(\frac{\theta_{\rm eff}/\theta_{\rm cl}}{\%}\right)= (2.08\pm 0.03) - (0.31\pm0.03) \log \mathrm{ \left(\frac{I_{\rm DT}}{\sigma_{\rm cl}} \right)}
\end{equation}
\noindent
with a scatter of $0.062\pm0.004$ dex. This indicates that the detection threshold of SpM images $I_{\rm DT}$ strongly depends on the resolution ratio: $I_{\rm DT} \propto (\theta_{\rm eff}/\theta_{\rm cl})^{-3}$.

With the data sample in this study, we find that the brightness distribution in the emission-free area of SpM images is similar to that of positive values in the CLEAN images. This suggests that the brightness distribution in SpM images consists of real dust emission and artificial positive noise patterns. Assuming that any noise can be characterized as a fine structure significantly smaller than the spatial resolution, the noise intensity (or $I_{\rm DT}$) expressed in a unit of $\rm Jy~arcsec^{-2}$ should increase with improved spatial resolution. We see that the situation is reflected in the above relationship.

Here we perform a sensitivity calculation using the point source injection method to determine the correspondence between $I_{\rm DT}$ and $\sigma_{\rm cl}$. We focus on the T Tau data, where the noise ratio ($I_{\rm DT}$/$\sigma_{\rm cl} \sim 100$) is the highest in our sample. We first inject a point source into the observed visibility of T Tau, setting the total flux of this point source to $100\sigma_{\rm cl}$ (4.0 mJy) and locating it at $0''.6$ north of the phase center. We then reconstruct the visibility data in both the SpM and CLEAN images and measure the peak ($I_{\rm p}$) of the reconstructed point source structure. The image sensitivity results show that $I_{\rm p, spm} / I_{\rm DT} = 40$ and $I_{\rm p, cl} / \sigma_{\rm cl} = 100$. Given the SNR relationship of $(I_{\rm p, cl} / \sigma_{\rm cl}) = X (I_{\rm p, spm} / I_{\rm DT})$, the factor $X$ is 2.5, indicating that the point source sensitivity of SpM imaging can be worse by up to several factors. For the T Tau N data, $(I_{\rm DT} / \sigma_{\rm cl})\sim100$, but the peak intensity of the point source in SpM is larger by a factor of $\sim 40$ probably due to the improvement in spatial resolution. Therefore, the point source sensitivity is not as bad as a factor of 100.

\section{Radial Visibility Profile}\label{appendix:vis_profile}

We describe methods to derive an azimuthally averaged visibility profile using a reconstructed image (e.g., an SpM image, a CLEAN model, and a CLEAN image) and to evaluate the goodness of fit with the reduced $\chi^{2}$.

In a preprocessing step, we set the observed visibility data $O$ by applying full-channel averaging (i.e., one channel per spectral window) to all spectral windows in the self-calibrated visibility data.  Next, we obtain the model visibility data $M$ in the same $uv$ sampling as the observations from the Fourier transform of the reconstructed image, using the Fast Fourier Transform algorithm $\tt fft$ in $\tt NumPy$ \citep{Harris2020} and the bivariate spline approximation $\tt RectBivariateSpline$ in $\tt Scipy$ \citep{Virtanen2020}.

The observation $O$ and the model $M$ are deprojected in the $uv$ plane \citep{Butler1999} using the disk inclination and position angle (see Table \ref{tab:stellar_properties}). We then bin the visibility data every $30~k\lambda$ on the deprojected $uv$ plane and take the average of the data in the interval. The real part of the visibility is used to calculate the weighted average $\hat V_{Re}$ within the bin: $\hat V_{Re} = \sum_{j=1}^K (V_{Re,j}  w_{j})/\sum_{j=1}^K w_{j}$, where  $V_{Re}$ is the real part of the visibility data, and $w$ is visibility weight, and $K$ is the number of data within the bin. The corresponding error $\hat\sigma$ is obtained from the error propagation of $w$ as follows $\hat\sigma = (\sum_{j=1}^K w_{j})^{-0.5}$.

To quantify the difference between the model and the observed visibility of the $i$-th bin, we define the normalized error (NR) as $\mathrm{NR}_{i} = (| \hat V_{Re,O, i} - \hat V_{Re,M,i}|/| \hat V_{Re,O, i}|)^{0.5}$. On the other hand, to evaluate the goodness of fit between the model and the observation, we apply the reduced $\chi^{2}$ ($\chi_{red}^{2}$) with the weighted visibility data. The formula is given by
\begin{equation}
\chi_{red}^{2}=\frac{1}{N} \sum_{i=1}^{N}\frac{f_{i}\left|\hat V_{Re,O, i} - \hat V_{Re,M,i} \right|^{2}}{\hat\sigma^{2}_{i}},
\end{equation}
\noindent
where $N$ is the total number of binned visibility data points and $f$ is the re-weighting factor, which is the ratio between the weight and the standard deviation (stddev) of the real part of the visibility data. Specifically, $f_i = \mathrm{stddev}^{-2}_{i} / \bar{w}_{i}$, where $\mathrm{stddev}_{i}$ and $\bar{w}_{i}$ are the standard deviation and the averaged weight of the data in the $i$-th bin. This approach ensures that the reduced $\chi^{2}$ converges to $\sim 1$ when the model correctly matches the observations. The re-weighting factors vary with the observed data and typically fall in the range of $0.1-0.3$, consistent with values reported in other disk observations ($0.2-0.3$; \cite{Hashimoto2021zztau, Hashimoto2021dmtau}).

\section{Relationship between the goodness of fit and the improvement of spatial resolution}\label{appendix:chisq_resolution}

We find that the goodness of fit of the visibility data measured by reduced $\chi^{2}$ values (Section \ref{sec:goodness_visfit}) shows a correlation with the improved spatial resolutions. Figure \ref{fig:resolutionratio_chisquare} shows the spatial resolution ratio between SpM and CLEAN as a function of the ratio of the reduced chi-square values of the two imaging methods. For CLEAN, we show two cases with CLEAN image and CLEAN model. In the case of CLEAN image versus SpM, the distribution has a high Pearson correlation ($\rho = -0.82$, $p-$value $<0.01$) in the logarithmic plane. We obtain the empirical relationship by linear regression as  
\begin{eqnarray}
\log \left(\frac{\theta_{\rm eff}/\theta_{\rm cl}}{\%}\right) &=&  (1.95 \pm 0.03) \nonumber \\ &-&(0.14\pm0.02) \log \left(\frac{\rm Red\chi^{2}_{\rm cl}}{\rm Red\chi^{2}_{\rm spm}} \right),
\end{eqnarray}
\noindent
with a scatter of $0.09 \pm 0.01$ dex. In the case of the CLEAN model versus SpM, the correlation is marginal ($\rho = -0.51$, $p-$value $<0.01$). The empirical relation by linear regression is\begin{eqnarray}
\log \left(\frac{\theta_{\rm eff}/\theta_{\rm cl}}{\%}\right) &=& (1.78 \pm 0.02) \nonumber \\
&-& (0.20 \pm 0.06) \log \left(\frac{\rm Red\chi^{2}_{\rm clmodel}}{\rm Red\chi^{2}_{\rm spm}} \right),
\end{eqnarray}
\noindent
with a scatter of $0.12\pm0.01$ dex.

\section{Radial Profiles Derived from CLEAN Model}\label{appendix:utility_clmodel}

The CLEAN algorithm generates a model consisting of a group of point sources (or multi-scale Gaussian distributions) that fit the observed visibility data. This ``CLEAN Model'' image has, by definition, a lot of artificial bumpy structures and does not resemble an actual disk. Therefore, it is smoothed with a beam to obtain a final ``CLEAN Image''. However, this smoothing process changes the visibility of the image, resulting in a visibility profile that does not match the observations. Here, we note that it is possible to consider azimuthal averaging as another way of ``smoothing''. The ``beam'' in this case is extended all over the azimuth while we do not convolve the image in the radial direction. The azimuthal average in the image therefore does not alter the radial visibility profile so it is possible to obtain the radial profile that matches with observed visibility from the CLEAN model.

The radial profiles produced by the CLEAN model are presented in Figures \ref{fig:radialprofile_1} and \ref{fig:radialprofile_2}. In many cases, we see that the radial profiles produced by CLEAN models and SpM match each other. We see better matches in the outer part of the disks because the larger area is covered when averaging, making it possible to get rid of artificial bumpy structures. Additionally, the locations of gaps and rings in the outer region of large disks are consistent with those obtained from the SpM images, while the widths and depths of the gaps are not always consistent. Therefore, the radial profiles obtained from the CLEAN model have the potential to constrain the structures in the outer part of the disks and identify substructure candidates.

\section{Evaluation of Sparse Modeling with Three Disks}\label{appendix:eval_spm}

\begin{table*}
\centering %
\tbl{Disks Sample and CLEAN/SpM Image Properties.
\label{tab:appendix_spmclean}}{%
\renewcommand{\arraystretch}{1.2}
\begin{tabular}{l@{\hspace{0.6cm}}r@{\hspace{0.5cm}}r@{\hspace{0.5cm}}r@{\hspace{0.5cm}}r@{\hspace{0.5cm}}r@{\hspace{0.5cm}}r@{\hspace{0.6cm}}c}
\hline\noalign{\vskip3pt} 
Name & Freq & $D_{\rm max}$ & $\theta_{\rm cl}$ & $\theta_{\rm eff}$ & CLEAN $I_{\rm peak},~\sigma_{\rm cl}$ & SpM $I_{\rm peak},~I_{\rm DT}$ & $\log(\Lambda_l, ~ \Lambda_{tsv})$ \\
Data & (GHz) & (m) & (mas, PA) & (mas, PA) & ($\rm mJy~beam^{-1}$) & ($\rm mJy~asec^{-2}$)  & \\
\hline
\multicolumn{8}{l}{RY Tau} \\ 
Data 1 &  225 &  3638 & $150 \times 106 (-11.4^{\circ})$ & $50 \times 40(-9.0^{\circ})$ &   19.51, 0.05 &  1548.4, 72.1 &   (5,10) \\
Data 2 &  225 & 16196 & $51  \times 30 (20.5^{\circ})$   &  $\cdot$                     &   2.75,  0.04 & $\cdot$       &   $\cdot$ \\
\hline
\multicolumn{8}{l}{DG Tau} \\ 
Data 1 &  245 & 3697 &  $129 \times 101 (-7.2^{\circ})$  &  $50 \times 40(-5.2^{\circ})$&   44.84, 0.04 & 5269.9, 46.3  &  (5,11) \\
Data 2 &  237 & 13815 & $36 \times 27(-5.7^{\circ})$     &   $\cdot$                    &   6.44, 0.01  & $\cdot$       & $\cdot$ \\
\hline
\multicolumn{8}{l}{CI Tau} \\ 
Data 1 & 225 & 3013 &  $150 \times 128  (4.9^{\circ})$   &   $130 \times 110 (5.2^{\circ})$ &  6.90, 0.03 & 397.5, 8.0  &  (5,13) \\
Data 2 & 230 & 11931 &  $58 \times 42 (26.2^{\circ})$    &   $\cdot$                    &   2.53, 0.01    & $\cdot$     &   $\cdot$ \\
\hline\noalign{\vskip3pt} 
\end{tabular}}
\begin{tabnote}
\textbf{Note.} Column description: (1) Name of the host star and data set. (2) observed frequency. (3) maximum baseline length $D_{\rm max}$ of observed visibilities. (4) CLEAN beam size $\theta_{\rm cl}$. $\tt{Brigss robust}$ parameter is set to be 0.5. (5) SpM beam size $\theta_{\rm eff}$. The beam size (effective spatial resolution) of each image is taken from a point source simulation. (6) RMS noise $\sigma_{\rm cl}$ of the CLEAN image. The value is derived from the emission-free area. (7) Detection threshold $I_{\rm DT}$ of the SpM image. The value is derived from the peak value of the emission-free area. (8, 9) Total flux $F_{\rm nu }$ of each CLEAN and SpM image. The flux is taken from the total value above $5\sigma_{\rm cl}$ for CLEAN or $I_{\rm DT}$ for SpM. (10) Two regularization parameters ($\log\Lambda_l$, $\log\Lambda_{tsv}$) adopted for each SpM image.
\end{tabnote}
\begin{tabnote}
ALMA project IDs.\\
RY Tau Data 1: 2016.1.01164.S. Data2: 2016.1.01164.S, 2017.1.01460.S. \\ 
DG Tau Data 1: 2016.1.00846.S. Data 2: 2015.1.01268.S, 2016.1.00846.S. \\
CI Tau Data 1: 2016.1.01164.S, and 2018.1.01631.S. Data 2: 2016.1.01164.S, 2016.1.01370.S, and 2018.1.01631.S.
\end{tabnote}
\end{table*}

\begin{figure*}[!ht]
\begin{center}
\includegraphics[width = 1. \textwidth]{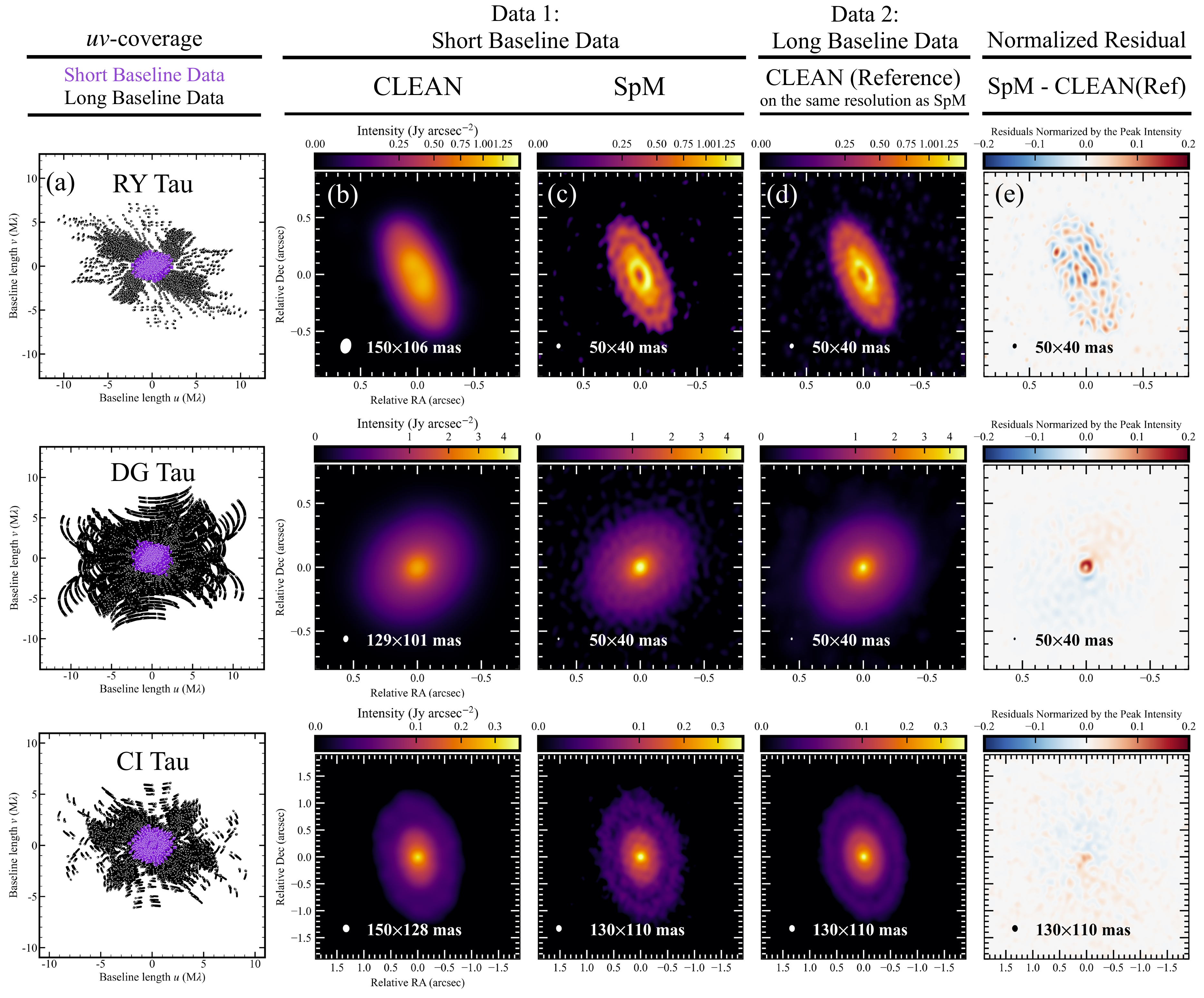}
\end{center}
\caption{Images of three disks (top: RY Tau, middle: DG Tau, and bottom: CI Tau). The same color scales are used. (a) the $uv$-coverage of the short baseline data (Data 1, purple) and the long baseline data (Data 2, black). (b) the CLEAN images of Data 1 with $\tt Briggs robust$ parameter of 0.5. (c) the SpM images of Data 1. (d) the CLEAN images of Data 2 (reference image) with $\tt Briggs robust$ parameter of 0.5. The Data 2 CLEAN model is finally convolved with the beam which is the same as the effective spatial resolution of the Data 1 SpM image. The filled white ellipses denote the spatial resolution of Data 1 SpM and Data 2 CLEAN in the bottom left corner. (e) the normalized residual map ((c) - (d)). The color scale is normalized by the peak intensity of the residual map. The total flux of the Data 1 SpM and CLEAN images are scaled to that of the Data 2 CLEAN image to minimize the effects of flux-calibration errors.}
\label{fig:spm_vs_clean_imagedomain}
\end{figure*}

\begin{figure}[!htbp]
\begin{center}
\includegraphics[width = 0.48 \textwidth]{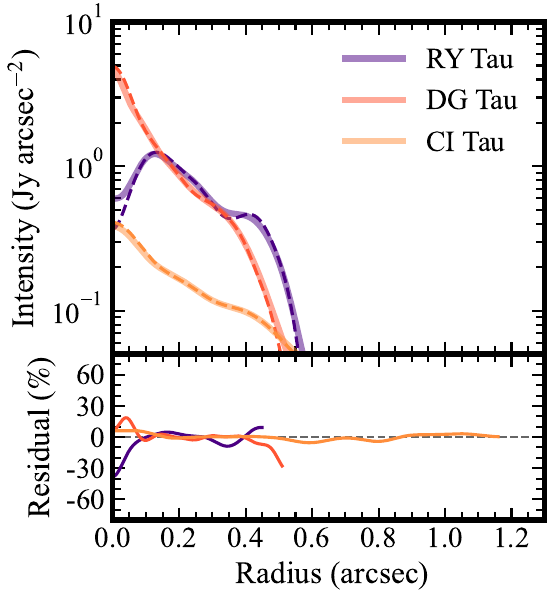}
\end{center}
\caption{Top: deprojected and azimuthally averaged radial intensity profiles of RY Tau (purple), DG Tau (orange), and CI Tau (yellow). The solid lines are the profiles obtained by the Data 2 CLEAN image (reference image), while the dashed lines are those obtained by the Data 1 SpM image. Bottom: the residual profiles between the radial profiles of Data 1 SpM image and Data 2 CLEAN image. The percentage of residual scale is calculated as (Data 1 - Data 2)/Data 2. The total flux density of the Data 1 SpM image is scaled to that of the Data 2 CLEAN image to minimize the effects of flux-calibration errors.}
\label{fig:spm_performance_radialintproflie}
\end{figure}

In this section, we evaluate the validity of SpM super-resolution imaging. Following the methods in \cite{Yamaguchi2020}, we use two data sets of short and long baselines on the same object.  We compare the SpM image of the short baseline data and the CLEAN image of the long baseline data to investigate whether the substructures in the SpM image are real. For this purpose, we use data from three objects: RY Tau, DG Tau, and CI Tau (see Table \ref{tab:appendix_spmclean}). We call the short baseline data set ``Data 1'' and the long baseline data set ``Data 2''.

The left panels of Figure \ref{fig:spm_vs_clean_imagedomain} show the $uv-$coverages of each data set, and the maximum baseline lengths between Data 1 and Data 2 differ by a factor of $\sim 3-4$ for all objects. The middle panels of Figure \ref{fig:spm_vs_clean_imagedomain} show the images obtained with SpM and CLEAN. The total flux density of the Data 1 SpM image is scaled to that of the Data 2 CLEAN image to minimize the effects of errors in the flux calibration. The spatial resolutions of Data 1 SpM are $35\%$, $39\%$, and $86\%$ of the CLEAN beam sizes for RY Tau, DG Tau, and CI Tau, respectively. We have verified that these spatial resolution improvement factors are consistent with the SNR at long baselines, as discussed in Section \ref{sec:spatial_resolution}. In the case of RY Tau and CI Tau disks, the Data 1 SpM images successfully reveal substructures such as gaps, inner holes, and asymmetries that appear in the Data 2 CLEAN images. We also note that SpM does not create artificial gaps or rings, as shown in the DG Tau disk, which had no gap.

Figure \ref{fig:spm_performance_radialintproflie} shows the azimuthally averaged radial intensity profiles for each image. The profiles of the Data 1 SpM image and the Data 2 CLEAN image agree within $\sim 10\%$, except in the inner cavity of the RY Tau disk, where the difference reaches $\sim 30\%$. This is because the Data 2 CLEAN image of RY Tau shows a faint inner disk at 30 mas spatial resolution, which is not present in Data 1 due to the lack of long baseline data.

\section{Comparison with Frankenstein approach
}\label{appendix:comparison_other_imaging}

\cite{Jennings2022taurus} present non-parametric one-dimensional (1D) fitting by the software Frankenstein ($\tt frank$) to the azimuthally average visibility data to derive the radial intensity profiles for 24 disks around single stars. Part of the targets are from the Taurus survey by \citet{Long2019} and overlap with ours. The $\tt frank$ approach utilizes a Gaussian process to optimize the model of the deprojected real-part visibility profile and has the potential to reconstruct substructures that are not recognized in standard CLEAN methods \citep{Jennings2020}.

From visual inspection, the published radial visibility and intensity profiles for several disks from their study qualitatively agree with our results. The gap/ring features (DS Tau, DR Tau, MWC 480, and UZ Tau E), inflections (DO Tau, Haro 6-13, IQ Tau, and V409 Tau), isolated rings (CIDA 9A and IP Tau), and inner dips (BP Tau) are consistent in many targets. However, there are some inconsistent results.  For instance, the $\tt frank$ approach reveals additional gap structures in the inner region of large disks (CI Tau, DL Tau, GO Tau, FT Tau). Conversely, some gaps and inflections are seen only in our analyses (DQ Tau, GI Tau, HP Tau, and V836 Tau). The different approaches to fitting visibility may result in different results, especially for weak features. We plan to make thorough comparisons between different techniques in a future publication.


\end{document}